	\documentclass[12pt,table]{article}
	\newcommand{\footnotesizelarge}{\fontsize{9pt}{11pt}\selectfont}
	\usepackage[top=1.5in, bottom=1.5in, left= 1.0in, right=1.0in]{geometry}

	\usepackage[margin= 16pt , footnotesize ,  sc , labelfont=bf , labelsep=period , format=plain , justification=centerfirst]{caption}
	\usepackage{floatrow}
		\usepackage{amsfonts}
		\usepackage{amsmath}
		\usepackage{amssymb}
		\usepackage{amsthm}
		\usepackage{array}
		\usepackage{booktabs,longtable}
		\usepackage{eurosym}
		\usepackage{floatrow}
		\usepackage{graphicx}
		\usepackage{hyperref}
		\usepackage{mathrsfs}
		\usepackage{longtable}

		\usepackage{color, colortbl}
		\usepackage[english]{babel}
		\usepackage{multirow}
		\usepackage{lscape}
		\usepackage{pdflscape}
		\usepackage{rotating}
		\usepackage{subfig}
		\usepackage{setspace}
		\usepackage{titlesec}
		\usepackage{verbatim}
		\usepackage{listings}
		\usepackage[pdftex,dvipsnames]{xcolor}
		\usepackage[textsize=tiny]{todonotes}
		\usepackage{natbib}
		\setcitestyle{authoryear,open={(},close={)}}

		\hypersetup{
		    colorlinks,
		    linkcolor={violet!70!black},
		    citecolor={blue!70!black},
		    urlcolor={blue!60!black}
		}

		\DeclareMathOperator{\E}{E}

		\newcommand{\AND}{\qquad \text{and} \qquad}

		\definecolor{orange_dark}   {HTML}{C70039}
		\definecolor{green_dark}    {HTML}{1E8449}
		\definecolor{blue_dark}    {HTML}{000088}

		\definecolor{Uniswap_color}   {HTML}{ff007a}
		\definecolor{Uniswap_v3_color}{HTML}{EE82EE}
		\definecolor{Binance_color}   {HTML}{f3ba2f}
		\definecolor{Kraken_color}    {HTML} {87CEEB}		
		
		\definecolor{Coinbase_color}  {HTML}{1652F0}
		\definecolor{Huobi_color}     {HTML}{EF5F00}
		\definecolor{Okex_color}      {HTML}{111111}

		\definecolor{Uniswap_color_dark}   {HTML}{ff007a}
		\definecolor{Uniswap_v3_color_dark}{HTML}{9C559C}
		\definecolor{Binance_color_dark}   {HTML}{F2AC00}
		\definecolor{Kraken_color_dark}    {HTML}{00A0E1}

		\definecolor{Coinbase_color_dark}  {HTML}{1652F0}
		\definecolor{Huobi_color_dark}     {HTML}{EF5F00}
		\definecolor{Okex_color_dark}      {HTML}{111111}

	\usepackage{tikz}
	\usetikzlibrary{decorations.pathreplacing}
		\definecolor{cinnabar}{rgb}{0.89, 0.26, 0.2}
		\definecolor{ceruleanblue}{rgb}{0.16, 0.32, 0.75}

	\usepackage{tabularx, booktabs}
	\newcolumntype{Y}{>{\centering\arraybackslash}X}

		\makeatletter
		
		\newcommand{\Rmnum}[1]{\expandafter\@slowromancap\romannumeral #1@}
		\makeatother

		\renewcommand{\thesubsection}{\thesection.\Alph{subsection}}

		\makeatletter
		\renewcommand{\baselinestretch}{1.10}

		\titleformat{\section}{\centering\large\bfseries}{\thesection.}{1em}{}
		\titleformat{\subsection}{\normalsize\itshape}{\thesubsection.}{1em}{}
		\titleformat{\subsubsection}{\normalsize\itshape}{\thesubsubsection.}{1em}{}

	\newcommand{\dx}{\Delta x}
	\newcommand{\dy}{\Delta y}
	
	\newcommand{\DP}{\Delta P}

	\newcommand{\ddx}{\varphi \Delta x}

	\newcommand{\figA}{Figure  \ref{fig:TC_all_coins} }
	\newcommand{\tabA}{Table   \ref{tab:TCs_all_sz} }

	\newcommand{\tabAc}{Table   \ref{tab:TCs_all_sz}}

	\newcommand{\CEXs}{Binance, Kraken, Coinbase, Huobi, and OKX}

	\makeatletter
	\pdfpageheight\paperheight
	\pdfpagewidth\paperwidth
	\usepackage{csquotes}
	\usepackage{subfig}
	\usepackage{ragged2e}
	\usepackage{silence}
		\newcommand{\figurecap}[4]{ 
			\begin{figure}[t]
				\centering
				\begin{minipage}[c]{1\textwidth}
					\begin{centering}
						\vspace{-1em}
						\includegraphics[width=#3\linewidth]{IMG/#2.pdf}
						\captionsetup{justification=justified, singlelinecheck=true, font=footnotesize} 
						\caption{\textbf{#1.}\footnotesizelarge #4}
						\label{fig:#2}
					\end{centering}
				\end{minipage}
			\end{figure}
		}

		\newcommand{\figurecaph}[4]{ 
			\begin{figure}[h]
				\centering
				\begin{minipage}[c]{1\textwidth}
					\begin{centering}
						\vspace{-1em}
						\includegraphics[width=#3\linewidth]{IMG/#2.pdf}
						\captionsetup{justification=justified, singlelinecheck=true, font=footnotesize} 
						\caption{\textbf{#1.}\footnotesizelarge #4}
						\label{fig:#2}
					\end{centering}
				\end{minipage}
			\end{figure}
		}

\begin{document}


		\title{
			\Large{ \bf On The Quality Of Cryptocurrency Markets }
			\\\vspace{0.3em}
			\large{ \bf Centralized Versus Decentralized Exchanges }
		}

		\date{
			September 2024
		}
		\author{Andrea Barbon, Angelo Ranaldo\thanks{\fontsize{8}{8}\selectfont 
			Andrea Barbon (\href{mailto:andrea.barbon@unisg.ch}{andrea.barbon@unisg.ch}) is an Assistant Professor of Finance at the Swiss Finance Institute, University of St. Gallen, Switzerland.
			Angelo Ranaldo (\href{mailto:angelo.ranaldo@unibas.ch}{angelo.ranaldo@unibas.ch}) is Professor of Finance and Financial Economics at the University of Basel, Swiss Finance Institute.
			Andrea Barbon is the corresponding author.
			We thank Agostino Capponi, Alfred Lehar, Thomas Moser, Engin Iyidogan, Laurence Daures, Igor Kozhanov, Peter O'Neill, Olga Klein,
			and participants of 
			the SNB-CIF Conference on Cryptoassets and Financial Innovation,
			the Central Bank Research Association (CEBRA) Annual Meeting,
			SNB Technology and Finance Webinars,
			the Bank of Canada finance seminars,
			the World Federation of Exchanges Webinars,
			the FiRe Research Seminars,
			the 13th Annual Hedge Fund Research Conference,
			the EFT Conference by the University of Edinburgh Business School
			for advice and stimulating discussions.
			The authors thank Charles Milliet for his excellent research assistance. Angelo Ranaldo acknowledges financial support from the Swiss National Science Foundation (SNSF grant 204721).
			All errors are our own.
			}
		}
		\maketitle
		\vspace{-1cm}
		\begin{abstract}
			We analyze the market quality of centralized crypto exchanges (CEXs) and decentralized blockchain-based venues (DEXs) using a unique and comprehensive dataset. 
			Focusing on two fundamental aspects, transaction costs and deviations from the no-arbitrage condition, we estimate the causal effect of ``gas fees'' on DEX market quality.
			We show that these fixed costs impose a significant burden on relatively small trades and cause persistent arbitrage deviations.
			Conversely, DEXs offer more competitive transaction costs for larger trades, offering a more favorable environment for institutional investors. 
			Furthermore, we provide causal evidence that innovations aimed at enhancing the flexibility of liquidity provision in DEX markets lead to sizeable improvements in market quality.
		\end{abstract}
		\newpage



	\section{Introduction}\label{sec:introduction}
        A pivotal question in finance concerns the creation of a high-quality market, characterized by liquidity and price efficiency. 
        The seminal work by \cite{glosten1994} offers profound insights into this matter, highlighting the merits of centralized electronic limit order books (LOBs) and predicting their broad adoption. 
        Over the past three decades, this forecast has materialized, as numerous asset classes are now traded on centralized exchanges (CEXs) employing electronic LOBs to match end-user orders transparently, efficiently, and centrally.
        Glosten's forecast also appears pertinent to emerging financial instruments like crypto\-currencies, with LOB markets extensively adopted for \emph{off-chain} trading on CEXs.\footnote{
        	Throughout the paper, ``on-chain'' refers to transactions that occur directly on the blockchain and are recorded on its ledger, while ``off-chain'' denotes trades on centralized platforms that are not immediately recorded on the blockchain.} 

        Nonetheless, driven by the surge in innovations enabled by blockchain technology, decentralized exchanges (DEXs) have surfaced as an alternative market structure for crypto assets.
		These platforms leverage smart contracts to facilitate \emph{on-chain} trading,\footnote{
			We refer to LOB-based centralized exchanges as CEXs, while we use ``DEXs'' for decentralized exchanges based on the \emph{automated market maker} (AMM) system. 
			%
			%
		}
		marking a significant shift from the traditional LOB mechanism. 
		DEXs have  garnered considerable attention and seen a surge in trading activity in recent years, surpassing CEXs in trading volume for exchange pairs listed on both types of platforms, as illustrated by Figure \ref{fig:volumes}.
		This pronounced trend prompts a critical question: 
		do DEXs provide superior market quality compared to their centralized counterparts?
  
        As the bid-ask spread is a key (inverse) indicator of market quality \citep{Bessembinder2010}, we begin by empirically analyzing the spreads required to transact on CEX and DEX markets, using a granular dataset covering the major centralized and decentralized crypto exchanges.\footnote{
        	Our list of exchanges is comprehensive, covering $70\%$ of the volume traded on CEXs and roughly $60\%$ on DEXs. 
        	The sample spans from March 2021 to February 2023 and includes all trading pairs across these exchanges.} For CEXs, we accurately measure the quoted half-spread plus the fees charged by the exchange. 
        For DEXs, we take the sum of the quoted half-spread, the transaction fees charged by the protocol, and the ``gas fees'', that is, the costs associated with performing on-chain transactions.\footnote{
			The gas fee attached to a blockchain transaction is the product of the required ``gas units'' (proportional to the computational complexity of the transaction) and the user-specified ``gas price'', which affects execution priority.
		}
        This additional component is expected to weigh on DEX transaction costs, potentially diminishing overall market quality.
        However, since gas fees are fixed costs, their relative impact should decrease when executing larger-sized trades.
        A frequency analysis of transaction sizes supports this, revealing a statistically significant prevalence of larger orders on DEX platforms.
        These observations lead to our first testable hypothesis: DEXs offer lower transaction costs, but only for sufficiently large trades.
        Our estimation of transaction costs clearly shows that it is less expensive to trade small and medium amounts on CEXs. 
        On the contrary, DEXs become more cost-effective for relatively large amounts (above \$10,000). 
        These findings strongly support our first hypothesis and highlight the crucial impact of gas fees on DEX markets, which are effectively diluted only when trading larger amounts. 
        Consequently, DEXs can offer superior liquidity in environments with low gas prices and, by the same token, more competitive transaction costs for large traders such as institutional investors.
        On the other hand, CEXs tend to be more attractive to investors executing smaller transactions, such as retail traders and arbitrageurs.
		%
		%

        The empirical support for the first hypothesis motivates our second hypothesis, which posits that CEX markets should feature higher price efficiency, enjoying smaller deviations from no-arbitrage conditions. 
        The rationale for this conjecture is that 
        arbitrage opportunities typically involve small dollar amounts to close price discrepancies. 
        However, our analysis indicates that smaller transactions incur higher costs on DEXs, discouraging arbitrage activity and thereby increasing the likelihood of mispricing.
        %
        %
        %
        %
        Specifically, we examine price efficiency by analyzing deviations from the \emph{law of one price} implied by the `triangular'' no-arbitrage condition.\footnote{
			A triangular price deviation is represented by the difference between the price of exchanging one currency for another directly (e.g., buying USDC against ETH) and the ``synthetic'' price that replicates this position by switching from another currency, implying two additional trades (e.g., selling ETH for USDT and then selling the latter to obtain USDC).
		} 
  		
        Our price-efficiency analysis proceeds in two steps. First, we perform panel regressions to determine whether gas fees systematically predict mispricing after controlling for other factors such as volatility and cross-market variables. Our panel regression results reveal a significant positive relationship between gas fees and arbitrage deviations.
        Second, we note that providing evidence that gas fees \emph{cause} price inefficiency is an empirical challenge due to the endogeneity of gas prices. 
        A shock to the level of information asymmetry, for example, may lead to increased trading activity and inflate gas prices due to network congestion. Simultaneously, such a shock may deteriorate market quality due to strategic responses by market makers. Hence, addressing this issue requires identifying the direct impact of gas fees on price efficiency, independent of the confounding effects of asymmetric information.
        We thus use instrumental variable regressions to ascertain whether gas fees \emph{cause} price inefficiency on top of the variation induced by changes in information asymmetry.
		Specifically, we exploit network congestion generated by minted Non Fungible Tokens (NFTs) to construct our instrumental variable.
		The resulting variation in gas prices is not driven by information shocks on the fundamental value of cryptocurrencies. 
		In the first-stage specification, the gas price is regressed on the volume of minted NFTs. 
		The positive and significance coefficient corroborates the relevance of the instrument, supporting the idea that the competition among NFT investors leads to inflated gas prices.
		In the second stage, DEX price deviations are regressed on the predicted values of gas prices. 
		%
        This instrumental variable analysis provides causal evidence that gas fees deteriorate DEX price efficiency, regardless of asymmetric information issues.
      	The magnitude of the causal effect is similar to that implied by the panel estimation, with a standard deviation increase in gas prices associated with an increase in the magnitude of price deviations between $18\%$ and $25\%$.

		Our price-efficiency analysis clearly indicates that CEXs exhibit tighter deviations from no-arbitrage conditions, while those observed on DEXs are an order of magnitude larger. 
		Our findings also provide causal evidence that gas fees  deteriorate price efficiency, which are borne out by our second hypothesis.
		%
  


        Nevertheless, it is important to acknowledge that DEXs represent an emergent technology within the financial ecosystem, 
        and they likely possess potential for further enhancement.
        These platforms have been undergoing updates to their market design, 
        driven by innovations in smart contract architecture and liquidity provision mechanisms.
        We posit that these innovations can have a material impact on the market quality of DEXs.
        The introduction of Uniswap v3 serves as a natural experiment to test this hypothesis, allowing us to analyze the impact of two key features while controlling for gas fees: 
        First, Multiple Fee Tiering (MFT), which allows LPs to choose between earning 1, 5, 30, or 100 basis points for their provided liquidity. 
        Second, the Discretionary Price Range (DPR) enables LPs to designate specific price ranges within which they allocate their liquidity, effectively giving them control over their risk exposure. Although different from the traditional LOB system, MFT and DPR can be considered mechanisms that enhance the flexibility of liquidity provision — a characteristic prevalent in CEXs based on electronic LOBs. These new DEX features, especially DPR, can be seen as step in the direction of an \emph{hybrid} market design that combines the best of the two worlds.
        We perform difference-in-differences analyses around the upgrade of Uniswap from v2 to v3, deployed in May 2021. This setting naturally controls for the effects of gas fees and other confounding factors.
        Our analysis provides causal evidence that the introduction of Uniswap v3 considerably improved market quality both in terms of liquidity and price efficiency. The impact of the upgrade is economically significant, leading to a $58\%$ reduction of transaction costs and a $75\%$ improvement in price efficiency. The cross-sectional heterogeneity in the effect suggests that the new MFT system fostered a reduction of exchange fees, with stablecoin pairs benefiting the most.

		The paper is organized as follows: After the literature survey, Section \ref{sec:AMM} introduces CEX and DEX systems. Section \ref{sec:Data} describes our dataset and preliminary results. Section \ref{sec:TCs} analyzes transaction costs. 
		Section \ref{sec:PE} studies triangular price deviations and the causal role of gas fees.
		Section \ref{sec:v3} provides causal evidence on the positive impact of Uniswap v3 on DEX market quality.
		Section \ref{sec:conclusion} concludes.

    \section{Related Literature}\label{sec:Literature}
		%
       
		The literature examining cryptocurrency market liquidity has, so far, focused either on CEXs or DEXs. Regarding CEXs, for example, \cite{brauneis2021measure} perform a horse-race comparison among low-frequency transactions-based liquidity measures. Other papers study market liquidity using LOB data \citep[see, e.g.,][]{marshall2019bitcoin,borri2018cryptomarket}.\footnote{Although liquidity is not the focus of their study, \cite{brauneis2018price} assess the market efficiency of a set of cryptocurrencies using unit root tests and by computing some liquidity proxies, finding that less liquid cryptocurrencies are less efficient.}

		Regarding DEXs, theoretical studies have primarily focused on constant-function Automated Market Makers (AMMs) like Uniswap v2 \citep[e.g.,][]{angeris2019analysis, capponi2021adoption, evans2020liquidity, evans2021optimal, Hasbrouck2022, lehar2021decentralized, park2021conceptual}.\footnote{For instance, \cite{capponi2021adoption} model the impact on utility for LPs and traders of the curvature of the pricing function on Uniswap, and \cite{park2021conceptual} provides conditions under which "sandwich attacks" (akin to front-running) can be profitable when AMM relies on the constant product rule.} Among them, \cite{aoyagi2021coexisting} examine the conditions for the coexistence of such CEX and DEX exchanges. Unlike the trader's endogenous choice between CEX and DEX trading venues as in \cite{aoyagi2021coexisting}, we empirically assess the market quality of DEX and CEX, thus analyzing the coexistence of DEXs and CEXs in equilibrium.\footnote{To conduct empirical research on the endogenous choice of traders, one would ideally need data revealing the identity of market participants on both CEX and DEX, which, to the best of our knowledge, are inaccessible. This limitation prevents a detailed examination of what features deter participants from simultaneously and efficiently trading between DEX and CEX, including aspects of interoperability. In recent work, \cite{han2021trust} analyze a quasi-natural experiment and establish the causal impact of Uniswap liquidity provision on the trading activity on Binance.} \cite{lehar2021decentralized} compare the price impact of Uniswap v2 and Binance, employing a measure akin to that of \cite{Amihud2002illiquidity}.\footnote{\cite{oneillAMM2022} also examines Uniswap v2 showing that it leans toward efficiency as liquidity flows into more profitable pools, while \cite{fukasawa2022weighted} examine how to hedge against impermanent losses. Impermanent Loss is analyzed in other papers \citep[e.g.,][]{aigner2021uniswap,ChenDelta2022,heimbach2022risks,heimbach2021behavior}.}

		Some more recent work considers Uniswap v3, which allows LPs to use MFT and DPR \citep[e.g.,][]{Adams2023, heimbach2022risks,LPZ}. For instance, \cite{Adams2023} analyze Uniswap v3 trades for USDC-ETH and PEPE-ETH, finding that gas costs are dominating for small trades. In addition to this, we offer in-depth comparative analyses between various CEXs and DEXs. \cite{heimbach2022risks} propose a theoretical framework for the decision-making complexity of liquidity providers in Uniswap v3, faced with the trade-off between returns and risks. \cite{LPZ} document fragmented liquidity in the sense that large (small) LPs prevail in high-fee (low-fee) pools. We extend their findings by showing that Uniswap v3 actually improves market quality by lowering transaction costs and increasing price efficiency. From this perspective, our empirical findings support the theoretical model of \cite{Hasbrouck2022}, which posits that the ability of LPs to charge higher fees results in lower transaction costs in equilibrium.
		Further, our findings on the causal role of gas fees complement those of \cite{caparros2023blockchain}, who provide causal evidence that gas fees constitute a friction harming the ability of LPs to reposition their liquidity.

		Concerning price efficiency, prior research focuses on Bitcoin and provides evidence against it. For instance, \cite{makarov2020trading} find large and persistent arbitrage deviations, and \cite{kruckeberg2020decentralized} also provide evidence of arbitrage spreads concentrating during certain periods.\footnote{Other papers focusing on Bitcoin include \cite{urquhart2016inefficiency}, \cite{bariviera2017inefficiency}, and \cite{nadarajah2017inefficiency}). \cite{nadarajah2017inefficiency} explore a large set of cryptocurrencies documenting wide price variation. \cite{dyhrberg2018investible} assess whether and when Bitcoin is investible and at what trading costs.} \cite{hautsch2018limits} stress that consensus protocols generate settlement latency, exposing arbitrageurs to price risk.

		We contribute to the literature by jointly studying transaction costs and price efficiency in CEXs and DEXs, utilizing comprehensive and granular data. In terms of market liquidity, we examine all the main transaction cost components, thus going beyond estimating price impact, and highlighting the crucial role of gas fees. 
		Importantly, our difference-in-differences analysis of the introduction of Uniswap v3 provides evidence that more flexible liquidity provision significantly reduces transaction costs on DEXs. 
		Regarding price efficiency, we analyze violations of the triangular no-arbitrage condition and identify factors that explain price inefficiency. Using instrumental variable regressions, we provide causal evidence that gas fees lead to mispricing, suggesting that gas fees adversely impact DEX price efficiency independently of private information on the fundamental value of cryptocurrencies. The overarching message of our study is that the DEX system represents a viable and competitive microstructure, despite gas fees being the principal friction for its market quality advancement.


	\section{CEX and DEX Markets}\label{sec:AMM}
		Contemporary financial markets primarily employ a central LOB system, wherein a central institution records buy and sell orders, with market prices set by the latest matched orders. The main advantage of the LOB system is the ability to provide a transparent and efficient price discovery process and liquidity clustering even in extreme situations \citep{glosten1994}.
		However, implementing a LOB exchange on the blockchain is challenging due to limited throughput -- a crucial resource for order-based exchanges -- and the cost of blockchain transactions (that is, gas fees).       
		Crypto exchanges like Binance or Kraken, based on LOB, are thus forced to operate \emph{off}-chain, sacrificing the benefits of decentralization.
        In contrast, DEXs use the Automated Market Maker (AMM) protocol to establish transaction and market prices based on liquidity pools deployed on-chain and funded by Liquidity Providers (LPs).\footnote{
			It is often asked whether it is possible to envision a LOB-based DEX. 
			Current technical constraints of blockchain technology, namely, limited transaction speed and high gas costs, make an \emph{on-chain} order book unviable. However, future iterations could make this possible, as exemplified by the range orders feature in Uniswap v3, which allows for specific limit orders. 
			Meanwhile, an AMM-based CEX is technically possible but, to the best of our knowledge, remains unexplored by major exchange providers.
		}
		%

        %

        %
        The AMM system relies on a \emph{conservation function}, which automatically regulates transaction prices and enables users to trade without interacting with a third party.\footnote{A trade on a DEX is also referred to as a \emph{swap} transaction.}
		The predominant conservation function, known as \emph{constant product}, mandates that the product of the available liquidity for the two currencies remains constant.
		As an incentive for users to supply liquidity to the pools, LPs earn an \emph{exchange fee} from each executed trade, proportional to the share of liquidity they own.
		 These fees are 30 basis points of the traded amount for Uniswap v2, while are variable for Uniswap v3 as discussed in Section \ref{sub:uni_v3} of the Internet Appendix.
		However, providing liquidity also carries risks. 
		Price divergence between provision and withdrawal can cause economic loss, as LPs receive more of the depreciating asset and less of the appreciating one. 
		The \emph{impermanent loss}, formally introduced in Section \ref{sub:IL} of the Internet Appendix, represents the relative loss compared to holding the two currencies, gross of transaction fee revenues. 
		Such risk is akin to adverse selection faced by market makers in the presence of asymmetric information, where losses occur only when flows generate a permanent price impact.

		We now discuss the main advantages and inconveniences of DEXs compared to LOB-based CEXs, which revolve around the trade-off between retaining control over funds and benefiting from LOB operational efficiency.
		Indeed, an important drawback of CEXs is that traders have to deposit their crypto assets into the exchange. 
		Withdrawals and deposits are costly. The former are subject to fees charged by the exchange, while the latter involve gas costs to submit the blockchain transaction.
		Further, to lower the risk of potential double-spending attacks, exchanges require several block confirmations (12 on Binance and 20 on Kraken) for deposits to be accepted, leading to delays of two to four minutes.
		Moreover, since CEXs are not tightly regulated, another risk that has arisen in informal discussions with crypto-asset investors is that the CEX operators themselves engage in arbitrage activities by leveraging their privileged position and inside information.
        Finally, if exchanges mix their own funds with user funds, this exposes clients to bankruptcy risk.\footnote{
		More generally, CEXs are subject to at least three sources of risk: 
		(i) unauthorized access to crypto wallets by hackers (e.g., the cases of Poly Network and Japan-based Liquid \citep{Ryder2022}); 
		(ii) misappropriation of client funds by CEX managers, see e.g., Thodex and BitConnect \citep{ORX2021}; 
		(iii) inefficient security management, for example, in the case where a coin exchange executive, such as Gerry Cotten of QuadrigaCX, unexpectedly dies, leaving the digital vault locked \citep{Mance2019}.
		}

		On DEXs, on the contrary, the custody of assets remains fully with the user, as no third party is required to execute and settle trades. 
		This benefit arising from the decentralized trust provided by blockchain technology has several important implications. 
		First, users can take full advantage of the censorship-resistant and trustless nature of their crypto assets \citep{pagnotta2018equilibrium}. 
        Second, crypto assets can be used in DeFi protocols to create value.\footnote{For instance, ERC20 tokens can be staked to earn interest, used as a means of payment, posted as collateral in decentralized lending protocols, and can provide access to airdrop events.}
		Third, it neutralizes the risk of hackers attacking the exchange and stealing assets. 
		Fourth, it allows users to save on the fees commonly associated with depositing and withdrawing assets in CEXs. 
		Finally, but very importantly, in DEXs trade and settlement coincide. 

		Another DEX innovation is that users have the option of passive liquidity provision. 
		This renders the market fairer, given that anyone can provide liquidity, including agents with any degree of sophistication and level of endowment \citep{LPZ}, and does not require investing in expensive hardware or developing complex algorithms.
		By contrast, in CEXs market makers are highly specialized, and entry costs are significant in terms of both sophistication and capital.
		Market makers need high-speed computers and state-of-the-art algorithms to update their quotes as quickly as possible and avoid being picked off by high-frequency traders \citep{foucault2017toxic}.\footnote{
			Market quality is a broad concept that includes concepts such as price efficiency, liquidity, and fairness in the sense that each agent has an equal chance of participating and obtaining a market price that reflects the fundamental value of financial security. This study focuses on the first two aspects, but given the aforementioned aspects, one can argue that the DEX setting is fairer.
		} 

		The DEX market design implies that the exchange fees charged to each transaction are distributed to LPs \citep{adams2020uniswap}.
		There is thus no welfare reduction stemming from profits accrued by the exchange itself, which translates into economically significant gains for both traders and LPs.
		
		In DEXs, the market can rapidly adapt and evolve according to participants' needs.
		Users can instantly quote any pair of ERC20 tokens without screening procedures. As a result, new tokens tend to become tradeable sooner on DEXs, while CEX approval processes can be time-consuming.
		Moreover, DEXs may enable trading tokens unavailable on CEXs. This expands investment opportunities, enhances diversification, and accelerates market completeness. However, it also exposes users to potentially malicious assets.

		Finally, since DEX transactions are processed by smart contracts and recorded on the blockchain, users bear the non-trivial cost of gas fees. 
		This also implies that transactions are subject to an execution delay, 
		the duration of which depends on 
		the speed of the underlying blockchain,
		the chosen gas price, and 
		the level of network congestion.
		It is important to highlight, however, that for DEXs execution and settlement coincide. 
		In contrast, trades on CEXs cannot be considered settled as long as the funds are inside the exchange.
		%

	\section{Data and Preliminary Results}\label{sec:Data}
		Since DEXs are based on smart contracts deployed on blockchains, records of every interaction with those contracts are available to the public.  
		This rich dataset includes, as primitives, the creation of exchange pairs, the addition and removal of liquidity by LPs, and swap transactions between two quoted tokens.
		Building on those, one can reconstruct liquidity levels, quoted prices, transaction prices, and trading volume at the pair level at any time.
		We leverage the application programming interface of \href{https://thegraph.com}{TheGraph.com}, 
		an indexing protocol enabling efficient querying of data from blockchains,
		to obtain data for Uniswap v2 from the Ethereum Mainnet.
		Equivalent data regarding Uniswap v3 is collected using custom queries on \href{https://dune.com/hagaetc/dex-metrics}{Dune.com}, a community-driven platform that facilitates querying of public blockchain data.
		For CEXs, by contrast, data are proprietary. We obtain minute-frequency LOB snapshots and Open, High, Low, Close, and Volume data from \href{https://tardis.dev/}{Tardis.dev}, a data provider specialized in cryptocurrencies
		%
		%
  
		In our empirical investigation of market quality, we analyze seven among the most prominent crypto exchanges. 
		On the DEX side we consider Uniswap v2 and Uniswap v3, while our list of CEXs includes \CEXs.  
		For both CEXs and DEXs, our sample period spans from March 2021 to February 2023.\footnote{
			Specifically, Uniswap v2 data ends in February 2022, while Uniswap v3 data is available from its introduction, in May 2021, to February 2023.
		}
		Our study covers all trading pairs at the intersection of the pairs quoted in the seven exchanges and available throughout our sample period.
		When we restrict to Uniswap v2, Uniswap v3, Binance, and Kraken, we extend the analysis to the larger set of currency pairs at the intersection of their quoted pairs\footnote{
			The list of pairs common to all the seven exchanges is: 
			ETH-USDC, ETH-USDT, ETH-BTC, LINK-ETH, BTC-USDC, DAI-ETH, MANA-ETH, and USDC-USDT.
			The larger set of pairs common to the four above-mentioned exchanges is: 
			ETH-USDC, ETH-USDT, ETH-BTC, LINK-ETH, BTC-USDC, DAI-ETH, MANA-ETH, USDC-USDT, 
			DAI-USDT, AAVE-ETH, BAT-ETH, BTC-DAI, CRV-ETH, GRT-ETH, KNC-ETH, REP-ETH, SNX-ETH, STORJ-ETH, UNI-ETH, and OMG-ETH.
        }.
        Our list of exchanges is highly representative, covering more than $70\%$ of the volume traded on CEXs and roughly $60\%$ on DEXs.
        Moreover, our sample includes all pairs at the intersection of these exchanges, enabling us to assess market quality on the maximal set of available currency pairs.
        This extended coverage is instrumental in performing a comparison of transaction costs and price efficiency across exchanges.

        There are two important moments during our sample period that allow us to carry out insightful diff-in-diff analyses. 
        %
        The first event is the launch of Uniswap v3 on May 5th, 2021. 
        This event enables a comparison of market quality between Uniswap v3 and its predecessor, Uniswap v2, using CEX markets as a control group.
        This quasi-natural experiment allows us to test our third hypothesis: whether innovation in decentralized exchange (DEX) market design leads to improved market quality.

        The second event is the collapse, on November 10th, 2022, of FTX, one of the largest LOB-based centralized exchanges for cryptocurrency. 
        The bankruptcy was triggered by a liquidity crisis pertaining to the exchange's proprietary token, FTT, and led to significant market volatility and multibillion-dollar losses. 
		Some evidence suggests the misuse of its clients' assets. Specifically, it has been reported that FTX allocated \$10 billion from its client accounts to bolster Alameda Research, a cryptocurrency trading entity operated by Sam Bankman-Fried, who is also the founder of FTX. 
		Such an action contravenes the stipulated terms of service of FTX, which explicitly maintains that clients' funds should not be appropriated for purposes beyond trading on the FTX platform.
		Furthermore, allegations have emerged contending that FTX employed a Ponzi-like scheme to misappropriate funds, involving the transfer of customer resources among different entities. 
		These claims, if substantiated, highlight a serious breach of fiduciary duty, underscoring the custody risks associated with centralized cryptocurrency exchanges.        
        Hence, this event provides us with an ideal laboratory to analyze the effects of custody risks embedded in LOB-based CEX markets. 

		\subsection{Summary Statistics on Trading Volume}\label{sub:summary_stats}
			The top panel of Figure \ref{fig:volumes} showcases the total volume generated by the twenty pairs included in our sample on CEXs and DEXs, expressed in million USD on a logarithmic scale.
			The bottom panel plots the time-series evolution of the share of total volume generated by DEXs, unveiling a significant upward trend during our sample period.
			DEX volume constitutes roughly 5\% of the total in March 2021, while it climbs above 50\% in the last part of the sample.\footnote{
				While these figures are remarkable, especially given that we are comparing two DEXs with five CEXs,
				we note that the \emph{total} traded volume on CEXs (including all quoted pairs) is between 5 and 10 times larger than the total volume generated by DEXs.
				In this regard, it is important to recall that only \emph{on-chain} assets can be traded on DEXs, while CEXs provide trading pairs involving the US Dollar and other fiat currencies.
			}
			A noticeable peak in the DEX volume share, reaching 70.6\%, is observable in November 2022 around the collapse of the FTX exchange. 
			In Subsection \ref{sub:TCs_additional} we will analyze more formally the impact of such an event on CEX and DEX volume.

            Table \ref{tab:summary_stats} provides summary statistics on the cross-section of pairs analyzed in our study, focusing on the trading activity.
			For each exchange, we display the cross-sectional distribution of the total volume generated during our sample period (Panel A) and the number of transactions (Panel B).
			Three main findings emerge.
				First, the eight \emph{main pairs} quoted at the intersection of all exchanges are on average more actively traded with respect to the larger set of twenty pairs (\emph{all pairs}).
				Second, the exchanges generating the most trading volume are Binance, Uniswap v3, and Huobi.
				Third, for DEXs and especially for Uniswap v3, the ratio between trading volume and number of trades is significantly higher than for CEXs, suggesting that DEX transactions are on average larger in terms of dollar value.
			This is confirmed in Figure \ref{fig:trade_sizes}, presenting the distribution of trade sizes in U.S. Dollars for transactions executed in CEXs and DEXs. 
			The histogram shows a sizeable difference between the two distributions, demonstrating that DEX trades are on average larger in size.
			In fact, 80\% of CEX transactions are smaller than \$ 1,000, versus only 35\% for DEXs,
			while, 33\% of DEX trades are larger than \$ 10,000, versus only 1.7\% for CEXs.
			Further, there are instances of very large trades surpassing $1$ million USD on DEXs, making up for $0.5\%$ of the observations, while there are virtually no such large trades executed on CEXs.
			These findings are economically significant and robust across all exchanges in our sample, in the aggregate and individually.
			Moreover, Table \ref{tab:trade_sizes_by_pair} of the Internet Appendix confirms that this pattern holds at the individual pair level.
			In the next Sections, we will provide evidence that such a difference is mainly driven by gas fees, constituting a fixed cost for DEX transactions and acting as an economically relevant friction for DEXs.

			Finally, Table \ref{tab:vol_pair_exch} presents the daily average trading volume for the pairs in our sample, expressed in millions of USD and broken down by exchange.
			The data show that there is significant heterogeneity across pairs and across exchanges, and that the first eight pairs generate the vast majority of the trading volume.

		\subsection{Gas Fees}\label{sub:gas_price}
			The term \emph{gas} refers to the unit of measure of the computational effort required to execute transactions on the Ethereum network \citep{buterin2013ethereum}.
			The aggregate gas fees for a given block, summing over all transactions, are collected by the miner validating the block.
			To trade on a DEX, as for any on-chain transaction, the user is required to pay a number of gas units proportional to the computational complexity of the transaction.
			The resulting dollar cost is the product between the units of gas used and the \emph{gas price}, which the user can choose to control the priority of execution.
			Miners select, among the set of pending transactions, those to include in the new block, prioritizing the most profitable transactions, that is, those with the highest gas price.
			Wallet interfaces automatically suggest an optimal gas price, depending on the current level of network activity and based on the trade-off between the probability of execution within the next few blocks and the cost.
			While users can edit the gas price according to their preferences, our data reveals that the vast majority of transactions on DEXs are executed at the median gas price of the block. 
			
			%
			Gas costs are undoubtedly important in the study of DEXs since all transactions are recorded \emph{on-chain}.
			An important feature of these costs is that they do not scale with the Dollar value of transactions but, rather, they constitute a fixed cost associated with trading on DEXs.
			We hence argue that gas fees may lead to a significant friction affecting the market quality of DEXs. 
			Our hypothesis is consistent with the distribution of DEX trade sizes presented in Figure \ref{fig:trade_sizes} and discussed in Subsection \ref{sub:summary_stats}, which is much more skewed toward large Dollar values, relative to CEXs. 
			This is consistent with DEX traders placing larger transactions to dilute the fixed cost arising from gas fees.

			%
			%
			%
			%
			%

			%
			Figure \ref{fig:gas_fees_uniswap} plots the evolution of the gas fees required to execute a swap transaction on Uniswap, equal to the product between gas units and gas price, in USD.
			The figure is based on an estimate of the gas units required for a swap on Uniswap v2, constant over time, and an estimate of the prevailing gas price on each hour.
			We estimate the gas units required for each pair by sampling swap transactions every 100 blocks over the entire sample period using a local Ethereum node and Infura APIs, resulting in roughly 330,000 transactions.
			While, in principle, those could vary across specific trading pairs depending on the implementation of the ERC-20 contracts for the two tokens, we empirically find only minor variations across pairs.
			%
			%
			We separately repeat the estimation for swaps performed on Uniswap v3, based on roughly 200,000 transactions, finding that the required gas amounts are also homogeneous across pairs, and roughly $10\%$ more costly with respect to Uniswap v2.
			Throughout the paper, we use the median values of 118,340 gas units for Uniswap v2 and 130,889 for Uniswap v3.
			Finally, we estimate the gas price attached to swaps using the above-mentioned set of Uniswap v2 and Uniswap v3 swap transactions by taking the median gas price on an hourly basis.\footnote{
				In a previous version, we used hourly mean values of gas prices across all transactions, even those unrelated to Uniswap. 
				While the new conditional estimate is more precise and slightly lower, the difference is immaterial.
			}
			Since the amount of required gas for such an operation is constant over time, the observed time-series variation arises from two factors: 
			(i) the price of ETH relative to the USD and 
			(ii) the prevailing gas price among Uniswap transactions, which may be influenced by network congestion.
			%
			The series presents substantial variability, ranging from less than $10\$$ in the first part and last part of the sample to $300\$$ between April and May 2022.\footnote{
				The exceptional gas price observed between April 30th and May 1st, 2022, was likely caused by the NFT drop of the ``Otherside''. 
				The highly awaited launch of the collection by ``Yuga Labs'', the company behind Bored Ape Yacht Club and ApeCoin, resulted in more than \$150 million spent on gas fees and a network-wide increase in gas prices.
			}\label{foot:NFT_drop}
			The two vertical lines mark the period from March 2021 to February 2023, in which our primary analysis of market quality is conducted.

	\section{Transaction Costs}\label{sec:TCs}
		Now that we possess a comprehensive understanding of the operational mechanisms of CEXs and DEXs, and have access to the relevant data, we can proceed to analyze the first key aspect of market quality - market liquidity.
		A common definition of market liquidity is the ease with which an asset can be traded at a price close to its consensus value \citep{foucault2013market}.
		We employ a widely accepted measure of market illiquidity, that is, the effective transaction cost associated with a single trade, expressed as a percentage of the traded amount.
		Transaction costs account for both the price impact associated with a given trade size and any kind of commissions charged by the protocol or the exchange.
		Due to their fundamentally different mechanics, CEX and DEX transaction costs are modeled using distinct methodologies.
		Nevertheless, the two measures are comparable and based on the same conceptual framework, as they are meant to capture the effective costs incurred by a trader transacting a given amount (in US Dollar terms), including slippage, fees, and settlement costs.

		Empirically, we measure transaction costs $TC_{XY}(\Delta x)$ for a trade $X\leftrightarrow Y$ at an hourly frequency for the twenty exchange pairs in our sample and different trade sizes $\Delta x \in (10^3, 10^4, 10^5, 10^6)$ expressed in USD, separately for DEXs and CEXs.

		\subsection{CEX Transaction Costs}\label{sub:TCs_CEX}
			For CEXs based on LOB, we measure transaction costs of a market order of a given dollar amount by considering two distinct components: 
			(i) the Bid/Ask spread implied by the depth of the LOB, 
			(ii) the exchange fees charged by the platform (taker fees).
			%
			%
			%
			%
			%
			%

			We start with the Bid/Ask spread associated with a market order which, since we observe the full depth of ask and bid quotes present in the order book at any point in time, can be computed directly using the volume-weighted bid and ask prices.
			More specifically, we define the volume-weighted bid price $B_{XY}$ for a sell order of size $\dx$ as
			$$ B_{XY}(\dx) = \frac{\sum_{i} v_i b_i}{\dx} \qquad \text{such that} \qquad \sum_{i} v_i = \dx\;, $$
			where $v_i$ and $b_i$ represent the volume and the price of each filled bid limit order $i$, respectively.
			The volume-weighted ask price $A_{XY}$ for a buy order of size $\dx$ is defined symmetrically as
			$$ A_{XY}(\dx) = \frac{\sum_{j} v_j a_j}{\dx} \qquad \text{such that} \qquad \sum_{j} v_j = \dx\;, $$
			where $v_j$ and $a_j$ represent the volume and the price of each filled ask limit order $j$, respectively.
			We then define the Bid/Ask spread\footnote{Note that our definition agrees with the standard ``volume-weighted quoted half-spread".} as
			\begin{equation}\label{eq:spread_LOB} 
				S_{XY}(\dx) = \frac{A_{XY}(\dx)-B_{XY}(\dx)}{A_{XY}(\dx)+B_{XY}(\dx)} \;.	
			\end{equation}
			%
			%
			%
			%
			%
			%
			%
			%
			%
			%
			Finally, we add the fees $f$ charged by the exchange, thus getting
			\begin{equation}\label{eq:TC_LOB} 
				TC_{XY}(\dx) = S_{XY}(\dx)\, +\, f.
			\end{equation}
			Notice that the first term in the above expression is time-varying and depends on the trade size, while exchange fees are constant at the exchange level.\footnote{
				CEXs may periodically revise their withdrawal fees. Using the \href{https://archive.org/web/}{WayBackMachine}, we reconstruct the fee time series for Binance and Kraken based on the available snapshots. Although this approach isn't flawless, given the infrequent updates to these fees, we contend that our methodology is sufficiently accurate.
				A similar examination of exchange fees reveals that they remained constant throughout our sample period on the CEXs covered by our sample.
			}

		\subsection{DEX Transaction Costs}\label{sub:TCs_DEX}
			For DEXs based on AMM, we measure transaction costs of a trade of a given dollar amount by considering three distinct components: 
			(i)   the Bid/Ask spread implied by the depth of the liquidity pools; 
			(ii)  the exchange fees charged by the exchange;
			(iii) the gas fees paid to submit the blockchain transaction.
			%
			%
			%
			As previously explained, the dollar value of gas fees depends on the computational complexity of the smart-contract function being called, the execution priority chosen by the trader, and the prevailing gas price at the execution time.
			For our purposes, we are interested in the gas required to execute a swap transaction; For instance, in Uniswap v2, this involves invoking the \texttt{swapExactTokensForTokens} function of the relevant router contract.\footnote{
				Depending on the nature of the token, the exact router function may be different. 
				For instance, for tokens featuring fee re-distribution like \href{https://bscscan.com/address/0x8076c74c5e3f5852037f31ff0093eeb8c8add8d3\#code}{SafeMoon}, the \texttt{swapExactTokensForTokensSupportingFeeOnTransferTokens} function must be used.
				Nevertheless, the amount of gas required is not significantly different.
			}
			We assume that the quantity of gas required to execute a swap transaction is constant across all currency pairs at $\Gamma = $ 118,340 gas units for Uniswap v2 and 130,889 for Uniswap v3, as discussed in Section \ref{sec:Data}.\footnote{
				Note that these figures are significantly larger than the gas required by a simple \emph{transfer} function on an ERC20 contract which costs 65,000 gas units, or a transfer of ETH (the native currency of the Ethereum blockchain), which costs 21,000 gas units.
			}
			We then approximate the gas cost $g$ of a swap during each hour of our sample period multiplying $\Gamma$ by the median gas price paid across the Uniswap v2 and Uniswap v3 transactions recorded during that hour, in dollar terms.

			The transaction costs for Uniswap v2 are thus computed as the sum of the Bid/Ask spread $S_{XY}$ (defined in equation \eqref{eq:spread_AMM} of the internet Appendix) averaged across both directions $X\to Y$ and $Y\to X$, 
			the constant exchange fee $f=30$ bps, 
			and the gas fee $g$ as a fraction of the trade size.
			
			When it comes to Uniswap v3 we apply a similar methodology, even though the calculation differs slightly due to the MFT system and the existence of multiple pools for any given exchange pair. 
			Specifically, we compute transaction costs for each pool separately, using the Bid/Ask spread derived in equation \eqref{eq:spread_V3} of the Internet Appendix, and considering the exchange fees associated with the pool. 
			For each trade size and hour, we then identify the most cost-effective pool for executing the trade, that is, the one resulting in the lowest overall transaction costs, including gas fees.
			Our approach considers that traders can effortlessly select the most advantageous pool, since the Uniswap v3 interface, through its router contract, automatically recommends it.\footnote{
				In fact, the Uniswap v3 router allows for the trade to be split across multiple pools, in case routing would result in the lowest possible transaction costs. Since we force a trade to be fully executed in a single pool, our measure should be considered as an upper bound for the real transaction costs.
			}
			Note that, as the optimal pool depends on both the offered fees and the available liquidity, it can vary across different trade sizes and over time.
			
			To summarize, DEX transaction costs for a trade of size $\dx$ are defined as
			\begin{equation}\label{eq:TC_AMM}
				TC_{XY}(\dx) = S_{XY}(\dx) + f + \frac{g}{\dx},
			\end{equation}
			where $S_{XY}$ is the spread, $f$ the exchange fees, and $g$ the gas fees associated to the trade. 
			The expression for $S_{XY}$, derived in the Internet Appendix, is described by equation \eqref{eq:spread_AMM} for Uniswap v2 and equations \eqref{eq:spread_V3_pre} and \eqref{eq:spread_V3} for Uniswap v3.
			The first and last terms in the above expression are time-varying for both DEXs.
			The exchange fees $f$ are constant for Uniswap v2, while in Uniswap v3 they depend on the optimal pool selected, as discussed above.

		\subsection{Main Results on Transaction Costs}
			Our analysis begins by examining the average transaction costs (TCs) across different trade sizes on DEXs and CEXs.
			The results, presented in \figA, illustrate log TCs calculated at an hourly frequency using equations \eqref{eq:TC_LOB} and \eqref{eq:TC_AMM}, and then averaged over the entire sample period.
			The plot reports detailed results for the eight \emph{main pairs}, that is, those in the intersection of all seven exchanges, and the average across the twenty pairs quoted only in the first four exchanges.
			Additionally, \tabA offers a detailed breakdown of the total TCs into their three components (bid-ask spread, gas fees, and exchange fees) across different trade sizes.

		    Two main findings stand out.
		    First, 
		    DEXs offer lower TCs for smaller transactions between \$1,000 and \$10,000, 
		    while Uniswap v3 offers more competitive TCs for larger trades between \$100,000 and \$1,000,000.
		    This holds on average, but also for virtually every single pair, with the exception of ETH-USDT, where Binance offers slightly lower TCs.
		    This pattern aligns with expectations as, according to \tabAc, gas fees on DEXs represent a substantial proportion of the traded amount for smaller transactions. 
		    On the one hand, that TCs offered by DEXs are decreasing in trade size is a natural consequence of the fixed-cost nature of gas fees.
		    On the other hand, the fact that Uniswap v3 is performing better than all CEXs for large transactions is a remarkable and non-trivial finding.\footnote{
				At first sight, our results are significantly different from those of \cite{lehar2021decentralized}, who estimate price impact for the pair USDC/ETH on Uniswap v2 to be below 0.1 bps, order of magnitudes lower than our estimates for transaction costs.
				There is no contradiction, though, as these differences arise from the use of distinct methodologies.
				First, they employ a proxy for price impact similar to the illiquidity ratio of \cite{Amihud2002illiquidity}. 
				On the contrary, we do not rely on a proxy, but we measure price impact directly based on liquidity pools and LOB data. 
				Moreover, our measure of transaction costs goes beyond price impact, as it also includes exchange fees and gas fees.
				Finally, their proxy is computed from observed transactions in their sample, averaging across different transaction sizes. 
				Conversely, we estimate transaction costs as a function of trade size, reporting results for a specified range of amounts.
			}

			The findings indicate that modern DEXs, such as Uniswap v3, may deliver superior liquidity compared to CEXs, particularly when the fixed costs associated with blockchain transactions are sufficiently diluted. 
			These results provide strong support for our first hypothesis, namely, that DEXs offer lower transaction costs only for transactions of a sufficiently large size.
			Further, our analysis suggests that future advancements in blockchain technology leading to lower gas costs, for instance through Layer 2 solutions, could make DEXs the best option also for smaller transactions.
		    Our findings complement those of \cite{caparros2023blockchain} and \cite{LPZ}, who provide evidence on the frictions imposed by gas fees on the ability of LPs to rebalance their positions.

		    Second,
		    the dominance of Uniswap v3 for large-sized trades is particularly strong for the stablecoin pair USDC-USDT, reaching a remarkably low rate of $2$ basis points for a \$1 million transaction.\footnote{
		    	The same holds true for the stablecoin pain DAI-USDT, even if not directly reported in the Tables.
		    }
		    Such low cost in traditional financial securities is attributable to the most liquid securities, for example, the EURUSD or USDJPY foreign exchange (FX) rates \cite[see, e.g.][]{Karnaukh_et_al_2015}. 
			This competitive figure -- significantly lower than the $12$ bps on Binance, the best performing CEX -- can be attributed to the innovative DPR and MFT systems and the stable nature of the assets.
			Since both tokens are pegged to the US Dollar, the exchange pair volatility is minimal, thereby implying a negligible expected impermanent loss for LPs.
			As predicted by equation \eqref{eq:IL3} of the Internet Appendix, LPs leverage their position by concentrating liquidity around the $1$ price point, resulting in extremely narrow spreads for trades around this price.
			Moreover, the MFT system fosters a reduction of exchange fees for this pair. Indeed, considering the high trading volume and the minimal expected losses, LPs can opt for the lowest fee tier of 1 bps while still anticipating a positive return.
			These findings are consistent with the theoretical models of \cite{capponi2021adoption} and \cite{hasbrouck2023economic}, predicting an inverse relationship between exchange rate volatility and liquidity.

		\subsection{Additional Results}\label{sub:TCs_additional}

			Our analysis is rounded off with four additional investigations to ensure the robustness of our findings.
			First, one may argue that our comparison of TCs of CEXs and DEXs is unfair, as the latter include also settlement costs. 
			Hence we re-run our analysis excluding DEX gas fees and report the findings in Figure \ref{fig:TC_all_coins_NG} of the Internet Appendix.
			Results show that, in the absence of gas fees, Uniswap v3 offers lower TCs for most of the pairs across all trade sizes. 
			This supports our argument on the main obstacle to DEX liquidity being rooted in the characteristics of the underlying blockchain technology, rather than in an economically-motivated friction.
			Second, Figure \ref{fig:TC_all_exchanges} in the Internet Appendix shows that our results are confirmed when adding Coinbase, Huobi, and OKX to the list of CEXs and restricting to the eight pairs in the intersection of all seven exchanges.
			Third, as Uniswap v3 was launched in May 2021, its TCs are averaged over a smaller sample size compared to other exchanges, which could potentially result in an imbalanced comparison.
			To address this issue, we re-run our analysis on the v3 subsample starting from May 2021. 
			The outcomes of this exercise, detailed in Figure \ref{fig:TC_all_coins_V3} of the Internet Appendix, exhibit no significant discrepancy compared to our previously reported results.
			%

			%
			Finally, one may argue that the high levels of liquidity offered DEXs may be related to the custody of assets. 
			Specifically, investors may choose to trade on these alternative venues in order to eliminate their exposure to custody risk, as discussed in Section \ref{sec:AMM}.
			In turn, this could increase the incentives to LPs to provide liquidity to DEXs. 
			For the same reason traders may decide to execute small transactions on DEXs, as indicated by Figure \ref{fig:trade_sizes}, even bearing higher transaction costs.
			If this is the main factor behind the success of DEXs, one may expect their utility to diminish in the future, if tighter regulation for CEXs will reduce the degree of custody risk.
			The collapse of the FTX exchange gives us an ideal laboratory to analyze the effects of the risks in granting CEXs the custody of assets.
			FTX was one of the world's largest cryptocurrency exchanges. It filed for bankruptcy on November 11, 2022, jeopardizing around $740\$$ million worth of crypto assets deposited on the FTX platform \citep{NYT}.
			We posit that this event may have triggered an erosion of trust due to the increased saliency of deposit risks and mismanagement of users' assets by CEX system operators.
			In addition, a simultaneous increase in trading volumes on DEXs would suggest a substitution effect from CEXs to DEXs, thus underpinning the importance of custody risk as a driver of DEX adoption. 
	        To gauge the importance of custody risk, or at least of its saliency, we perform difference-in-differences regression analyses of the trading activity on CEXs and DEXs around the FTX bankruptcy. 
	        The sample encompasses a period of two months before (September 9, 2022) and after (January 9, 2023) the event. 
	        The dependent variable is the daily trading volume on CEXs and DEXs, which is regressed on a treatment dummy (DEX) indicating DEXs, a time dummy (FTX) indicating the period after the FTX collapse, and their interaction. 
	        We check and confirm the validity of the parallel trend assumption.
	        
	        The evolution of aggregate trading volumes on CEXs and DEXs around the event, depicted in Figure \ref{fig:FTX_DD_bars} of the Internet Appendix, clearly illustrates a sharp decrease in CEX volumes. 
	        The regression results reported in Table \ref{tab:FTX_DD} indicate a significant collapse in trading volumes on CEXs and a stable holding of volumes on DEXs.
	       	The effect partially reverts within six months from the event, suggesting it could be driven by a temporary increased saliency of custody risk. 
	       	Our results are robust to removing Binance, thus suggesting that the decrease in CEX volume is not driven by its role in the bankruptcy and it potential exposure to FTX\footnote{See, for instance, \href{https://www.investopedia.com/what-went-wrong-with-ftx-6828447}{this article} for more details on the role played by Binance.}.
	       	Further, the observed decrease in volume cannot be explained by the cessation of cross-exchange arbitrage activity alone.\footnote{
	       		Indeed, the total trading volume on FTX averaged at most 10\% of the Binance volume, according to CoinMarketCap. 
	       		Since the cross-exchange arbitrage volume is by definition a subset of the total, its cessation following the collapse of FTX can at best explain only a relatively tiny fraction of the observed drop in volume.}
	        All in all, these findings suggest that the absence of custody risk is not a significant driver of the volume generated on DEXs.
	        %

	\section{Price Efficiency}\label{sec:PE}
		We now address our second testable hypothesis, that is, CEXs offer superior price efficiency compared to DEXs.
		Our conjecture is motivated by the results of the previous section, showing that DEXs are not ideal for small-sized transactions due to the burden of gas fees.
		We thus expect this friction to limit arbitrage forces, allowing deviations from efficient prices to persist and blurring the informativeness of transaction prices.
		The underlying mechanism is that arbitrage opportunities with smaller dollar capacity are not profitable net of transaction costs, thus removing the incentive for arbitrageurs to restore efficient prices.
		Empirically, we explore deviations from the law of one price by focusing on triangular arbitrage. 
		Performed in only one specific market and nearly risk-free, this no-arbitrage condition is the ideal laboratory for identifying market-specific frictions and for comparing the price efficiency of different market venues.\footnote{
			We do not account for the possibility of cross-exchange arbitrage opportunities arising from price differences on the same pair quoted in distinct exchanges.
			A triangular deviation on exchange $A$ may be arbitraged away by a cross-exchange transaction between $A$ and second exchange $B$.
			The observed triangular deviations on $A$ could thus lead to an over-estimation of its price efficiency, relative to a theoretical situation in which it operates in the vacuum.
			Our results can therefore be interpreted as an upper bound on the intrinsic price efficiency of each exchange.
		}
		A triangular arbitrage opportunity arises when the law of one price is violated for a closed triplet of currency pairs $X\leftrightarrow Y$, $Y\leftrightarrow Z$, and $Z\leftrightarrow X$.
		A direct measure of the deviation from the law of one price is
		\begin{equation}\label{eq:theta}
			\theta = \, P_{XY} \, P_{YZ} \, P_{ZX} - 1\,,
		\end{equation}
		where $P_{AB}$ is the quoted price of $A$ in units of $B$.
		%
		%
		A triangular trade is profitable only if the magnitude of $\theta$ is sufficiently large or,
		more precisely, the gross profits from the triangular trade is higher than the costs of executing the three transactions.

		To estimate the magnitude of price deviations, we sample $\mid\theta\mid$ at an hourly frequency for each exchange-triplet available in our sample,
		%
		employing different proxies for quoted prices depending on the exchange.
		For CEXs, we use the mid-price, that is, the average between the best ask and the best bid;
		For Uniswap v2, we use the ratio between the reserves of the two tokens, as in equation \eqref{eq:quoted_price};
		For Uniswap v3 we retrieve historical quoted prices from \href{https://dune.com}{Dune.com}\footnote{
			These have been obtained by calling the \emph{slot0} function of the relevant liquidity pool contract and saving the response in an archive database.
		}.
		We then average the resulting hourly deviations by exchange-triplet, over the sample period from March 2021 to February 2023.

		\subsection{Main Results on Price Efficiency}\label{sub:AB_results}
			%
			%
			Figure \ref{fig:price_efficiency_2024} presents the results, displaying the average price deviations across our sample, on a logarithmic scale.
			The main takeaway is that DEXs are far less price-efficient than their centralized counterparts.
			For most triplets, Uniswap v2 price deviations average between $10$ and $30$ bps, while they rise above $600$ bps for the less liquid BTC-ETH-DAI.
			Uniswap v3 performs better than its predecessor, with average deviations ranging between $5$ and $50$ bps.
			These estimates are significantly higher than those for CEXs, which are below $5$ bps for all triplets.
			Binance dominates regarding price efficiency, with average deviations below $3$ basis points.
			%

			%
			Figure \ref{fig:Bounds} displays the time-series evolution of hourly price deviations for the ETH-USDC-USDT triplet, with each dot representing an observation.
			The top panel presents the series for CEXs, defined as the price deviation with the minimum absolute value, on each hour, between those of \CEXs.
			Similarly, the bottom panel plots the hourly price deviation with the minimum absolute value among those recorded on Uniswap v2 and Uniswap v3.
			To visualize the introduction of v3 in May 2021, dots on this panel are colored in pink when the minimum is achieved in v2, and in violet when the minimum is achieved in v3.
			On both panels, the solid lines represent the top decile of the distribution of absolute deviations, estimated on a 7-day rolling window.
			The main takeaway of the figure is that the dominance of CEXs regarding price efficiency is stable across the sample period, with deviations an order of magnitude smaller than in DEXs.
			It also shows, however, that the introduction of Uniswap v3 is followed by a significant improvement in DEX price efficiency, decreasing by more than 50\% from the start of the sample.
			This phenomenon will be confirmed through a systematic analysis in Section \ref{sec:v3}.

			These results lend strong support to our second hypothesis, confirming that CEXs offer superior price efficiency.
			In the following sections, we investigate the causal role of gas prices in driving this observed outcome.

		\subsection{Gas Fees and Price Efficiency}
			We posit that the suboptimal price efficiency of DEXs can be attributed primarily to two factors: the absence of custodial delegation and the significant impact of gas fees.
			To see this, note that CEX arbitrageurs are likely to have their capital readily available in the exchange.
			The reason is that moving capital from a non-custodial wallet to a CEX is costly and takes a significant amount of time.\footnote{
				Deposits from a non-custodial wallet to a CEX take one to five minutes to be executed.
				The reason is that the exchange initially freezes the assets and requires the user to wait for a predefined number of blocks to be validated on the blockchain (12 for Binance, 20 for Kraken) before the funds are accessible. This measure is in place to decrease the probability of a double-spending attack that would result in a net loss for the exchange.
			}
			In a competitive environment, arbitrageurs are thus incentivized to delegate the custody of their arbitrage capital to the exchange, accepting increased counterparty risk.
			As a result, their transactions are not subject to gas costs associated with the deposit and withdrawal of funds.
			Since gas costs are independent of trade size, avoiding those allows CEX arbitrageurs to transact any amount of capital -- even very small dollar amounts -- whenever a triangular arbitrage opportunity arises.
			DEX arbitrageurs, on the contrary, do not have the option to deposit their arbitrage capital in the exchange to avoid gas costs.
			As a first-order consequence, they bear the gas cost of each transaction, thus making their arbitrage trades more costly.
			Further, a second-order implication is that, as is the case with models that feature entry costs, arbitrageurs face a trade-off between diluting gas costs and limiting price impact. 
			%
			%
			%
			Therefore, DEX arbitrageurs do not profit from arbitrage opportunities with small dollar capacity, and intervene only when the available liquidity allows for large enough arbitrage transactions.
			Gas fees thus act as a friction for price efficiency, harming arbitrageurs' ability to restore the law of one price.\footnote{
				The enhanced price efficiency of CEXs may partly stem from their fee structures. Unlike DEXs, where all participants incur fees, CEXs can reduce fees for high-volume arbitrageurs or even conduct arbitrage internally, thus bypassing fees. 
				Given that triangular arbitrage involves three transactions, the cumulative impact of exchange fees becomes a crucial factor in the trade's net profitability.
			}

			To corroborate our argument regarding the significance of gas fees, we conduct a series of econometric tests.
			First, we examine the relationship between price deviations and gas fees through a panel regression.
			Second, to address potential endogeneity, we perform a similar exercise using a VAR model.
			Finally, we estimate the causal effect of gas fees on price efficiency through an instrumental variable regression, leveraging exogenous shocks to gas prices driven by fluctuations in NFT market activity.

		\subsubsection{Panel Regression and VAR}
			The panel regression is run at the triplet-hour level, with DEX price deviations as the dependent variable and the Ethereum gas price as the main regressor.
			We add controls for other potential determinants of DEX price deviations, including:
				(i) the contemporaneous price deviations on CEXs, to account for the possibility of simultaneous price deviations driven by common factors;
				(ii) the total percentage spreads between CEX and DEX prices for the pairs involved in each triplet, to account for cross-exchange mispricing;
				(iii) the average return volatility over the past 24 hours across the pairs of each triplet, serving as a risk measure possibly due to asymmetric information problems \citep{aoyagi2021coexisting};
				(iv) the absolute return of the USD-denominated price of ETH, to capture differences in market regimes;
				(v) hourly DEX transaction costs, averaged across the pairs composing each triplet, to test if gas prices have an incremental explanatory power on top of spreads and exchange fees.

	        The estimation results, reported in Table \ref{tab:DEX_deviations_reg}, show positive and highly significant coefficients on gas prices across all specifications, indicating a positive relationship between gas costs and the magnitude of DEX price deviations.
			The estimated coefficients imply that one standard deviation increase in gas prices is associated with a 4.2 basis points increase in DEX price deviations, corresponding to an 18\% increase relative to the unconditional average.
			
			%
			To address potentially endogenous dynamics, we estimate a VAR model incorporating DEX price deviations, gas prices, and other controls as endogenous variables. The VAR estimation, reported in Figure \ref{fig:IRF_VAR_gas_dev} of the Internet Appendix, supports our findings.

		\subsubsection{Causal Impact of Gas Fees on Price Efficiency}
			%
			A concern with our analysis is that there could be omitted variables that affect both gas levels and price deviations, thus potentially biasing our regression estimates.
			A natural candidate as an omitted variable is the degree of information asymmetry, potentially related to the release of fundamental news on the relative value of cryptocurrency pairs.
			On the one hand, a rise in information asymmetry may lead to increased activity on DEXs and on the Ethereum network more broadly, leading to congestion and driving up gas prices.
			On the other hand, information asymmetry may induce LPs to decrease their liquidity provision in an effort to reduce their impermanent-loss exposure, leading to higher transaction costs. 
			These in turn act as a friction limiting arbitrageurs activity, finally resulting in more sizeable deviations from the law of one price.
			
			%
			To address this concern, we instrument gas fees with NFT minting activity.
			NFTs are non-fungible tokens, usually linked to digital assets, that can be minted and traded on-chain \citep{barbon2023non}.
			NFT collections are typically released in limited supply, meaning that only a predefined number of tokens can be created through minting, typically at a fixed price per unit.
			Since highly anticipated collections are often sold-out within minutes of the sale opening, potential buyers compete against each other by bidding higher gas prices associated with their minting transactions.
			Typically, after the collection is sold-out in the primary market, a large number of on-chain transactions are performed in the secondary market.
			Thus, these minting events often result in network congestion and elevated gas prices during the subsequent minutes.\footnote{
				As an example, consider the minting of the ``Otherside'' collection, mentioned in a previous Footnote. The highly anticipated launch of the collection in April 2022 resulted in more than \$150 million spent on gas fees and a network-wide increase in gas prices, clearly visible in Figure \ref{fig:gas_fees_uniswap}.
			}
			
			%
			We thus use the hourly total dollar value of minted NFT as an instrument for the level of gas fees.
			Given the highly idiosyncratic nature of minting events and the fact that NFTs are not traded on DEXs, they should only have an effect on DEX dynamics through the channel of increased gas prices.
			We thus argue that such an instrument satisfies the exclusion restriction since it induces variations in gas prices which are exogenous to the level of information asymmetry on the exchange rate of cryptocurrency pairs\footnote{
				To address the concern that both NFT mints and DEX trading dynamics may be driven by the general sentiment prevailing in crypto markets, we check if mint volume is correlated to the \href{https://alternative.me/crypto/fear-and-greed-index/}{Crypto Fear \& Greed Index}. As expected, given that NFT mint dates are usually scheduled weeks in advance, a simple regression analysis shows that there is no significant correlation between our instrument and the sentiment index. Moreover, the results described below are robust to controlling for the Crypto Fear \& Greed Index in the first or in the second stage.
			}.
			The data for this exercise is gathered from a custom query written on \href{https://dune.com/queries/3378748/5965209}{Dune Analytics}.

			%
			As reported in Panel A of Table \ref{tab:DEV_NFT_reg}, the instrument has a strong first stage, showing a highly significant positive correlation with gas price levels.
			This is robust to controlling the absolute hourly return of the ETH/USD exchange rate, proxying for asymmetric information, and adding day fixed effects.
			The second stage regression estimates, reported in Panel B of Table \ref{tab:DEV_NFT_reg}, depict a highly significant positive correlation of instrumented gas prices with the degree of price deviations.
			These results are robust to adding a battery of controls and saturating the regression model with day and triplet fixed-effects.
			The findings provide causal evidence of a direct impact of gas fees on DEX price efficiency. 
			This effect is independent from a potential relationship driven by variations in the level of information asymmetry.

		\subsubsection{Implications}
			All in all, our findings support the idea that gas fees constitute an economically relevant friction dampening DEX price efficiency, even after controlling for other factors and accounting for potentially endogenous dynamics.
			The identification of a direct channel linking gas fees to price efficiency has important economic consequences.
			It implies that DEX price efficiency is not limited only by market design and economic factors like information asymmetry.
			Rather, it can be improved by a reduction of gas prices driven by future refinements of blockchain technology.
			
			Three key technological advancements that are expected to drive such a transformation are data sharding, layer 2 solutions, and data blobs.
			Data sharding is a process that divides a blockchain's database into smaller, more manageable pieces, or "shards," each capable of processing transactions independently. This significantly increases the network's capacity and throughput, directly impacting gas fees by alleviating network congestion.
			Layer 2 solutions are built on top of the base blockchain and facilitate transaction processing off-chain before recording the final state on-chain. This approach drastically reduces the load on the main blockchain, thereby decreasing gas costs.
			Finally, data blobs allow the inclusion of large amounts of data in a blockchain transaction in a compact and efficient form. This method can be beneficial for DEXs in particular, as it reduces the gas cost associated with arbitrage trades involving multiple pools.
			At the time of writing, multiple layer 2 solutions are available and, according to \href{https://l2beat.com/scaling/tvl}{L2Beat.com}, the total value locked has been increasing steadily and surpassed \$ 35 billion.
			While a partial implementation of blobs and data-sharding went live on Mainnet with the Cancun-Deneb (Dencun) network upgrade in March 2024, the full implementation will take several years, according to Ethereum developers.

			To conclude this section, we note that, while CEXs enjoy higher levels of price efficiency, they may not offer the best experience from an operational point of view.
			In fact, taking the perspective of a CEX user keeping her funds in a non-custodial wallet, the significant time delay affecting deposits induces a trade-off between price efficiency and speed of execution.
			Conversely, even if DEX quoted prices are further away from their efficient levels, the user can settle a transaction with a significantly shorter time delay.\footnote{
				Assuming the chosen gas price is reasonable compared to the one prevailing on the network, the transaction is settled in a block, which is significantly less than the 12 blocks required by Binance or the 20 blocks required by Kraken.
			}

    \section{Market Design and Uniswap v3}\label{sec:v3}
    	Uniswap v3 introduced two key innovative features, MFT and DPR.
    	In summary, the MFT system enables LPs to select the level of exchange fees, while DPR grants them control over the price range within which their liquidity is accessible\footnote{These two innovations are described in detail in Section \ref{sub:uni_v3} of the Internet Appendix.}.
    	The MFT system provides a way to choose the appropriate level of compensation for risk, thus facilitating the cross-sectional allocation of liquidity to pairs with diverse exchange rate volatility.
    	The DPR system increases capital efficiency, allowing LPs to concentrate their liquidity to specific price intervals rather than forcing it to be spread across the entire spectrum.
    	A natural parallel can be drawn between the flexibility provided by DPR and the quote updating process in a LOB, as both mechanisms enable LPs to adjust their exposure to price volatility risk.
    	As showed in Equation \eqref{eq:IL3} of the Internet Appendix, widening the price interval in Uniswap v3 results in reduced exposure to price movements, similar to the widening of bid/ask spreads in the LOB setting.
    	Further, the DPR allows LPs to post so called \emph{range orders}, which are very similar to limit orders.
    	The DPR feature thus mimics some of the core functionalities of CEXs that were not present in previous iterations of DEXs.
    	All-in-all, the enhanced flexibility in liquidity provision, coupled with advanced risk management tools, should theoretically strengthen the incentives for LPs to supply liquidity to DEX platforms, thereby improving their market quality.

    	Supporting this view, the results outlined in the previous sections show that Uniswap v3 exhibits superior performance relative to its predecessor, Uniswap v2.
    	Consequently, in our third hypothesis, we conjecture that innovation in DEX market design can lead to higher levels of market quality.
    	The observed improvements, however, could be attributed to the migration of a substantial portion of liquidity and users towards the newer protocol, subsequent to its launch.
    	To address potential biases arising from this migration phenomenon or simply due to changing market conditions, we exploit our cross-section of exchanges and trading pairs to provide empirical evidence supporting the causal effect of the introduction of Uniswap v3 on DEX market quality.

		\subsection{Transaction Costs}\label{sub:v3_cost}
	    	To investigate the causal impact of Uniswap v3 on DEX market liquidity, we run difference-in-differences regressions concerning transaction costs.
	    	The idea is to test whether the introduction of v3 led to a significant decrease in trading costs on DEXs, using the corresponding CEX metrics as the control group.
	    	This approach naturally controls for potential confounding factors contemporaneous to the introduction of v3, that may have caused a global improvement in the quality of cryptocurrency markets.
	    	
	    	Considering a window of six months around the deployment of Uniswap v3 (May 5th, 2021), we construct an hour-pair-level panel of the best transaction costs offered by DEXs and CEXs.\footnote{
				For each pair, the two time series are computed by taking the minimum between the transaction costs for Uniswap v2 and v3 (DEXs), and among \CEXs $\,$(CEXs).
				Taking the minimum between Uniswap v2 and v3 is motivated by the fact that the v3 interface automatically suggests to users the best routing across the two exchanges, minimizing transaction costs.
				We apply the same methodology to CEXs for consistency but verify that all results based on this panel are robust to taking the average, the median, or the maximum among CEXs.
	    	} 
	    	We verify that the Parallel Trend Assumption is met and implement a difference-in-differences regression.
			In this configuration, the treatment group comprises each pair quoted on DEXs, while the control group includes the same pairs quoted on CEXs.
	    	Table \ref{tab:diff_n_diffs_table} reports results from the exercise, where the dependent variable is regressed on 
	    	a time dummy ``V3'' indicating the period following v3's deployment, and 
	    	a treatment dummy ``DEX'' indicating decentralized exchanges, 
	    	and their interaction.
	    	The coefficient on the interaction term from specification (1) is negative and highly statistically significant, suggesting that v3 caused a decline in DEX transaction costs.
	    	The point estimates indicate that the effects are economically significant, with a $56$ bps decrease in DEX transaction costs relative to CEXs (a $58\%$ reduction).
	    	Specification (2) shows that these results are robust to the inclusion of time and pair fixed effects.

			Our results highlight the improvement in DEX transaction costs stemming from the novel market design of Uniswap v3, hence supporting our third hypothesis.
			In particular, the introduction of MFT and the resulting fragmentation of liquidity across different pools for the same exchange pair seem to play an important role, especially for sizeable transactions.
			As shown in Table \ref{tab:fees_heterogeneity}, the best execution for \$1,000,000 transactions is often provided by pools with higher exchange fees but offering greater levels of liquidity in the relevant price range. 
			For instance, while trades below \$10,000 on the ETH-USDT pair are almost always optimally executed in the 5-bps pool, 
			the best routing changes to the 30-bps pool 
			11\% of the time for \$100,000 and
			30\% of the time for \$1,000,000.
			These findings are consistent with the high degree of liquidity fragmentation found by \cite{LPZ} on Uniswap v3, both theoretically and empirically. 
			Additionally, our results indicate that liquidity fragmentation has a positive impact on the transaction-cost dimension of market quality.
			Our findings offer empirical support for the idea that the ability of LPs to charge higher fees leads to enhanced liquidity and reduced transaction costs in equilibrium, as outlined by \cite{Hasbrouck2022}.
	
		\subsection{Price Efficiency}\label{sub:v3_efficiency}
	    	Similar to the above exercise on market liquidity, we now perform a difference-in-differences regression to assess whether the introduction of Uniswap v3 led to an improvement in market quality. 
	    	We take advantage of this quasi-natural experiment to identify a potential causal effect on our measure of price efficiency. 

	        Similarly to the previous setting, this approach uses the corresponding CEX metrics as the control group and it naturally controls for potential confounding, common factors.\footnote{
	        	We consider a window of six months around the deployment of Uniswap v3 (May 5th, 2021), and construct an hour-triplet-level panel of the smallest price deviations recorded on DEXs and CEXs
	        }
	        We construct an hour-triplet-level panel with the minimum absolute deviations by exchange category.\footnote{
	    		The choice of taking the minimum between Uniswap v2 and v3 is instrumental for our analysis, aimed at estimating the potential improvement brought by v3 in terms of price efficiency.
	    		We apply the same methodology to CEXs for consistency but verify that all results based on this panel are robust to taking the average, the median, or the maximum between CEXs.
	        }
	    	Table \ref{tab:diff_n_diffs_table} report results from difference-in-differences regressions, 
	    	where the dependent variable measuring price errors is regressed on 
	    	a time dummy ``V3'' indicating the period following v3's deployment, and 
	    	a treatment dummy ``DEX'' indicating decentralized exchanges, and
	    	their interaction.
	    	The coefficient on the interaction term from specification (3) is negative and highly statistically significant, suggesting that v3 caused a decline in price deviations.
	    	The estimated coefficients indicate that the effects are economically significant, 
	    	with a $89$ bps decrease in DEX price deviations relative to CEXs (a $75\%$ reduction).
	    	Specification (4) shows that results are robust to the inclusion of time and triplet fixed effects. 

	    	This analysis provides evidence of a sizeable increase in price efficiency caused by innovations in market design, thus supporting our third hypothesis.
	    	In particular, this is fostered by the implementation of the DPR and MFT systems, making it possible for LPs to take leverage on their liquidity positions, as detailed in Section \ref{sub:IL} of the Internet Appendix, and to choose the level of exchange fees.
	    	Interestingly, these enhancements arise from innovations which take inspiration from the market design of LOB-based exchanges.
	    	Thanks to the DPR system, in fact, LPs in v3 can dynamically manage their risk exposure, for instance reducing their leverage when exchange rate volatility increase. 
	    	Even though this improvement is not enough to surpass the bar of market quality set by CEX systems, it highlights how improving the market design of DEXs can significantly enhance their market quality, thus supporting our third hypothesis.

	\section{Conclusion}\label{sec:conclusion}
        This paper addresses the question of which market design best supports market quality for cryptocurrencies. 
		Ex ante, the relative market quality of decentralized exchanges (DEXs) compared to centralized exchanges (CEXs) remains uncertain.
		While CEXs have historically dominated, DEXs have shown rapid growth, recently surpassing CEXs in trading volumes for some of the major currency pairs.
		Determining which market design offers superior quality is an empirical question, and we
		tackle it leveraging a unique and granular dataset encompassing both exchange types.
		Our analysis evaluates market quality across two key dimensions: liquidity and price efficiency.
        Regarding market liquidity, we measure all transaction cost components for CEXs and DEXs. 
        For price efficiency, we examine deviations from the triangular no-arbitrage condition. 
        Moreover, we aim to identify the most economically important frictions, if any, dampening market quality.

        Employing panel regressions, difference-in-differences analysis, and an instrumental variable approach, our main findings are as follows: 
        %
        First, CEXs are significantly less expensive for small and medium trades, while DEXs become more cost-effective for relatively large amounts due to the dilution of fixed gas fees.
        Consequently, DEXs are currently most convenient for institutional investors or high-net-worth individuals.
        %
        Second, transaction costs adversely affect price efficiency, with gas fees being a critical determinant. 
        Our instrumental variable analysis provides causal evidence that gas fees deteriorate price efficiency in DEX systems, regardless of asymmetric information issues.
        Our estimation of the causal effect suggests that a potential exogenous reduction of gas fees (due to advancements in layer 2 solutions and other expected improvement in blockchain technology) may materially improve the market quality of DEXs in the future.
        %
        Third, Uniswap v3 has substantially reduced transaction costs in DEXs, making them competitive with CEXs, especially for stablecoin pairs like USDC-USDT.
        Fourth, Uniswap v3 has also improved price efficiency in DEXs, though they still lag behind CEXs, with gas fees being a major deterrent. 
        These findings support the idea that future updates to DEX market design could lead to significant improvements in market quality.

        All in all, our analysis suggests that while the innovative market structure of DEXs holds promise, their full potential is yet to be realized. 
        Future improvements could arise from reductions in gas fees, driven by expected advancements in blockchain technology, including data sharding, layer 2 solutions, and data blobs.
 		If realized, these developments could allow the DEX market design to be applied to traditional asset classes, challenging the dominance of limit order book systems in terms of market quality.\\
        Ultimately, the presumed  inevitability of the LOB system, as argued by \cite{glosten1994}, may warrant reconsideration.

		\newpage
		\renewcommand\baselinestretch{1.0}
		\small
		\bibliographystyle{apalike}
		\bibliography{Biblio}

		\clearpage
		\section*{Tables and Figures}

	
		%
			\begin{table}[ht]
				\centering
				\resizebox{0.93\textwidth}{!}{\begin{tabular}{lrrrrrrrrr}
\toprule
\multicolumn{10}{c}{Panel A: Volume (Million USD)}\\
\midrule
Exchange 	&    Sample     	&  
$\quad$ N & $\quad$ Mean & $\quad$ Std & $\quad$  1\% & $\quad$ 10\% & $\quad$ 50\% & $\quad$ 90\% & $\quad$ 99\% \\
\midrule

\color{Uniswap_color_dark}$\blacksquare\;$\color{black}Uniswap v2  	
& Main Pairs        & 8  & 8854.63 & 11859.55 & 148.16 & 1255.92 & 3141.73 & 11980.77 & 29260.94 \\
& All Pairs       	& 20 & 3888.23 & 8353.41 & 15.29 & 133.16 & 482.91 & 2716.47 & 29260.94 \\[0.5em]

\color{Uniswap_v3_color_dark}$\blacksquare\;$\color{black}Uniswap v3 	
& Main Pairs        & 8 & 85227.50 & 137528.45 & 2922.24 & 11081.52 & 47344.69 & 69502.28 & 417644.26 \\
& All Pairs       	& 20 & 34439.69 & 93697.89 & 3.39 & 90.64 & 758.57 & 19034.86 & 417644.26 \\[0.5em]

\color{Binance_color_dark}$\blacksquare\;$\color{black}Binance 		
& Main Pairs        & 8 & 172271.96 & 397862.21 & 1.55 & 6544.04 & 21620.12 & 76273.26 & 1152478.83 \\
& All Pairs       	& 20 & 72706.08 & 263001.33 & 0.01 & 0.09 & 0.33 & 9302.61 & 1152478.83 \\[0.5em]

\color{Kraken_color_dark}$\blacksquare\;$\color{black}Kraken 			
& Main Pairs        & 8 & 2697.85 & 3373.78 & 0.07 & 253.83 & 1536.38 & 3695.86 & 9581.25 \\
& All Pairs       	& 20 & 1086.32 & 2452.95 & 0.00 & 0.02 & 0.03 & 421.27 & 9581.25 \\[0.5em]

\color{Coinbase_color_dark}$\blacksquare\;$\color{black}Coinbase 		
& Main Pairs       	& 8 & 7401.96 & 8125.49 & 0.35 & 970.06 & 1982.36 & 15044.61 & 17801.68 \\[0.5em]
\color{Huobi_color_dark}$\blacksquare\;$\color{black}Huobi 			
& Main Pairs       	& 8 & 66462.87 & 158339.84 & 0.06 & 210.40 & 1079.46 & 5571.90 & 389628.17 \\[0.5em]
\color{Okex_color_dark}$\blacksquare\;$\color{black}OKX 				
& Main Pairs       	& 8 & 29879.12 & 75061.40 & 0.23 & 62.52 & 4219.26 & 7265.49 & 215466.88 \\

\midrule\\

\multicolumn{10}{c}{Panel B: Trades (Thousands)}\\
\midrule
Exchange 	&    Sample     	&  
$\quad$ N & $\quad$ Mean & $\quad$ Std & $\quad$  1\% & $\quad$ 10\% & $\quad$ 50\% & $\quad$ 90\% & $\quad$ 99\% \\
\midrule

\color{Uniswap_color_dark}$\blacksquare\;$\color{black}Uniswap v2 
& Main Pairs        & 8 & 860.50 & 1230.33 & 41.00 & 119.00 & 210.50 & 1126.75 & 2843.00 \\
& All Pairs       	& 20 & 374.40 & 851.52 & 10.00 & 29.75 & 52.50 & 180.00 & 2843.00 \\[0.5em]
\color{Uniswap_v3_color_dark}$\blacksquare\;$\color{black}Uniswap v3
& Main Pairs        & 8 & 1236.88 & 1649.72 & 97.00 & 202.00 & 646.50 & 1240.25 & 4916.00 \\
& All Pairs       	& 20 & 516.10 & 1169.76 & 1.00 & 21.50 & 62.00 & 308.25 & 4916.00 \\[0.5em]
\color{Binance_color_dark}$\blacksquare\;$\color{black}Binance 
& Main Pairs        & 8 & 133423.49 & 264911.40 & 5505.89 & 9660.23 & 28961.76 & 73886.37 & 775874.90 \\
& All Pairs       	& 20 & 56885.16 & 178295.47 & 81.53 & 714.91 & 2698.40 & 16891.24 & 775874.90 \\[0.5em]
\color{Kraken_color_dark}$\blacksquare\;$\color{black}Kraken 	
& Main Pairs        & 8 & 4741.23 & 8567.93 & 91.80 & 777.43 & 1422.51 & 3303.29 & 25478.89 \\
& All Pairs       	& 20 & 1963.00 & 5697.87 & 42.78 & 74.25 & 137.43 & 1023.69 & 25478.89 \\[0.5em]

\color{Coinbase_color_dark}$\blacksquare\;$\color{black}Coinbase 
& Main Pairs       	& 8 & 7803.98 & 5756.48 & 1243.57 & 2372.89 & 9578.16 & 11861.43 & 15337.49 \\[0.5em]
\color{Huobi_color_dark}$\blacksquare\;$\color{black}Huobi 	
& Main Pairs       	& 8 & 71492.68 & 131672.25 & 846.05 & 13941.68 & 19895.80 & 32840.34 & 339206.24 \\[0.5em]
\color{Okex_color_dark}$\blacksquare\;$\color{black}OKX 		
& Main Pairs       	& 8 & 32634.42 & 69654.82 & 654.24 & 4255.81 & 6762.86 & 15569.05 & 204336.46  \\
\bottomrule

\end{tabular}



}
				\captionsetup{justification=justified, singlelinecheck=on, font=footnotesize}
				\caption{\textbf{Summary Statistics.}\footnotesizelarge  
			The table presents summary statistics on the twenty trading pairs analyzed in our study, 
			focusing on the trading activity across the seven exchanges during our sample period, from March 2021 to February 2023.
			Panel A displays the distribution of the total traded volume by exchange and trading pair, expressed in million USD.
			Similarly, Panel B reports the total number of transactions (in thousands).
			For each of the seven exchanges included in our sample, we provide the distributions for the set of eight pairs at the intersection of all the exchanges (\emph{Main pairs}),
			while for Uniswap v2, Uniswap v3, Binance, and Kraken we display the distributions also for the set of twenty pairs at the intersection of those exchanges (\emph{All pairs}).
		}\label{tab:summary_stats}
			\end{table}

		\vspace{1em}

			\begin{table}[ht]
				\centering
				\resizebox{0.93\textwidth}{!}{\begin{tabular}{lrrrrrrr}
\toprule
Pair        & 
$$    \color{Uniswap_color_dark}      $\blacksquare\;$ \color{black}Uniswap v2 &
$$    \color{Uniswap_v3_color_dark}   $\blacksquare\;$ \color{black}Uniswap v3 &
$\,$  \color{Binance_color_dark}      $\blacksquare\;$ \color{black}Binance   &
$\;\;$  \color{Kraken_color_dark}     $\blacksquare\;$ \color{black}Kraken &
$$    \color{Coinbase_color_dark}     $\blacksquare\;$ \color{black}Coinbase &
$\;\;\;$   \color{Huobi_color_dark}   $\blacksquare\;$ \color{black}Huobi &
$\quad\;$   \color{Okex_color_dark}   $\blacksquare\;$ \color{black}OKX \\
\midrule

ETH-USDC & 29260.94 & 417644.26 & 33358.42 & 2402.77 & 13950.82 & 8034.45 & 839.63 \\[0.25em]
ETH-USDT & 26053.11 & 83515.76 & 1152478.83 & 5493.40 & 17801.68 & 215466.88 & 389628.17 \\[0.25em]
ETH-BTC & 3939.36 & 64831.12 & 597023.75 & 1836.26 & 0.53 & 0.50 & 0.66 \\[0.25em]
LINK-ETH & 2344.10 & 5162.84 & 155521.97 & 2357.65 & 0.35 & 0.23 & 0.06 \\[0.25em]
BTC-USDC & 190.23 & 13054.42 & 65683.65 & 3096.68 & 16138.41 & 7009.18 & 1319.28 \\[0.25em]
DAI-ETH & 7289.99 & 36976.21 & 9881.82 & 338.36 & 1939.59 & 2440.67 & 839.63 \\[0.25em]
USDC-USDT & 1611.15 & 57713.16 & 108042.07 & 9581.25 & 1982.36 & 6081.04 & 6989.44 \\[0.25em]
MANA-ETH & 567.24 & 469.82 & 0.30 & 0.03 & 0.22 & 0.24 & 0.05 \\[0.25em]
DAI-USDT & 148.16 & 2922.24 & 8723.40 & 670.00 &  &  &  \\[0.25em]
AAVE-ETH & 840.17 & 572.00 & 0.33 & 0.02 &  &  &  \\[0.25em]
BAT-ETH & 187.96 & 276.06 & 0.08 & 0.02 &  &  &  \\[0.25em]
BTC-DAI & 15.29 & 79.87 & 3238.54 & 143.47 &  &  &  \\[0.25em]
CRV-ETH & 653.58 & 945.15 & 0.05 & 0.02 &  &  &  \\[0.25em]
GRT-ETH & 196.74 & 94.23 & 0.13 & 0.01 &  &  &  \\[0.25em]
KNC-ETH & 88.17 & 6.94 & 0.09 & 0.01 &  &  &  \\[0.25em]
REP-ETH & 52.36 & 3.39 & 0.05 & 0.00 &  &  &  \\[0.25em]
SNX-ETH & 398.58 & 208.18 & 0.01 & 0.02 &  &  &  \\[0.25em]
STORJ-ETH & 52.88 & 3.57 & 0.33 & 0.01 &  &  &  \\[0.25em]
UNI-ETH & 3833.55 & 4306.03 & 0.03 & 0.03 &  &  &  \\[0.25em]
OMG-ETH & 40.96 & 8.66 & 0.18 & 0.02 &  &  &  \\[0.25em]

\bottomrule
\end{tabular}}
				\captionsetup{justification=justified, singlelinecheck=on, font=footnotesize}
				\caption{\textbf{Trading Volume by Pair.}\footnotesizelarge  
			The table reports the daily average trading volume from March 2021 to February 2023, expressed in millions of USD, for the twenty trading pairs and the seven exchanges considered in our empirical analysis.
			For each of the seven exchanges included in our sample, we provide statistics for the set of eight pairs at the intersection of all the exchanges,
			while for Uniswap v2, Uniswap v3, Binance, and Kraken we display statistics also for the set of the additional twelve pairs at the intersection of those exchanges.
		}\label{tab:vol_pair_exch}
			\end{table}
			
		\clearpage\newpage

		%
			\begin{table}[ht]
				\centering
				\resizebox{0.93\textwidth}{!}{\begin{tabular}{lrrrrrrrrrr}
\toprule
Pair &&
\multicolumn{1}{c}{ETH}  & 
\multicolumn{1}{c}{ETH}  & 
\multicolumn{1}{c}{ETH}  & 
\multicolumn{1}{c}{LINK} & 
\multicolumn{1}{c}{BTC}  & 
\multicolumn{1}{c}{DAI}  & 
\multicolumn{1}{c}{MANA} & 
\multicolumn{1}{c}{USDC} & 
\multicolumn{1}{c}{\textbf{Average}}  \\
     && 
\multicolumn{1}{c}{USDC} & 
\multicolumn{1}{c}{USDT} & 
\multicolumn{1}{c}{BTC} & 
\multicolumn{1}{c}{ETH} & 
\multicolumn{1}{c}{USDC} & 
\multicolumn{1}{c}{ETH} & 
\multicolumn{1}{c}{ETH} & 
\multicolumn{1}{c}{USDT} & 
\multicolumn{1}{c}{\textbf{(20 pairs)}} \\

\midrule\\[-0.5em]
\multicolumn{11}{c}{\color{Uniswap_color_dark}$\blacksquare\;$ \color{black}\textbf{Panel A: Uniswap v2}}\\
\midrule
Exchange Fees &  & 30.00 & 30.00 & 30.00 & 30.00 & 30.00 & 30.00 & 30.00 & 30.00 & 30.00 \\
\midrule\multirow[c]{4}{*}{B/A Spread} & 1,000 & 0.13 & 0.30 & 0.92 & 4.15 & 39.35 & 0.78 & 31.91 & 0.84 & 39.35 \\
 & 10,000 & 1.29 & 3.04 & 9.15 & 41.32 & 381.21 & 7.78 & 310.48 & 8.35 & 381.21 \\
 & 100,000 & 12.87 & 30.30 & 90.82 & 399.36 & 3,156.29 & 77.39 & 2,606.28 & 83.01 & 3,156.29 \\
 & 1,000,000 & 127.85 & 296.22 & 848.20 & 3,253.15 & 22,969.04 & 738.46 & 18,878.72 & 786.74 & 22,969.04 \\
\midrule\multirow[c]{4}{*}{Gas Fees} & 1,000 & 204.67 & 204.67 & 204.67 & 204.67 & 204.67 & 204.67 & 204.67 & 204.67 & 204.67 \\
 & 10,000 & 20.47 & 20.47 & 20.47 & 20.47 & 20.47 & 20.47 & 20.47 & 20.47 & 20.47 \\
 & 100,000 & 2.05 & 2.05 & 2.05 & 2.05 & 2.05 & 2.05 & 2.05 & 2.05 & 2.05 \\
 & 1,000,000 & 0.20 & 0.20 & 0.20 & 0.20 & 0.20 & 0.20 & 0.20 & 0.20 & 0.20 \\
\midrule\multirow[c]{4}{*}{Total TCs} & 1,000 & 234.80 & 234.98 & 235.59 & 238.82 & 274.02 & 235.45 & 266.58 & 235.51 & 274.02 \\
 & 10,000 & 51.76 & 53.50 & 59.62 & 91.78 & 431.68 & 58.25 & 360.95 & 58.82 & 431.68 \\
 & 100,000 & 44.92 & 62.35 & 122.87 & 431.41 & 3,188.34 & 109.43 & 2,638.32 & 115.06 & 3,188.34 \\
 & 1,000,000 & 158.05 & 326.42 & 878.40 & 3,283.36 & 22,999.24 & 768.67 & 18,908.93 & 816.95 & 22,999.24 \\

\midrule\\[-0.3em]
\multicolumn{11}{c}{\color{Uniswap_v3_color_dark}$\blacksquare\;$ \color{black}\textbf{Panel B: Uniswap v3}}\\
\midrule
Exchange Fees &&     Variable & Variable & Variable & Variable & Variable & Variable & Variable & Variable & Variable \\
\midrule\multirow[c]{4}{*}{B/A Spread} & 1,000 & 0.25 & 0.17 & 0.09 & 2.14 & 2.33 & 0.60 & 12.37 & 0.00 & 2.92 \\
 & 10,000 & 0.36 & 1.43 & 0.40 & 2.30 & 5.78 & 4.19 & 116.32 & 0.01 & 9.93 \\
 & 100,000 & 1.49 & 8.10 & 3.13 & 18.84 & 7.57 & 22.14 & 460.04 & 0.03 & 42.67 \\
 & 1,000,000 & 10.08 & 50.89 & 14.79 & 80.56 & 38.02 & 80.49 & 583.88 & 0.13 & 155.28 \\
\midrule\multirow[c]{4}{*}{Gas Fees} & 1,000 & 154.54 & 154.54 & 154.54 & 154.54 & 154.54 & 154.54 & 154.54 & 154.54 & 154.54 \\
 & 10,000 & 15.45 & 15.45 & 15.45 & 15.45 & 15.45 & 15.45 & 15.45 & 15.45 & 15.45 \\
 & 100,000 & 1.55 & 1.55 & 1.55 & 1.55 & 1.55 & 1.55 & 1.55 & 1.55 & 1.55 \\
 & 1,000,000 & 0.15 & 0.15 & 0.15 & 0.15 & 0.15 & 0.15 & 0.15 & 0.15 & 0.15 \\
\midrule\multirow[c]{4}{*}{Total TCs} & 1,000 & 159.04 & 159.71 & 159.73 & 183.15 & 167.03 & 160.14 & 196.91 & 156.28 & 184.18 \\
 & 10,000 & 20.46 & 22.04 & 21.10 & 46.40 & 34.51 & 25.79 & 163.11 & 17.21 & 56.72 \\
 & 100,000 & 8.01 & 17.41 & 10.01 & 48.53 & 26.82 & 36.06 & 494.63 & 3.32 & 82.30 \\
 & 1,000,000 & 13.57 & 63.48 & 25.79 & 92.17 & 51.47 & 89.34 & 614.09 & 2.03 & 205.12 \\

\midrule\\[-0.3em]
\multicolumn{11}{c}{\color{Binance_color_dark}$\blacksquare\;$ \color{black}\textbf{Panel C: Binance}}\\
\midrule
Exchange Fees &  & 10.00 & 10.00 & 10.00 & 10.00 & 10.00 & 10.00 & 10.00 & 10.00 & 10.00 \\
\midrule\multirow[c]{4}{*}{B/A Spread} & 1,000 & 8.93 & 1.18 & 0.78 & 10.43 & 12.36 & 7.35 & 31.52 & 1.00 & 12.25 \\
 & 10,000 & 9.09 & 1.23 & 0.80 & 19.73 & 10.59 & 10.98 & 99.33 & 1.01 & 17.20 \\
 & 100,000 & 17.37 & 2.93 & 4.84 & 273.71 & 14.25 & 49.66 & 2,166.77 & 1.04 & 171.35 \\
 & 1,000,000 & 225.97 & 15.07 & 48.14 & 3,971.68 & 141.78 & 5,799.07 & 6,803.65 & 1.97 & 5,707.90 \\
\midrule\multirow[c]{4}{*}{Total TCs} & 1,000 & 18.93 & 11.18 & 10.78 & 20.43 & 22.36 & 17.35 & 41.52 & 11.00 & 22.25 \\
 & 10,000 & 19.09 & 11.23 & 10.80 & 29.73 & 20.59 & 20.98 & 109.33 & 11.01 & 27.20 \\
 & 100,000 & 27.37 & 12.93 & 14.84 & 283.71 & 24.25 & 59.66 & 2,176.77 & 11.04 & 181.35 \\
 & 1,000,000 & 235.97 & 25.07 & 58.14 & 3,981.68 & 151.78 & 5,809.07 & 6,813.65 & 11.97 & 5,717.90 \\

\bottomrule

\end{tabular}
}
				\captionsetup{justification=justified, singlelinecheck=on, font=footnotesize}
				\caption{\textbf{Transaction Costs -- Breakdown by Trade Size.}\footnotesizelarge 
			The table displays the average transaction costs for the eight pairs present at the intersection of all exchanges covered by our sample.
			On Panel A, B, C, and D, the last column reports the average transaction costs across all twenty pairs present in the intersection of Uniswap v2, Uniswap v3, Binance, and Kraken.
			On Panel E, F, and G, the last column reports the average transaction costs across the eight pairs present in the intersection of Uniswap v2, Uniswap v3, \CEXs.
			The transaction costs are presented for different trade sizes denominated in USD, 
			based on equation \eqref{eq:TC_AMM} for DEX and equation \eqref{eq:TC_LOB} for CEXs, and are expressed in basis points.
			They are computed at an hourly frequency and then averaged across the sample period, from March 2021 to February 2023.
			They are decomposed into three components: 
			(i) exchange fees, paid directly to the exchange or protocol;
			(ii) B/A spread, computed as the percentage difference between the mid-price and the execution price (based on LOB quotes for CEXs, and on liquidity pools for DEXs);
			(iii) gas fees to submit the transaction on the Ethereum blockchain (only for DEXs).
		}\label{tab:TCs_all_sz}
			\end{table}
		
		\begin{table*}[ht]
			\centering
			\resizebox{0.93\textwidth}{!}{\begin{tabular}{lrrrrrrrrrr}

\toprule
Pair &&
\multicolumn{1}{c}{ETH}  & 
\multicolumn{1}{c}{ETH}  & 
\multicolumn{1}{c}{ETH}  & 
\multicolumn{1}{c}{LINK} & 
\multicolumn{1}{c}{BTC}  & 
\multicolumn{1}{c}{DAI}  & 
\multicolumn{1}{c}{MANA} & 
\multicolumn{1}{c}{USDC} & 
\multicolumn{1}{c}{\textbf{Average}}  \\
     && 
\multicolumn{1}{c}{USDC} & 
\multicolumn{1}{c}{USDT} & 
\multicolumn{1}{c}{BTC} & 
\multicolumn{1}{c}{ETH} & 
\multicolumn{1}{c}{USDC} & 
\multicolumn{1}{c}{ETH} & 
\multicolumn{1}{c}{ETH} & 
\multicolumn{1}{c}{USDT} & 
\multicolumn{1}{c}{\textbf{(20 pairs)}} \\

\midrule\\[-0.5em]
\multicolumn{11}{c}{\color{Kraken_color_dark}$\blacksquare\;$ \color{black}\textbf{Panel D: Kraken}}\\
\midrule
Exchange Fees &  & 26.00 & 26.00 & 26.00 & 26.00 & 26.00 & 26.00 & 26.00 & 26.00 & 26.00 \\
\midrule\multirow[c]{4}{*}{B/A Spread} & 1,000 & 6.78 & 4.14 & 3.21 & 22.46 & 7.06 & 19.99 & 31.64 & 1.71 & 37.04 \\
 & 10,000 & 8.93 & 5.37 & 4.47 & 40.77 & 9.22 & 24.52 & 77.99 & 2.13 & 77.95 \\
 & 100,000 & 30.72 & 11.25 & 7.95 & 401.52 & 24.71 & 81.31 & 3,485.00 & 5.29 & 2,455.26 \\
 & 1,000,000 & 940.42 & 156.83 & 47.97 & 6,530.44 & 380.06 & 12,183.95 & 13,089.10 & 1,598.82 & 13,253.72 \\
\midrule\multirow[c]{4}{*}{Total TCs} & 1,000 & 32.78 & 30.14 & 29.21 & 48.46 & 33.06 & 45.99 & 57.64 & 27.71 & 63.04 \\
 & 10,000 & 34.93 & 31.37 & 30.47 & 66.77 & 35.22 & 50.52 & 103.99 & 28.13 & 103.95 \\
 & 100,000 & 56.72 & 37.25 & 33.95 & 427.52 & 50.71 & 107.31 & 3,511.00 & 31.29 & 2,481.26 \\
 & 1,000,000 & 966.42 & 182.83 & 73.97 & 6,556.44 & 406.06 & 12,209.95 & 13,115.10 & 1,624.82 & 13,279.72 \\

\bottomrule\\[3.0em]

\toprule
Pair &&
\multicolumn{1}{c}{ETH}  & 
\multicolumn{1}{c}{ETH}  & 
\multicolumn{1}{c}{ETH}  & 
\multicolumn{1}{c}{LINK} & 
\multicolumn{1}{c}{BTC}  & 
\multicolumn{1}{c}{DAI}  & 
\multicolumn{1}{c}{MANA} & 
\multicolumn{1}{c}{USDC} & 
\multicolumn{1}{c}{\textbf{Average}}  \\
     && 
\multicolumn{1}{c}{USDC} & 
\multicolumn{1}{c}{USDT} & 
\multicolumn{1}{c}{BTC} & 
\multicolumn{1}{c}{ETH} & 
\multicolumn{1}{c}{USDC} & 
\multicolumn{1}{c}{ETH} & 
\multicolumn{1}{c}{ETH} & 
\multicolumn{1}{c}{USDT} & 
\multicolumn{1}{c}{\textbf{(8 pairs)}} \\

\midrule\\[-0.0em]
\multicolumn{11}{c}{\color{Coinbase_color_dark}$\blacksquare\;$ \color{black}\textbf{Panel E: Coinbase}}\\
\midrule
Exchange Fees &  & 10.00 & 10.00 & 10.00 & 10.00 & 10.00 & 10.00 & 10.00 & 10.00 & 10.00 \\
\midrule\multirow[c]{4}{*}{B/A Spread} & 1,000 & 11.44 & 6.82 & 3.58 & 24.83 & 26.50 & 19.83 & 47.34 & 1.67 & 19.83 \\
 & 10,000 & 11.03 & 7.02 & 5.14 & 94.72 & 19.54 & 29.42 & 198.83 & 2.43 & 19.54 \\
 & 100,000 & 15.87 & 11.61 & 12.44 & 1,148.65 & 13.24 & 1,392.26 & 2,561.75 & 182.50 & 182.50 \\
 & 1,000,000 & 114.22 & 236.85 & 95.19 & 7,279.33 & 74.96 & 9,213.66 & 15,825.19 & 7,753.72 & 7,279.33 \\
\midrule\multirow[c]{4}{*}{Total TCs} & 1,000 & 21.44 & 16.82 & 13.58 & 34.83 & 36.50 & 29.83 & 57.34 & 11.67 & 29.83 \\
 & 10,000 & 21.03 & 17.02 & 15.14 & 104.72 & 29.54 & 39.42 & 208.83 & 12.43 & 29.54 \\
 & 100,000 & 25.87 & 21.61 & 22.44 & 1,158.65 & 23.24 & 1,402.26 & 2,571.75 & 192.50 & 192.50 \\
 & 1,000,000 & 124.22 & 246.85 & 105.19 & 7,289.33 & 84.96 & 9,223.66 & 15,835.19 & 7,763.72 & 7,289.33 \\

\midrule\\[1.0em]
\multicolumn{11}{c}{\color{Huobi_color_dark}$\blacksquare\;$ \color{black}\textbf{Panel F: Huobi}}\\
\midrule
Exchange Fees &  & 20.00 & 20.00 & 20.00 & 20.00 & 20.00 & 20.00 & 20.00 & 20.00 & 20.00 \\
\midrule\multirow[c]{4}{*}{B/A Spread} & 1,000 & 14.21 & 1.93 & 1.60 & 44.28 & 12.92 & 142.80 & 72.55 & 1.13 & 10.98 \\
 & 10,000 & 43.06 & 2.88 & 3.91 & 732.04 & 25.70 & 605.27 & 1,426.77 & 1.38 & 17.06 \\
 & 100,000 & 1,320.77 & 6.67 & 18.59 & 9,016.60 & 316.52 & 847.41 & 14,368.14 & 20.75 & 90.10 \\
 & 1,000,000 & 16,059.75 & 35.00 & 208.73 & 13,799.28 & 5,124.96 & 313.44 & 19,983.70 & 4,374.30 & 2,982.98 \\
\midrule\multirow[c]{4}{*}{Total TCs} & 1,000 & 34.21 & 21.93 & 21.60 & 64.28 & 32.92 & 152.80 & 92.55 & 21.13 & 30.98 \\
 & 10,000 & 63.06 & 22.88 & 23.91 & 752.04 & 45.70 & 615.27 & 1,446.77 & 21.38 & 37.06 \\
 & 100,000 & 1,340.77 & 26.67 & 38.59 & 9,036.60 & 336.52 & 857.41 & 14,388.14 & 40.75 & 110.10 \\
 & 1,000,000 & 16,079.75 & 55.00 & 228.73 & 13,819.28 & 5,144.96 & 323.44 & 20,003.70 & 4,394.30 & 3,002.98 \\

\midrule\\[1.0em]
\multicolumn{11}{c}{\color{Okex_color_dark}$\blacksquare\;$ \color{black}\textbf{Panel G: OKX}}\\
\midrule
Exchange Fees &  & 10.00 & 10.00 & 10.00 & 10.00 & 10.00 & 10.00 & 10.00 & 10.00 & 10.00 \\
\midrule\multirow[c]{4}{*}{B/A Spread} & 1,000 & 15.56 & 0.69 & 2.57 & 46.03 & 7.77 & 265.77 & 65.16 & 1.24 & 7.77 \\
 & 10,000 & 32.96 & 1.81 & 4.56 & 433.48 & 14.38 & 1,181.12 & 1,249.52 & 1.31 & 14.38 \\
 & 100,000 & 1,198.09 & 5.70 & 16.53 & 4,547.56 & 198.58 & 4,302.55 & 9,695.99 & 1.69 & 63.15 \\
 & 1,000,000 & 8,513.80 & 39.40 & 212.55 & 14,464.59 & 6,969.19 & 7,413.23 & 12,781.03 & 5.06 & 1,819.75 \\
\midrule\multirow[c]{4}{*}{Total TCs} & 1,000 & 25.56 & 10.69 & 12.57 & 56.03 & 17.77 & 275.77 & 75.16 & 11.24 & 17.77 \\
 & 10,000 & 42.96 & 11.81 & 14.56 & 443.48 & 24.38 & 1,191.12 & 1,259.52 & 11.31 & 24.38 \\
 & 100,000 & 1,208.09 & 15.70 & 26.53 & 4,557.56 & 208.58 & 4,312.55 & 9,705.99 & 11.69 & 73.15 \\
 & 1,000,000 & 8,523.80 & 49.40 & 222.55 & 14,474.59 & 6,979.19 & 7,423.23 & 12,791.03 & 15.06 & 1,829.75 \\

\bottomrule
\end{tabular}
}
		\end{table*}
		\clearpage\newpage

		%
			\begin{table}[ht]
				\centering
				\resizebox{0.93\textwidth}{!}{\begin{tabular}{lcccccc}
\toprule
                       & (1)           & (2)           & (3)           & (4)           & (5)           & (6)           \\
\midrule
 Dep. Variable         & DEX Price & DEX Price & DEX Price & DEX Price & DEX Price & DEX Price \\
                       & $\quad$Deviation$\quad$ & $\quad$Deviation$\quad$ & $\quad$Deviation$\quad$ & $\quad$Deviation$\quad$ & $\quad$Deviation$\quad$ & $\quad$Deviation$\quad$ \\[1em]
 Gas Price           & 4.153***      & 4.764***      & 4.771***      & 4.343***      & 4.358***      & 4.402***      \\
                     & (6.817)       & (4.043)       & (4.113)       & (3.280)       & (3.109)       & (2.908)       \\[1em]
 DEX TCs (no gas)    &               & 2.830***      & 2.793***      & 2.605***      & 2.609***      & 2.507***      \\
                     &               & (6.764)       & (5.610)       & (4.614)       & (4.468)       & (7.417)       \\[1em]
 CEX Deviation       &               &               & 0.248         & -0.175        & -0.173        & -0.061        \\
                     &               &               & (0.358)       & (-0.338)      & (-0.338)      & (-0.301)      \\[1em]
 CEX-DEX Spread      &               &               &               & 1.770**       & 1.768**       & 2.087***      \\
                     &               &               &               & (2.258)       & (2.238)       & (4.063)       \\[1em]
 Volatility          &               &               &               &               & 0.070         & 0.097         \\
                     &               &               &               &               & (0.194)       & (0.227)       \\[1em]
 Absolute ETH Return &               &               &               &               &               & -0.658        \\
                     &               &               &               &               &               & (-0.325)      \\[1em]
 Intercept           & 23.354***     &               &               &               &               &               \\
                     & (36.749)      &               &               &               &               &               \\[1em]
 Observations        & 115,450       & 115,450       & 115,450       & 115,450       & 115,450       & 115,450       \\
 R-squared           & 0.366         & 0.368         & 0.368         & 0.370         & 0.370         & 0.370         \\
 Triplet FE            & Yes           & Yes           & Yes           & Yes           & Yes           & Yes           \\
\bottomrule
\end{tabular}
}
				\captionsetup{justification=justified, singlelinecheck=on, font=footnotesize}
				\caption{\textbf{Price Efficiency and Gas Price.}\footnotesizelarge 
			The table presents the findings of panel regression analyses on triangular price deviations on DEXs and their relationship with the median gas price on the Ethereum blockchain, based on data sampled at an hourly frequency by triplet.
			The dependent variable is the absolute value of price deviations $\mid\theta\mid$, defined in equation \eqref{eq:theta}, and constructed by taking the minimum price deviation between Uniswap v2 and v3 by triplet and hour.
			In specification (2) we control for hourly DEX transaction costs excluding gas fees, averaged across the pairs composing each triplet, to test if gas prices have an incremental explanatory power on top of spreads and exchange fees.
			In specifications (3)-(6) we incrementally add controls for other potential determinants of DEX price deviations, including 
			the contemporaneous price deviation on CEXs (minimum between Binance and Kraken), 
			the sum of the absolute percentage spreads between CEX and DEX prices for the pairs involved in each triplet, 
			the realized return volatility over the past 24 hours averaged across the pairs of each triplet, 
			and the absolute return of the USD-denominated price of ETH. 
			We include triplet fixed effects in all specifications, to account for the different average level of price deviations across the triplets.
			T-stats are reported in parentheses and are based on robust standard errors double-clustered by triplet and hour. 
			Asterisks denote significance levels (***$ =1\%$, **$ =5\%$, *$ =10\%$).
		}\label{tab:DEX_deviations_reg}
			\end{table}

			\begin{table}[ht]
				\centering
				\resizebox{0.93\textwidth}{!}{\begin{tabular}{lcccc}
\\
\multicolumn{5}{c}{\textbf{Panel A: First Stage}}\\[0.2em]
\toprule
                     & (1)              & (2)              & (3)              & (4)               \\
\midrule
 Dep. Variable       $\qquad\qquad\qquad$
                     & Median Gas Price & Median Gas Price & Median Gas Price & Average Gas Price \\[1em]
 NFT Mint Volume (USD) & 5.873***         & 6.192***         & 6.208***         & 7.991***          \\
                     & (5.792)          & (4.023)          & (4.036)          & (4.882)           \\[1em]
 ETH Absolute Return &                  &                  & 3.549***         & 4.143***          \\
                     &                  &                  & (7.704)          & (7.628)           \\[1em]
 Intercept           & -9.685           &                  &                  &                   \\
                     & (-0.885)         &                  &                  &                   \\[1em]
 Observations        & 14,878           & 14,878           & 14,878           & 14,878            \\
 R-squared           & 0.014            & 0.437            & 0.439            & 0.411             \\
 Day FEs             &                  & Yes              & Yes              & Yes               \\
\bottomrule\\[2em]

\multicolumn{5}{c}{\textbf{Panel B: Second Stage}}\\[0.2em]
\toprule
                        & (1)             & (2)             & (3)             & (4)             \\
\midrule
 Dep. Variable          $\qquad\qquad\qquad$
                        & Price Deviation & Price Deviation & Price Deviation & Price Deviation \\[1em]
 Instrumented Gas Price & 4.400***        & 2.071***        & 2.084***        & 1.445***        \\
                        & (5.192)         & (3.142)         & (3.166)         & (3.419)         \\[1em]
 Intercept              & 11.865***       &                 &                 &                 \\
                        & (6.031)         &                 &                 &                 \\[1em]
 Observations           & 109,983         & 109,983         & 109,983         & 109,983         \\
 R-squared              & 0.035           & 0.101           & 0.201           & 0.234           \\
 Day FEs                &                 & Yes             & Yes             & Yes             \\
 Triplet FEs            &                 &                 & Yes             & Yes             \\
 Controls               &                 &                 &                 & Yes             \\

\bottomrule
\end{tabular}

}
				\captionsetup{justification=justified, singlelinecheck=on, font=footnotesize}
				\caption{\textbf{Price Efficiency and NFT Shocks to Gas Price.}\footnotesizelarge 
			The table presents results from an instrumental variable regression, and it is divided into two panels. 
			Panel A presents the first-stage regression results, based on data at the hourly frequency. 
			In the first specification, the median gas price of the hour is regressed on the log volume (in USD) of NFTs minted during that hour, to demonstrate the relevance of the instrument. 
			In specification (2) we saturate the model with time fixed effects, and in specification (3) we control for the absolute value of percentage price changes in ETH-USD. Finally, in the last specification, we use average gas prices as the dependent variable.
			Panel B displays the second-stage regression results, based on a panel at the triplet-hour level.
			The magnitude of price deviations measured in Uniswap v3 is regressed on the predicted values of the gas price from column (i) of the first stage, to estimate the causal effect of gas prices on DEX price efficiency.
			In specifications (2) and (3) we saturate the model with day and triplet fixed effects, while in columns (4) we add controls, including 
			the contemporaneous price deviation on CEXs (minimum between Binance and Kraken), 
			the sum of the absolute percentage spreads between CEX and DEX prices for the pairs involved in each triplet, 
			the realized return volatility over the past 24 hours averaged across the pairs of each triplet, 
			and the absolute return of the USD-denominated price of ETH.
			T-stats are reported in parentheses and are based on robust standard errors double-clustered by triplet and day. 
			Asterisks denote significance levels (***$ =1\%$, **$ =5\%$, *$ =10\%$).
		}\label{tab:DEV_NFT_reg}
			\end{table}

		%
			\begin{table}[ht]
				\centering
				\resizebox{0.86\textwidth}{!}{\begin{tabular}{lcccc}
\toprule
 & (1) & (2) & (3) & (4) \\
\midrule
Dep. Variable $\qquad\qquad$  & $\quad$Transaction   $\quad$ & $\quad$Transaction$\quad$  
                              & Price           & Price          \\
& Costs          & Costs       
                              & $\quad$Deviations$\quad$     & $\quad$Deviations$\quad$     \\[0.5em]
$\text{DEX} \times \text{V3} \qquad$ 
                              & -55.895*** & -55.921*** & -88.626*** & -86.562***  \\
                              & (-3.231)   & (-3.228)   & (-8.685)   & (-8.355)    \\[1.25em]
$\text{DEX}$                  & 96.378***  & 103.706*** & 119.313*** & 118.224***  \\
                              & (6.286)    & (6.708)    & (12.945)   & (12.651)    \\[1.25em]
$\text{V3}$                   & 10.231     &            & -0.945**   &             \\
                              & (1.314)    &            & (-2.160)   &             \\[1.25em]
Intercept                     & 62.469***  &            & 2.695***   &             \\
                              & (21.868)   &            & (6.511)    &             \\[2.0em]
Observations                  & 150,654    & 150,654    & 34,488     & 34,488      \\
$\text{R}^2$                  & 0.031      & 0.212      & 0.100      & 0.214       \\[1.25em]
Week FEs                      & No         & Yes        &  No        & Yes         \\
Pair/Triplet FEs              & No         & Yes        &  No        & Yes         \\
\bottomrule
\end{tabular}
}
				\captionsetup{justification=justified, singlelinecheck=on, font=footnotesize}
				\caption{\textbf{Introduction of Uniswap v3.}\footnotesizelarge 
			The table reports results from a difference-in-differences regression around the introduction of Uniswap v3, deployed on the Ethereum Mainnet on May 5th, 2021.
			The regression is run on an hour-pair-level panel of the best transaction costs and smallest price deviations offered by CEXs (\CEXs) and DEXs (Uniswap v2, Uniswap v3). 
			The measure of transaction costs is defined in equation \eqref{eq:TC_LOB} and \eqref{eq:TC_AMM}, 
			while price deviations are defined in equation \eqref{eq:theta}.
			In the first two specifications, the dependent variable is the lowest transaction costs offered by CEXs and DEXs, respectively, constructed by taking the minimum within each exchange category, by pair and hour.
			In the remaining specifications, the dependent variable is the lowest absolute price deviations on CEXs and DEXs, respectively, constructed by taking the minimum within each exchange category, by triplet and hour.
			The \emph{treatment} dummy "DEX" is equal to $1$ for pairs and triplets on DEXs, while it equals zero for CEXs.
			The \emph{post} dummy "V3" equals one after the introduction of Uniswap v3.
			We saturate the model with triplet and week fixed effects.
			T-stats are reported in parentheses and are based on standard errors double-clustered by week and pair or by week and triplet. 
			Asterisks denote significance levels (***$ =1\%$, **$ =5\%$, *$ =10\%$).
		}\label{tab:diff_n_diffs_table}
			\end{table}

			\begin{table}[ht]
				\centering
				\resizebox{0.86\textwidth}{!}{\begin{tabular}{llrrrrrrrrr}
\midrule
Pair &&
\multicolumn{1}{c}{ETH}  & 
\multicolumn{1}{c}{ETH}  & 
\multicolumn{1}{c}{ETH}  & 
\multicolumn{1}{c}{LINK} & 
\multicolumn{1}{c}{BTC}  & 
\multicolumn{1}{c}{DAI}  & 
\multicolumn{1}{c}{MANA} & 
\multicolumn{1}{c}{USDC} & 
\multicolumn{1}{c}{$\quad$\textbf{Average}$\quad$}  \\
&& 
\multicolumn{1}{c}{USDC} & 
\multicolumn{1}{c}{USDT} & 
\multicolumn{1}{c}{BTC} & 
\multicolumn{1}{c}{ETH} & 
\multicolumn{1}{c}{USDC} & 
\multicolumn{1}{c}{ETH} & 
\multicolumn{1}{c}{ETH} & 
\multicolumn{1}{c}{USDT} & 
\multicolumn{1}{c}{\textbf{(20 pairs)}} \\

\midrule
Size & Fees$$ & \multicolumn{8}{c}{Percentage Utilization}  \\[0.25em]
\midrule
\multirow[t]{3}{*}{\$1,000} 
 & 1 bps & 18.740 & 0.000 & 0.713 & 0.000 & 0.095 & 0.000 & 0.000 & 81.428 & 6.504 \\ 
 & 5 bps & 81.260 & 100.000 & 98.769 & 14.128 & 79.242 & 100.000 & 0.000 & 18.572 & 31.962 \\ 
 & 30 bps & 0.000 & 0.000 & 0.518 & 85.872 & 20.663 & 0.000 & 100.000 & 0.000 & 51.499 \\ 
 & 100 bps & 0.000 & 0.000 & 0.000 & 0.000 & 0.000 & 0.000 & 0.000 & 0.000 & 9.577 \\

\midrule
\multirow[t]{3}{*}{\$10,000} 
 & 1 bps & 9.711 & 0.000 & 0.713 & 0.000 & 25.725 & 0.000 & 0.000 & 81.364 & 7.188 \\ 
 & 5 bps & 90.148 & 99.391 & 98.196 & 5.394 & 37.040 & 95.406 & 0.000 & 18.636 & 26.482 \\ 
 & 30 bps & 0.141 & 0.609 & 1.091 & 94.606 & 37.235 & 4.594 & 98.091 & 0.000 & 56.862 \\ 
 & 100 bps & 0.000 & 0.000 & 0.000 & 0.000 & 0.000 & 0.000 & 1.909 & 0.000 & 9.009 \\

\midrule
\multirow[t]{3}{*}{\$100,000} 
 & 1 bps & 3.340 & 0.000 & 4.249 & 0.000 & 40.407 & 0.000 & 0.000 & 81.305 & 7.636 \\ 
 & 5 bps & 96.237 & 88.962 & 93.752 & 7.425 & 2.345 & 70.503 & 0.000 & 18.695 & 24.153 \\ 
 & 30 bps & 0.423 & 11.038 & 1.999 & 92.575 & 57.248 & 29.497 & 95.660 & 0.000 & 55.947 \\ 
 & 100 bps & 0.000 & 0.000 & 0.000 & 0.000 & 0.000 & 0.000 & 4.340 & 0.000 & 11.805 \\

 \midrule
\multirow[t]{3}{*}{\$1,000,000} 
 & 1 bps & 64.269 & 0.000 & 12.297 & 0.000 & 41.098 & 0.000 & 0.000 & 81.469 & 10.359 \\ 
 & 5 bps & 32.137 & 70.249 & 62.342 & 74.171 & 19.177 & 85.249 & 0.000 & 18.531 & 28.536 \\ 
 & 30 bps & 3.594 & 29.751 & 25.361 & 25.829 & 39.726 & 14.751 & 99.909 & 0.000 & 48.869 \\ 
 & 100 bps & 0.000 & 0.000 & 0.000 & 0.000 & 0.000 & 0.000 & 0.091 & 0.000 & 11.777 \\

 \midrule
\end{tabular}}
				\captionsetup{justification=justified, singlelinecheck=on, font=footnotesize}
				\caption{\textbf{Fees Heterogeneity on Uniswap v3.}\footnotesizelarge 
			The table provides empirical evidence on the usefulness of the new Multiple Fee Tiering (MFT) system introduced by Uniswap v3,
			presenting the heterogeneity in exchange fees for the same exchange pair, across different pools.
	        More specifically, for each trade size and for each pair-hour we select the pool providing the lowest total transaction costs on Uniswap v3, as defined in equation \eqref{eq:TC_AMM}.
	        The table reports the frequency with which each pool is selected, in percentage terms, over the sample period.
	        The first eight columns present the results for the eight pairs in the intersection of Uniswap v2, Uniswap v3, \CEXs.
	        The last column presents the average across the twenty pairs in the intersection of Uniswap v2, Uniswap v3, Binance, and Kraken.
		}\label{tab:fees_heterogeneity}
			\end{table}


		%
		\figurecap{Aggregate Trading Volume}{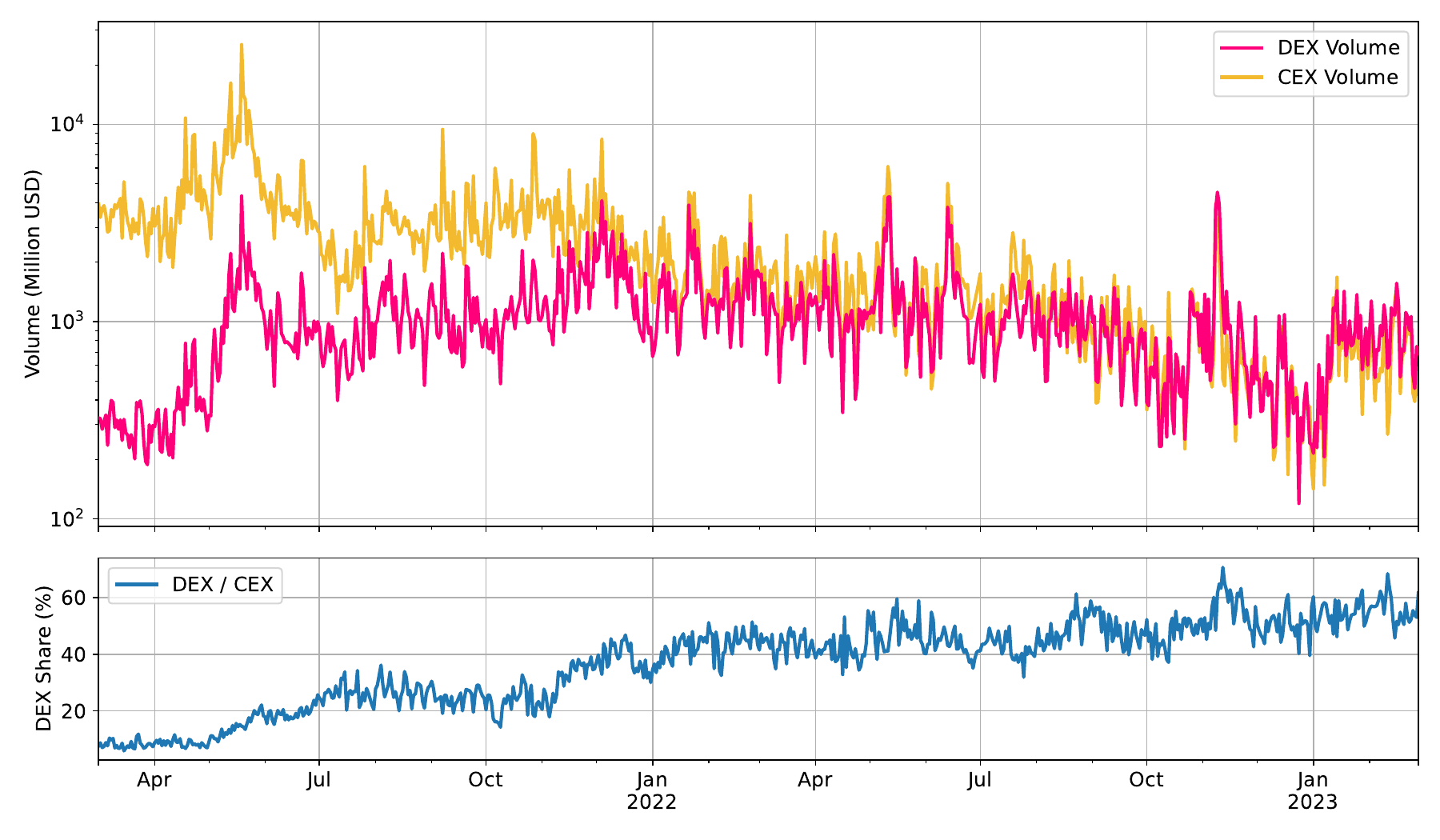}{0.95}{
			The top panel presents the aggregate trading volumes for the CEXs (\CEXs) and DEXs (Uniswap v2, and Uniswap v3) in our sample,
			from March 2021 to February 2023.
			The reported metric is the total trading volume summed across the 20 exchange pairs covered by our sample.
			The vertical axis uses a logarithmic scale and is expressed in million USD.
			The bottom panel displays the share of DEX volume relative to the total volume generated by CEXs and DEXs, defined as in the top panel, expressed in percentage points.
		}

		%
		\figurecap{Distribution of CEX and DEX Trade Sizes}{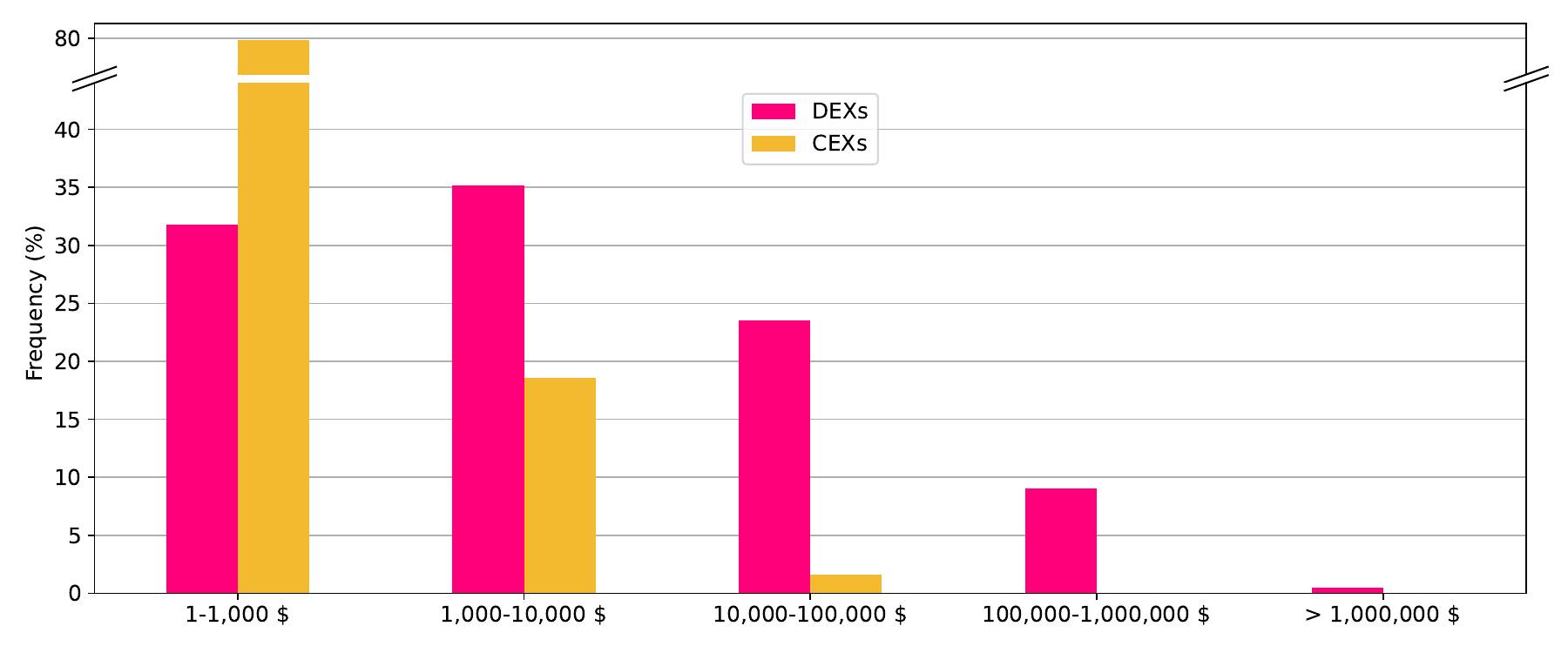}{0.93}{
			The figure presents the frequency distribution of trade sizes in US Dollars for transactions executed on CEXs and DEXs,
			during our sample period and restricting to the twenty pairs featured in our study.
			Transactions with a Dollar value below $0.1$ are excluded, and gas costs are not included.
			The data for CEX is retrieved from tardis,
			while for DEXs, the data is retrieved from a Dune Analytics query based on the `\lstinline{uniswap_v2_ethereum.trades}' and `\lstinline{uniswap_v3_ethereum.trades}' tables.
		}
		\clearpage\newpage

		\figurecap{Gas Fees}{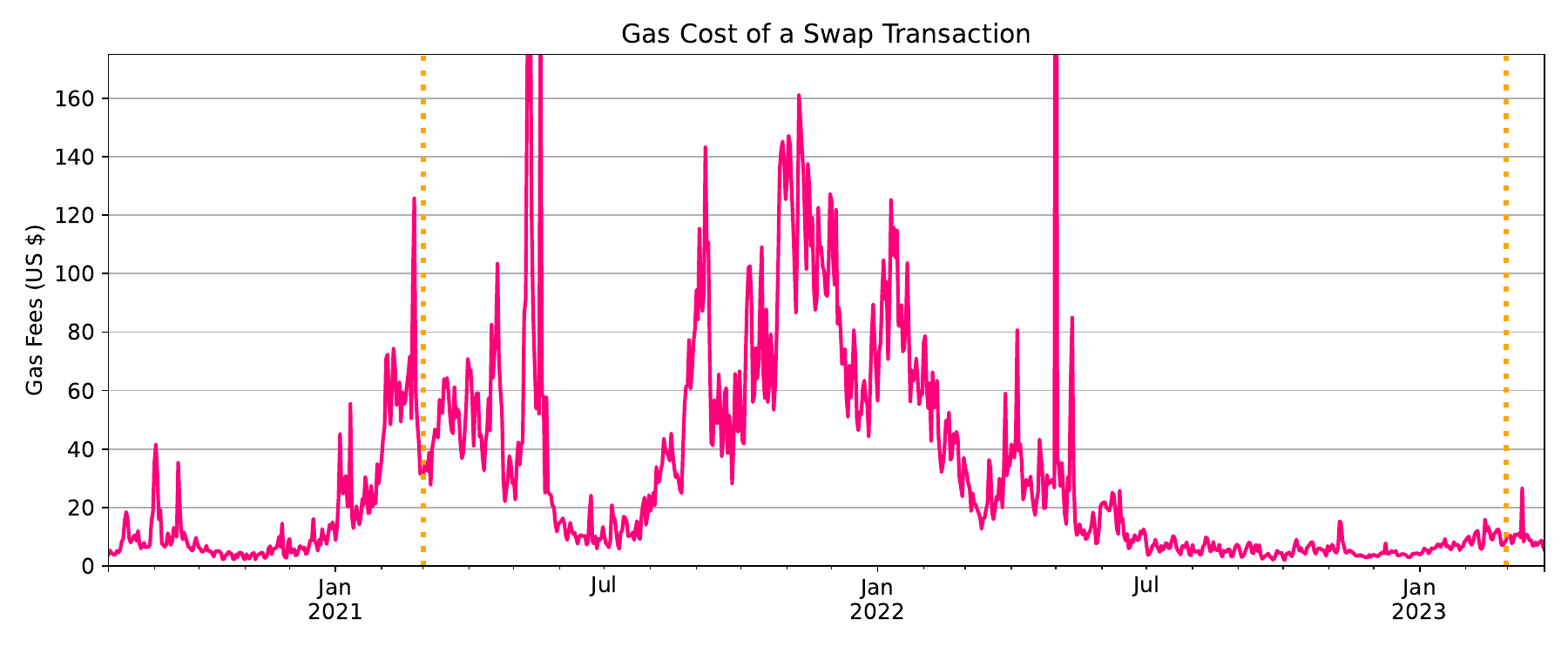}{0.95}{
			The figure presents the time series evolution of the gas costs of a swap transaction in our sample in USD.
			This is computed at the hourly frequency, multiplying the units of gas required to execute a swap, numbering roughly 120,000, by the average gas price associated with transactions in the blocks validated during each hour in USD.
			Since the number of gas units is constant over time, the time series variation comes from oscillating gas prices in ETH and the fluctuation of the USD/ETH exchange rate.
			The two orange vertical lines indicate the period on which our main market quality analysis is conducted, that is, from the beginning of March 2021 to the end of February 2023.
		}	


		\vspace{4em}

		\figurecap{Transaction Costs}{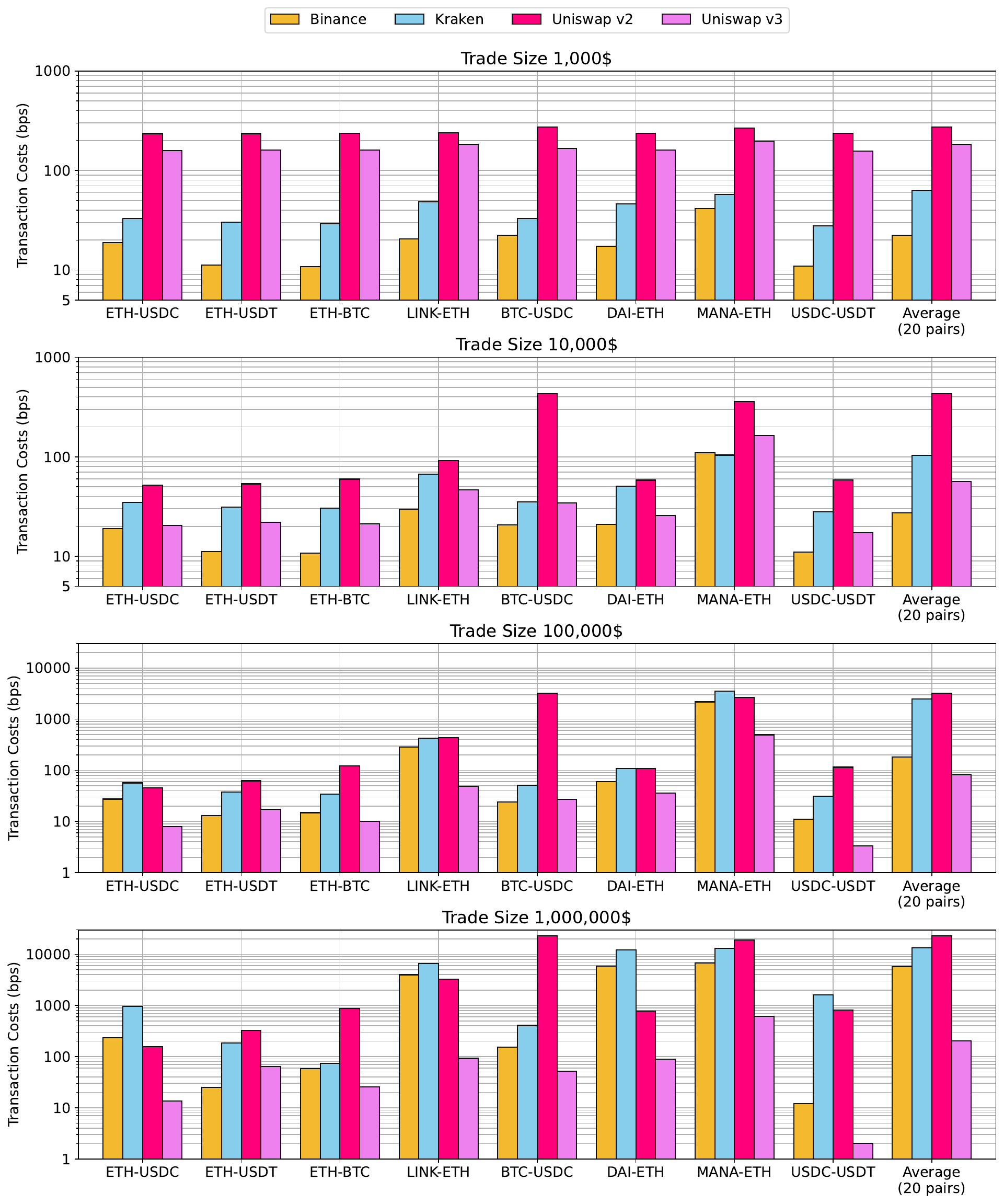}{0.95}{
			The figure presents transaction costs, computed as in equation \eqref{eq:TC_LOB} for the LOB-based Binance and Kraken, and on equation \eqref{eq:TC_AMM} for the AMM-based Uniswap v2 and Uniswap v3.
			We report results for the eight pairs in the intersection of Uniswap v2, Uniswap v3, \CEXs. 
			We also include the average across the twenty pairs in the intersection of Uniswap v2, Uniswap v3, Binance, and Kraken.
			The transaction costs are computed at an hourly frequency for different trade sizes ($10^3, 10^4, 10^5$, and $10^6$ US dollars), then averaged from March 2021 to February 2023.
			As discussed in Section \ref{sec:TCs}, the displayed transaction costs include B/A spreads, exchange fees 
			(30 basis points for Uniswap v2, variable for Uniswap v3, 10 basis points for Binance, and 26 basis points for Kraken), 
			and gas fees for Uniswap v2 and Uniswap v3.
			For Uniswap v3, daily and for each trade size, we consider the liquidity pool offering the lowest transaction costs among those available for the specific exchange pair.
			The vertical axis employs a logarithmic scale and is reported in basis points.
		}

		%
		\figurecap{Average Price Deviations}{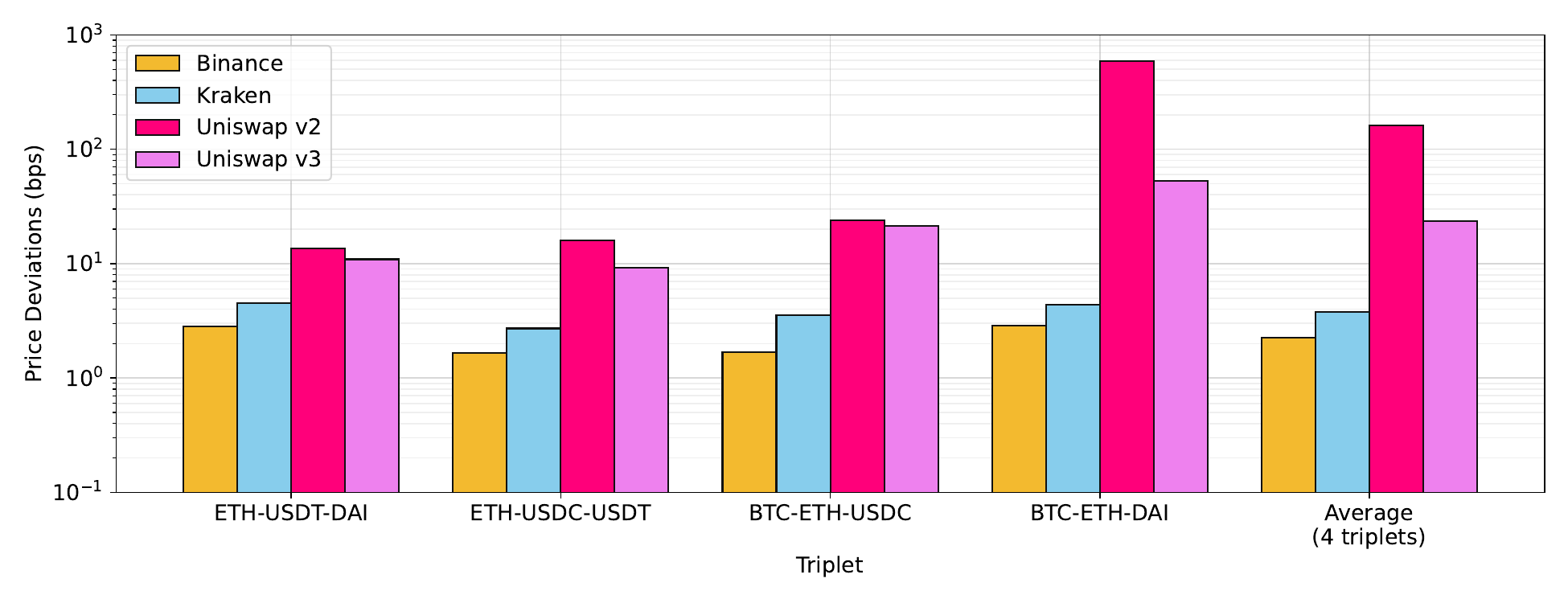}{0.95}{
			The figure displays price inefficiency levels, proxied by the absolute value of price deviations from the law of one price, defined by equation \eqref{eq:theta}.
			These are computed separately for each trading triplet and for each exchange, namely for CEXs (Binance and Kraken), and for DEXs (Uniswap v2 and Uniswap v3).
			Absolute price deviations are first estimated at the hourly frequency for the five triplets in our sample, then averaged from March 2021 to February 2023.
			On the rightmost part of the figure, we also report the average price deviations across the four triplets.
			For Uniswap v3 the estimation is based on a shorter sample starting on May 5th, 2021, when the exchange was launched.
			For the pairs involving USDC in Binance, the estimation is based on a truncated sample that ends on September 29th, 2022, when the exchange ceased quoting trading pairs in USDC.
			The vertical axis employs a logarithmic scale and is reported in basis points.
		}

		\figurecap{Price Deviations}{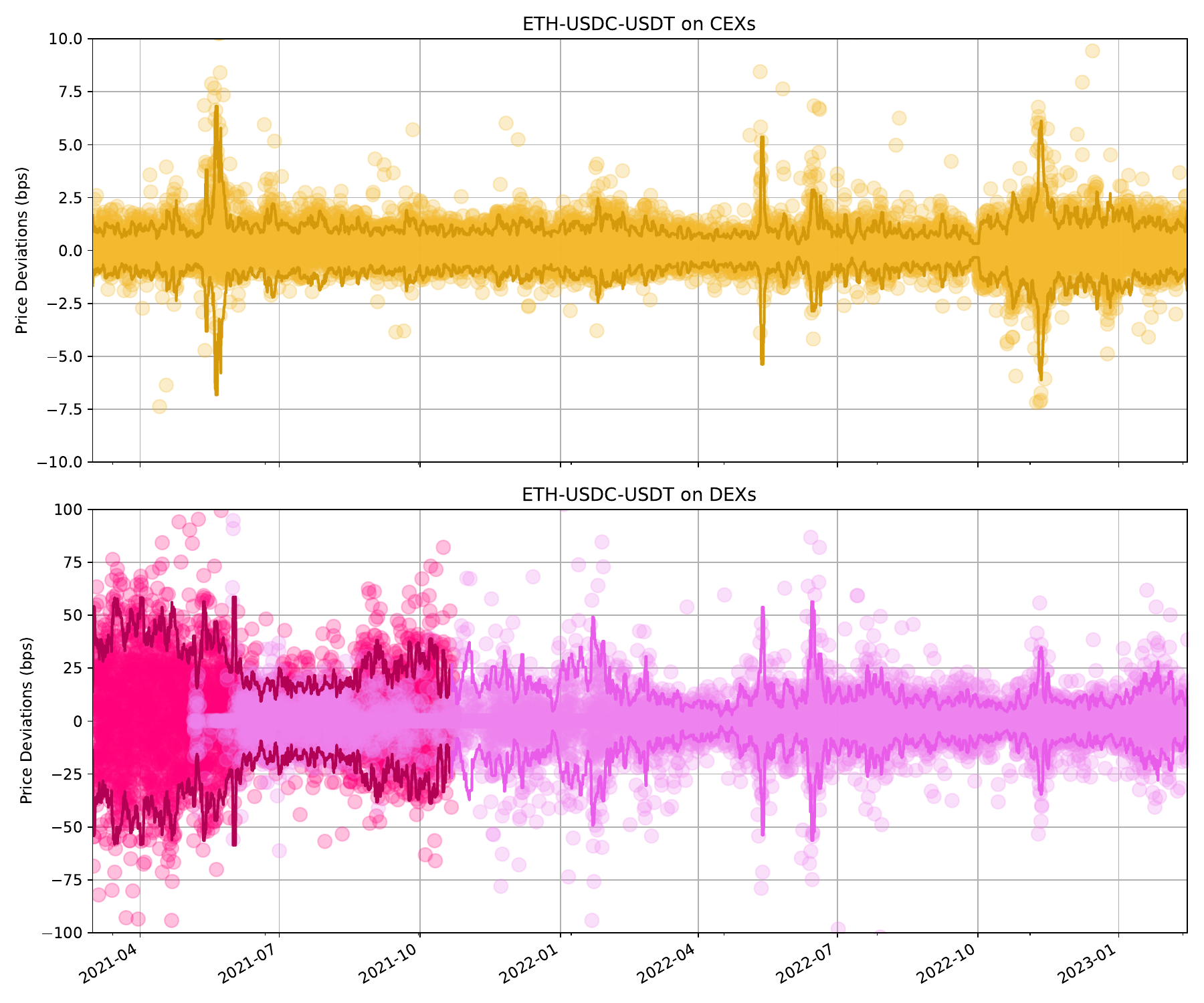}{0.93}{
			The figure displays the estimated price deviations from the law of one price $\theta$ for the triplet USDC-USDT-ETH, at an hourly frequency over the period from March 2021 to February 2023.
			The top panel illustrates deviations for CEXs. The time series is derived by selecting the minimum deviation (in absolute value) between Binance and Kraken for each hour.
			The bottom panel displays the deviations for DEXs. 
			This time series is generated by selecting the smallest absolute deviation between Uniswap v2 and Uniswap v3 at each hourly interval. 
			If the minimum deviation is found in v2, the corresponding data point is marked in pink, whereas if v3 has the minimum deviation, the data point is highlighted in violet.
			Please note that the vertical axes of the two panels have different scales.
			On both panels, the solid lines represent the top decile of the distribution of absolute deviations, estimated on a 7-day rolling window.
		}
		\clearpage

		\newpage\clearpage
		\renewcommand{\thesubsection}{\Alph{subsection}}
		\renewcommand\baselinestretch{1.1}
		\setstretch{1.5}
		\section*{Internet Appendix}\label{appendix}
		\subsection{Mathematical Foundations of AMM Markets}
	During both our sample period and at the time of writing, the majority of AMMs relied on the \emph{constant product rule}, which enables an algebraic determination of market price and transaction price based on the available reserves \citep{adams2021uniswap}.\footnote{
		There exist AMMs based on similar algebraic rules (e.g., \emph{constant sum}).
		Nevertheless, according to monthly historical snapshots of the \href{https://www.coingecko.com/en/dex}{CoinGecko DEX ranking}, the market share of constant-product AMMs has been above 60\% in the period from January 2021 and June 2023. We include Uniswap v3 in such a category even though, formally, the constant-product rule applies only locally within each tick \citep{adams2021uniswap}.
	}
	The leading example is Uniswap, developed and deployed on November 2018 by Hayden Adams, a former mechanical engineer at Siemens.
	In subsequent sections, we first offer a concise overview of the mathematical framework underpinning pure constant-product AMMs, employed by Uniswap v2, Sushiswap, Pancakeswap, and numerous other DEXs. 
	Next, we outline the functioning of the more recent Uniswap v3, which can be viewed as a generalization of the former.
	Finally, we introduce and formalize the concept of \emph{impermanent loss}.

	\subsubsection{Uniswap v2}\label{sub:uni_v2}
		Let $X$ and $Y$ be two crypto tokens. Consider the exchange pair $X\leftrightarrow Y$, and the associated liquidity pool containing $x$ units of $X$ and $y$ units of $Y$.
		The amount of tokens in the pool determines the current market price $P_{XY}$ of $X$ in terms of $Y$ and its inverse $P_{YX}$, which can be expressed as
		\begin{equation}\label{eq:quoted_price}
			P_{XY} = \frac{y}{x} \AND P_{YX} = \frac{x}{y}
		\end{equation}
		Let us denote as $f$ the percentage exchange fees charged by the DEX, and let $\varphi = 1-f$.
		These fees are immediately applied to the traded amount $\dx>0$, so that the net quantity of token $X$ that goes into the swap transaction is $\ddx$.
		Each trade (swap transaction) is automatically regulated by the constant product rule, which states that the product of the reserves must remain constant before and after any transaction.
		Hence, when trading an amount $\dx > 0$ of token $X$ in exchange for token $Y$, the output quantity $\dy$ is mathematically determined by the following equation
		$$ xy = k = (x+\ddx)(y-\dy)\;, $$
		where $k$ is the constant product invariant. 
		Solving for $\dy$, one obtains that the output amount is given by
		\begin{equation}\label{eq:amount_out}
			\dy = y \frac{\ddx}{x+\ddx}.
		\end{equation}
		The transaction price is, therefore, lower than the quoted price and is given by 
		$$ T_{XY}(\dx) \;=\; \frac{\dy}{\varphi\dx} \;=\; \frac{y}{x + \ddx} $$
		and the quoted half-spread (as a percentage of the quoted price) can be computed as
		\begin{equation}\label{eq:spread_AMM}
			S_{XY}(\dx) = \frac{P_{XY} - T_{XY}}{P_{XY}} = \frac{\ddx}{x + \ddx} \;.
		\end{equation}
		Note that the above is an increasing but concave function of the transaction volume $\dx$, implying that larger volumes have a larger impact on prices but with a marginally decreasing effect. 
		\textit{Ceteris paribus}, a purchase could have a greater impact than a sale, as shown in \cite{aoyagi2021coexisting}. 
		Throughout the paper, we compute the quoted half-spread for both directions ($X\to Y$ and $Y\to X$) for the same traded amount in terms of dollars and consider the average of the two measures.
		In the following, we refer to this metric as \emph{Bid/Ask Spread} or \emph{B/A Spread} for short.

	\subsubsection{Uniswap v3}\label{sub:uni_v3}
		Uniswap v3, released on May 5th, 2021, is based on a generalization of the constant-product AMM model.
		The upgrade, deployed through a new set of smart contracts,\footnote{
			A comprehensive list of the address of each deployed contract is available on the official Uniswap documentation at \href{https://docs.uniswap.org/protocol/reference/deployments}{https://docs.uniswap.org/protocol/reference/deployments}.
		} 
		includes two innovations that are highly relevant to our market quality analysis, namely:
			(i)  the Multiple Fee Tiering (MFT) system, that is, the possibility for LPs to choose the level of exchange fees;
			(ii) the Discretionary Price Ranges (DPR) system, allowing LPs to post liquidity on a specific price interval.
		%

		The MFT system is based on the capability of the new protocol to deploy several liquidity pools for the same exchange pair, each with a different level of exchange fees attached to it.
		More specifically, the available fee levels -- in basis points -- have support in the discrete set \{1, 5, 30, 100\}.
		This implies, in particular, that there can be at most four liquidity pools for the same exchange pair.
		LPs are free to decide their allocation of liquidity across pools, and, likewise, traders can decide in which pool they want to trade.\footnote{
			Given a trade size and a trading pair, The Uniswap v3 interface automatically suggests the best possible route to follow to minimize transaction costs. The optimal solution may be achieved by splitting the trade across different pools.
		}
		Each of those pools works as a standard v2 pool and is independent of its siblings. 
		It is endowed with pool-specific levels of liquidity $x_{f}$ and $y_{f}$, aggregating the liquidity supplied by LPs at fee level $f$.
		This implies that different pools for the same currency pair may display different quoted prices and Bid/Ask spreads.
		In equilibrium, however, arbitrage activity should limit price deviations from the law of one price.
		From a theoretical standpoint, liquidity allocation depends on the risk-return trade-off faced by LPs, as described in the simple model of liquidity provision proposed in Section \ref{sub:liquidity_model} of the Internet Appendix.
		In particular, relevant factors determining the allocation of liquidity across different fee levels should include:
			(i) the expected volatility of the exchange rate, proportional to the expected impermanent loss. For pairs with a higher expected impermanent loss, LPs should require higher exchange fees as a risk compensation;
			(ii) the expected trading volume in the pair, proportional to the expected profit from liquidity provision. For pairs with a larger expected volume, LPs should accept lower exchange fees;
			(iii)
			the level of liquidity already present in the pool, since less crowded pools provide higher returns on capital to the marginal LP.
		Further, the allocation of liquidity between high- and low-fee pools may depend on the heterogeneity of capital endowment across LPs, as theorized by \cite{LPZ}.

		MFT can enhance market quality by lowering the effective transaction costs for traders. With the support of Uniswap's automatic router, traders can select the optimal pool for a specific trade size by balancing the trade-off between Bid/Ask spreads and fees.
		%
		%

		The second innovative feature of v3 is DPR, which opens up the possibility for LPs to confine their liquidity provision to a specific price interval, allowing LPs to offer liquidity in a more proactive manner.
			Technically, the protocol uses a discrete set of price ticks to divide the full price range of a trading pair into a discrete number of intervals.
			An LP can provide liquidity to a custom range, opening a so-called \emph{liquidity position}, by specifying the lower and upper ticks in addition to the supplied quantity.
			Aggregating over all LPs' positions, one obtains the distribution of liquidity over the entire price range, which can take any arbitrary shape.

			To convey some intuition, let us briefly discuss the case of a single liquidity position delimited by two subsequent ticks.
			Formally, consider a liquidity pool for the exchange pair $X\leftrightarrow Y$ and a liquidity position concentrated on the interval $[P_a, P_b]$.
			Following \cite{adams2021uniswap}, the quantities $\tilde{x}$ and $\tilde{y}$ supplied to the pool are referred to as \emph{real reserves}.
			Let $x$ and $y$ be the corresponding \emph{virtual reserves}, defined on the current interval as
			\begin{equation}\label{eq:virtual}
				x = \tilde{x} + L / \sqrt{P_b} \AND y = \tilde{y} + L \sqrt{P_a}
			\end{equation}
			where $L=\sqrt{xy}$ is referred to as the \emph{virtual liquidity} attached to the interval.\footnote{
				Note that real and virtual reserves coincide in Uniswap v2, where liquidity can be allocated only on the full price range, that is, $P_a\to 0$ and $p_b\to \infty$.
			}
			Locally on the interval, all relevant quantities are defined similarly as in Uniswap v2, but in terms of virtual reserves rather than real reserves. 
			In particular, the quoted price is $P = y/x$, and the constant product rule reads $xy=k$.\footnote{
				This implies that, differently from Uniswap v2, the dollar values of the real reserves $\tilde{x}$ and $\tilde{y}$ are not necessarily equal but depend on the relative position of the quoted price within the interval $[P_a, P_b]$.
 			If the quoted price is outside of the interval, the supplied liquidity is composed of only one of the two tokes. In this case, the liquidity position constitutes a \emph{range order}, similar to a traditional limit order \citep{adams2021uniswap}.
			}

			In other words, when a swap transaction occurs within a single price interval, the pricing rule is exactly the same as in Uniswap v2, driven by the constant-product formula applied to the virtual liquidity available on that interval.
			If the price impact of the trade pushes the quoted price across one or multiple ticks, the constant-product formula applies locally based on the virtual liquidity attached to each interval.
			
			The exchange fees collected on each interval are divided among LPs based on the share of liquidity they deposited in that specific interval. 
			In particular, a liquidity position earns fees only if swap transactions are performed on its support.
			The DPR system aims at reducing transaction costs mainly for stable-coins pairs, enjoying very low levels of exchange-rate volatility.
			For instance, on the USDC-USDT pair, where both stablecoins are pegged to the US dollar, LPs are incentivized to concentrate their liquidity within a few intervals around the price of $1$, thus leading to lower Bid/Ask spreads in that region.
			%

			%
			The Bid/Ask spread formula for Uniswap v3 is less straightforward than its v2 counterpart, described in \eqref{eq:spread_AMM}, as it depends on the liquidity distribution across price intervals.
			To derive it, consider a liquidity pool for the exchange pair $X\leftrightarrow Y$ with quoted price $P$.
			Assume the trade size is $\dx$ and denote by $P'$ the final quoted price after the transaction.
			Let us start with the simple case where a trade of size $\dx$ is fully executed within a single price interval. 
			To make the notation lighter, we ignore exchange fees in this derivation.
			Recall that the liquidity of the interval is $L=\sqrt{xy}$, where $(x,y)$ are the virtual reserves for the interval before the trade, and $(x',y')$ denote the virtual reserves after the trade.
			Since $L$ is invariant to trades, we have
			\begin{equation}
				x = L / \sqrt{P}, \qquad y = L \sqrt{P}, \qquad x' = L / \sqrt{P'}, \qquad y' = L \sqrt{P'}.
			\end{equation}
			It thus follows that the relationships between $\dx$, $\dy$, and the quoted prices are
			\begin{equation}
				\dx = L\left( 1/\sqrt{P'} - 1/\sqrt{P} \right) \AND
				\dy = L\left( \sqrt{P} - \sqrt{P'} \right)
			\end{equation}
			The transaction price can therefore be written as $\dy/\dx = \sqrt{PP'}$, showing that it equals the geometric mean of $P$ and $P'$.
			To derive a general expression for the transaction price in the case of a trade spanning multiple intervals, 
			define the price intervals $I_t = [P_t, P_{t-1}]$ for $t\in \mathbb{Z}$ such that $P\in I_0$ and let $\tau$ be a positive integer such that $P'\in I_\tau$.
			Further, let $\dx_t$ denote the amount traded within $I_t$, and let $T_t$ be the corresponding transaction price.
			The effective transaction price can thus be written as the volume-weighted average of the \emph{local} transaction prices for each of the relevant intervals as a function of the trade size and the distribution of liquidity:
			\begin{eqnarray}\label{eq:spread_V3_pre}
				T &=&
				\frac{1}{\dx}\left[
					\dx_0 T_{0} + 
					\sum_{t=1}^{\tau-1} \dx_{t} T_{t} + 
					\dx_{\tau} T_{\tau}
				\right]
				\\[2.0em]
				&=& 
				\frac{1}{\dx}\left[
 				L_0\left( \sqrt{P} - \sqrt{P_0} \right) +
 				\sum_{t=1}^{\tau-1} L_{t}\left( \sqrt{P_{t-1}} - \sqrt{P_t} \right) + 
 				L_{\tau}\left( \sqrt{P_{\tau-1}} - \sqrt{P'} \right)
 			\right]
			\end{eqnarray}
			Finally, the spread equals the percentage deviation from the quoted price, that is
			\begin{equation}\label{eq:spread_V3}
				S_{XY}(\dx) = \frac{P-T}{P}
			\end{equation}

	\subsubsection{Impermanent Loss}\label{sub:IL}
		Similarly to liquidity provision in LOB markets, providing liquidity to AMM-based DEXs involves a trade-off between expected profits and adverse selection risk.
		On the one hand, LPs are compensated by pocketing the transaction fees applied to the trading volume generated by liquidity takers through swaps.
		On the other hand, a permanent price change leads to an \emph{impermanent loss} (IL) for the LP.
		This loss arises from the fact that, gross of fees, providing funds to a liquidity pool is less profitable than simply holding the tokens \citep{loesch2021impermanent}.
		In the following, we derive a mathematical expression for the IL in Uniswap v2, followed by a brief discussion of the same issue in the context of Uniswap v3.

		Consider a liquidity pool on Uniswap v2 for the exchange pair $X\leftrightarrow Y$, containing $x$ and $y$ units of the two tokens at time $t=0$. 
		Assume an LP owns a share $\psi$ of the pool, and the current quoted price is $P = y/x$. 
		At $t=0$, the value of her position in units of $y$ is 
		$ W = \psi (x P + y) = 2 \psi y $.
		Denote the new reserves at $t=1$ as $(x', y')$ and the new price as $P'$, so that the value of her position in units of $y$ changes to
		$ W' = \psi (x' P' + y') = 2 \psi y'$.
		The gross percentage change in the value of the deposited liquidity can therefore be expressed as $ R_{LP} = W'/W = y'/y$.
		Given the constant product rule $xy = k = x'y'$ and the definition of the quoted price, we can write
		$ y = \sqrt{kP} $ and
		$ y' = \sqrt{k P'}\,.$
		Hence, the change in value of the LP’s liquidity position depends solely on the square root of the gross price change $\DP = P'/P$ between $t=0$ and $t=1$:
		$$ R_{LP} = \frac{y'}{y} = \frac{\sqrt{kP'}}{\sqrt{kP}} = \sqrt{\DP}\,.$$

		On the other hand, the gross return $R_H$ from holding the tokens is simply the average of the returns arising from holding each individual token. This equals to $1$ for $Y$ -- the accounting unit -- and $\DP$ for $X$. The total holding return is thus
		$$ R_H = \frac{1}{2}(\DP + 1)\,.$$
		The impermanent loss, that is, the net opportunity cost from providing liquidity in Uniswap v2 instead of holding the tokens, is therefore given by\footnote{
			We define the IL as the \emph{difference} between $R_{LP} $ and $ R_H$, as in \cite{aigner2021uniswap} and \cite{fukasawa2022weighted}.
			The IL can alternatively be defined in percentage terms \citep{ChenDelta2022,heimbach2022risks,heimbach2021behavior} by
			$IL_2 = R_{LP} / R_H - 1 = 2 \sqrt{\DP} / (\DP + 1) -1$.
		}
		\begin{equation}\label{eq:IL2}
			IL_2 = R_H - R_{LP} =  \frac{1}{2}\left(\DP + 1\right) - \sqrt{\DP}\,.
		\end{equation}
		%
		%
		%
		By taking the first order derivative with respect to $\DP$, one can easily see that $IL$ has a global minimum of $0$ for $\DP=1$, while it is strictly positive otherwise.
		Hence $IL$ represents a cost and highlights that, gross of pocketing the fees, LPs providing liquidity are always worse off than token holders.
		We note that $IL$ can be seen as a measure of the level of adverse selection faced by LPs, similar to that faced by market makers in LOB markets.
		In fact, for any given horizon, $IL=0$ if the order flow is uninformed and gives only rise to a temporary price impact ($\DP=1$), 
		while it increases in magnitude in the presence of informed order flow, causing a permanent price change ($\DP\neq1$).
		Intuitively, the provision of liquidity in the AMM framework can be remunerative if the LP provides immediacy primarily to liquidity traders while, at the same time, it may involve net losses when facing a higher fraction of informed traders.
		Quantitatively, \cite{fukasawa2022weighted} show that the IL in constant-product AMMs can be hedged through weighted variance swaps.
		The introduction of the DPR system in Uniswap v3 complicates the calculation of IL for a liquidity position, as it hinges on the position's price range. 
		Although a comprehensive discussion is beyond this paper's scope, we provide a derivation of IL in a special case, assuming liquidity is provided on a single interval centered on the quoted price and that the final price stays in the interval.
		Formally, consider a liquidity position on the interval $[P_a, P_b]$, with virtual reserves $(x,y)$ corresponding to real reserves $(\tilde{x}, \tilde{y})$. 
		Assume the initial quoted price $P_0$ equals the geometric mean of the interval so that the dollar value of the real reserves is balanced.
		It follows that the initial value of the position in units of $Y$ is given by
		$ W = \tilde{x} P + \tilde{y} = 2\tilde{y} $
		and the $t=1$ value resulting from holding the two tokens is
		$ W_{H}' = \tilde{x} P' + \tilde{y} = (1+\Delta P)\tilde{y} . $
		Hence, the return from holding the tokens is $R^H = (1+\Delta P)/2$, as in Uniswap v2.
		Assuming the new price $P'$ does not exit the interval, we can easily find the new real reserves ($\tilde{x}', \tilde{y}'$) and the resulting value of the liquidity position at $t=1$, which is given by
		$ W_{LP}' = \tilde{x}' P' + \tilde{y}'. $
		Applying the constant product rule to virtual reserves and transforming them back to real reserves, using the relation expressed in \eqref{eq:virtual}, we get
		\begin{equation}
			\tilde{x}' = L (1/\sqrt{P'} - 1 / \sqrt{P_b})
			\AND
			\tilde{y}' = L (\sqrt{P'} - \sqrt{P_a})
			.
		\end{equation}
		Similarly, the initial value of the position can be re-written as 
		$ 2\tilde{y}_0 = 2L (\sqrt{P_0}-\sqrt{P_a}) $.
		Hence, with some algebraic manipulations, IL in Uniswap v3 can be expressed as
		\begin{subequations}
		\begin{eqnarray}\label{eq:IL3}
			IL_3 &=& R^{H} - R^{LP} 
			= \frac{(1+\Delta P)}{2} - \frac{\tilde{x}' P' + \tilde{y}'}{2\tilde{y}} \\[1em]
			&=& \frac{(1+\Delta P)}{2} - \frac{ L( 2 \sqrt{P'} - (1+\Delta P)\sqrt{P_a} ) } { 2L(\sqrt{P}-\sqrt{P_a})} \\[1em]
			&=& \frac{\sqrt{P}}{\sqrt{P}-\sqrt{P_a}} \left(\frac{1}{2}\left(\DP + 1\right) - \sqrt{\DP}\right) \\[1.75em]
			&=& \lambda \, IL_2
		\end{eqnarray}
		\end{subequations}
		
		where $\lambda = \sqrt{P}/(\sqrt{P}-\sqrt{P_a})$ can be interpreted as a leverage factor that increases as the interval narrows.
		Note that the impermanent loss in Uniswap v3 converges to that of Uniswap v2 if the interval covers the entire price range and $\lambda \to 1$.
		In summary, the introduction of the DPR system in Uniswap v3 opens the possibility for LPs to take a leveraged liquidity position, where both the earned exchange fees and the IL are increasing in the leverage factor.

\subsection{A Simple Model of Liquidity Provision}\label{sub:liquidity_model}
			We model a marginal LP in Uniswap v2,
			who faces the problem of providing the optimal quantity of liquidity to the exchange pair $X/Y$.
			We assume that the LP is risk-neutral and that the market is perfectly competitive as in, e.g., \cite{glosten1985bid}.
			At time $t=0$ the total liquidity in the pools is equal to $x$, and
			the LP can add or remove a quantity $\xi$ of liquidity.
			At time $t>0$, users start to trade on the pair until the trading stops at $t=1$.
			Let the random variables $V$ denote the total traded volume (in units of $X$)
			and let $\DP = P_1 / P_0$ denote the gross percentage change in the quoted price, respectively, between $t=0$ and $t=1$.
			As discussed in Section \ref{sec:AMM}, the profits and losses of the LP depend on two factors: the fees arising from liquidity takers' trading volume and the impermanent loss due to changes in quoted prices.
			Let $\E[V]$ denote the expected unsigned trading volume and let $\E[IL]$ denote the expected $IL$, 
			both estimated at time $t=0$.\footnote{
				For the sake of simplicity, we take $\E[V]$ as exogenous, abstracting from its potential correlation with volatility and impermanent loss.
			}
			The expected fees paid by liquidity takers amount to the product of the exchange fees $f$ and the expected volume $\E[V]$, expressed in units of $X$.
			Since this amount is distributed by the protocol to the participating LPs on a pro-rata basis, the marginal liquidity provider depositing $\xi$ units of additional liquidity gains in expectation 
			$\frac{\xi}{x+\xi} f\E[V]$ 
			units of $X$, corresponding to a percentage profit of $\frac{f\E[V]}{x+\xi}$.
			Hence, accounting for both fees and the expected impermanent loss, the net expected percentage profit $\E[R]$ from providing an additional amount $\xi$ of liquidity is equal to 
			$$\E[R] = \frac{f}{x+\xi}\E[V] - \E[IL] \,.$$
			The assumption of perfect competition results in zero expected profits for the LP,
			hence the equilibrium level of total liquidity $x^* = \xi + x$ is given by
			\begin{equation}\label{eq:equilibrium}
				x^* = \frac{f \E[V]}{\E[IL]}
			\end{equation}
			showing that total liquidity increases with the expected trading volume and (percentage) exchange fees remunerating the LP, while it decreases with the expected IL.
			The equilibrium condition \eqref{eq:equilibrium} has a clear economic interpretation that is conceptually related to standard microstructure models featuring market makers.
			First, the level of liquidity $x^*$ provided by LPs determines the quoted spread available to traders, as in \eqref{eq:spread_AMM}.
			Second, as noted above, the expected IL can be thought of as a proxy for the level of adverse selection risk faced by LPs.
			Thus \eqref{eq:equilibrium} says that spreads are increasing in the level of adverse selection; in other words, LPs require compensation for the losses caused by informed trading.
			We use daily liquidity data to test the predictions of our model,
			proxying for $\E[V]$ with the rolling average of daily traded volume
			and for $\E[IL]$ with the rolling average of the daily IL, 
			estimated over the previous two weeks.
			We regress daily log values of empirically observed liquidity on the ones predicted by \eqref{eq:equilibrium}, 
			for $100$ exchange pairs from February 2021 to February 2023.
			Results are reported in Table \ref{tab:model_fit_new} and Figure \ref{fig:model_fit_new}, showing a highly significant positive correlation between predicted and observed liquidity levels, with a
			remarkable $R^2$ coefficient equal to or higher than $92\%$.
			The results are robust to the inclusion of pair- and time-fixed effects.
			Our results are also robust to changing the size of the rolling windows used to estimate $\E[IL]$ and $\E[V]$.
			In particular we use 5 days and 20 days, corresponding to the median and average duration of liquidity positions reported in \cite{oneillAMM2022}, and we find that the resulting R-squared for the baseline specification is 91.38\% and 92.50\%, respectively.
			We thus conclude that our partial equilibrium model is empirically relevant, as it is able to capture the main economic trade-off faced by LPs in AMM-based DEXs. 
			At the same time, our findings suggest that LPs behave rationally in the aggregate.

\subsection{Gas Prices and DEX Price Deviations -- VAR Estimate}
	To account for potentially endogenous dynamics of gas-fees and price deviations, we estimate a VAR model with 24 lags featuring DEX price deviations, gas prices, and other controls as endogenous variables. Namely, these are:
		the average gas price on the Ethereum network,
		the DEX average price deviation for the triplet,
		the CEX average price deviation for the triplet,
		the average DEX transaction costs of the three pairs,
		the average CEX-DEX spread of the three pairs,
		the average volatility of the three pairs, and
		the absolute value of ETH/USD return.
	We estimate the model for each triplet separately and verify that, taking the first difference, all variables pass the Augmented Dickey–Fuller test for stationarity.
	Figure \ref{fig:IRF_VAR_gas_dev} displays the cumulative impulse response function, averaged across the triplets, depicting the effect on DEX price deviations of a unitary shock in the gas price.
	The results indicate that a standard deviation shock to the gas price has a positive and statistically significant impact on price deviations, which dissipates after five to eight hours.

\clearpage\newpage	
\subsection{Tables and Figures}

			\begin{table}[ht]
				\centering
				\resizebox{0.93\textwidth}{!}{\begin{tabular}{llrrrrr}
Pair &  & $\qquad$ 1-1,000 \$ & 1,000-10,000 \$ & 10,000-100,000 \$ & 100,000-1,000,000 \$ & Above 1,000,000 \$ \\
\toprule
\multirow[t]{2}{*}{AAVE-ETH} 
& CEX & 92.42 & 7.38 & 0.19 & 0.00 & 0.00 \\
& DEX & 31.50 & 34.31 & 32.14 & 2.02 & 0.03 \\[0.25em]
\multirow[t]{2}{*}{BAT-ETH} 
& CEX & 93.05 & 6.75 & 0.19 & 0.00 & 0.00 \\
& DEX & 37.60 & 41.95 & 19.71 & 0.74 & 0.00 \\[0.25em]
\multirow[t]{2}{*}{BTC-DAI} 
& CEX & 70.24 & 28.36 & 1.39 & 0.00 & 0.00 \\
& DEX & 57.32 & 33.26 & 9.11 & 0.32 & 0.00 \\[0.25em]
\multirow[t]{2}{*}{BTC-USDC} 
& CEX & 74.91 & 23.46 & 1.63 & 0.01 & 0.00 \\
& DEX & 17.45 & 28.56 & 38.97 & 14.84 & 0.17 \\[0.25em]
\multirow[t]{2}{*}{CRV-ETH} 
& CEX & 93.87 & 5.97 & 0.15 & 0.00 & 0.00 \\
& DEX & 31.89 & 42.73 & 24.09 & 1.26 & 0.03 \\[0.25em]
\multirow[t]{2}{*}{DAI-ETH} 
& CEX & 81.71 & 17.62 & 0.66 & 0.00 & 0.00 \\
& DEX & 29.31 & 32.93 & 30.94 & 6.64 & 0.18 \\[0.25em]
\multirow[t]{2}{*}{DAI-USDT} 
& CEX & 75.16 & 22.81 & 1.98 & 0.05 & 0.00 \\
& DEX & 35.44 & 40.05 & 20.25 & 4.06 & 0.21 \\[0.25em]
\multirow[t]{2}{*}{ETH-BTC} 
& CEX & 77.53 & 20.97 & 1.48 & 0.02 & 0.00 \\
& DEX & 19.09 & 21.50 & 40.13 & 18.45 & 0.83 \\[0.25em]
\multirow[t]{2}{*}{ETH-USDC} 
& CEX & 78.79 & 19.77 & 1.43 & 0.01 & 0.00 \\
& DEX & 32.25 & 33.77 & 21.00 & 12.24 & 0.74 \\[0.25em]
\multirow[t]{2}{*}{ETH-USDT} 
& CEX & 74.40 & 23.49 & 2.07 & 0.03 & 0.00 \\
& DEX & 35.81 & 39.09 & 20.17 & 4.81 & 0.12 \\[0.25em]
\multirow[t]{2}{*}{GRT-ETH} 
& CEX & 92.28 & 7.63 & 0.10 & 0.00 & 0.00 \\
& DEX & 43.17 & 45.88 & 10.82 & 0.13 & 0.00 \\[0.25em]
\multirow[t]{2}{*}{KNC-ETH} 
& CEX & 94.01 & 5.83 & 0.16 & 0.00 & 0.00 \\
& DEX & 37.19 & 52.96 & 9.81 & 0.05 & 0.00 \\[0.25em]
\multirow[t]{2}{*}{LINK-ETH} 
& CEX & 91.55 & 8.20 & 0.25 & 0.00 & 0.00 \\
& DEX & 22.66 & 31.93 & 40.67 & 4.71 & 0.03 \\[0.25em]
\multirow[t]{2}{*}{MANA-ETH} 
& CEX & 92.36 & 7.32 & 0.32 & 0.00 & 0.00 \\
& DEX & 36.15 & 41.39 & 21.93 & 0.52 & 0.00 \\[0.25em]
\multirow[t]{2}{*}{OMG-ETH} 
& CEX & 90.31 & 9.43 & 0.26 & 0.00 & 0.00 \\
& DEX & 46.66 & 47.44 & 5.81 & 0.08 & 0.01 \\[0.25em]
\multirow[t]{2}{*}{REP-ETH} 
& CEX & 91.18 & 8.76 & 0.07 & 0.00 & 0.00 \\
& DEX & 63.29 & 36.35 & 0.16 & 0.20 & 0.00 \\[0.25em]
\multirow[t]{2}{*}{SNX-ETH} 
& CEX & 92.12 & 7.65 & 0.23 & 0.00 & 0.00 \\
& DEX & 27.31 & 57.94 & 14.34 & 0.41 & 0.00 \\[0.25em]
\multirow[t]{2}{*}{STORJ-ETH} 
& CEX & 91.54 & 8.43 & 0.03 & 0.00 & 0.00 \\
& DEX & 42.31 & 51.53 & 6.16 & 0.01 & 0.00 \\[0.25em]
\multirow[t]{2}{*}{UNI-ETH} 
& CEX & 94.02 & 5.73 & 0.25 & 0.00 & 0.00 \\
& DEX & 30.82 & 29.23 & 35.21 & 4.68 & 0.06 \\[0.25em]
\multirow[t]{2}{*}{USDC-USDT} 
& CEX & 79.90 & 17.62 & 2.27 & 0.20 & 0.01 \\
& DEX & 28.04 & 38.25 & 24.00 & 8.39 & 1.31 \\[0.25em]
\bottomrule
\end{tabular}
}
				\captionsetup{justification=justified, singlelinecheck=on, font=footnotesize}
				\caption{\textbf{Distribution of Trade Size.}\footnotesizelarge 
		The table reports the frequency distribution of trade sizes in US Dollars for transactions executed on CEXs and DEXs during our sample period, for the twenty pairs featured in our study.
		For CEXs, we pool transactions from all the five CEX exchanges, 
		while for DEXs we pool transactions from Uniswap v2 and v3.
		Transactions with a Dollar value below $0.1$ are excluded, and gas costs are not included.
		The data for CEX is retrieved from tardis,
		while for DEXs it is retrieved from a Dune Analytics query based on the `\lstinline{uniswap_v2_ethereum.trades}' and `\lstinline{uniswap_v3_ethereum.trades}' tables.
	}\label{tab:trade_sizes_by_pair}
			\end{table}

	%
			\begin{table}[ht]
				\centering
				\resizebox{0.93\textwidth}{!}{\begin{tabular}{lcccc}
\toprule
{} & (1) & (2) & (3) & (4) \\
\midrule
Dep. Variable           & $\qquad\;$Volume$\qquad\;$         & $\qquad\;$Volume$\qquad\;$         
                        & $\qquad\;$log(Volume)$\qquad\;$    & $\qquad\;$log(Volume)$\qquad\;$    \\
&  &  &  &  \\
DEX $	\times$ FTX     & 6.903**    & 6.903**    & 0.421***   & 0.421***   \\
                        & (2.421)    & (2.426)    & (2.629)    & (2.635)    \\[1em]
DEX                     & -19.499*** & -19.499*** & -2.966***  & -2.966***  \\
                        & (-7.401)   & (-7.417)   & (-21.700)  & (-21.746)  \\[1em]
FTX                     & -6.903**   & -7.578***  & -0.421***  & -0.426***  \\
                        & (-2.421)   & (-5.327)   & (-2.629)   & (-5.328)   \\[1em]
Intercept               & 19.499***  &            & 2.966***   &            \\
                        & (7.401)    &            & (21.700)   &            \\
                        &            &            &            &            \\
Observations            & 236        & 236        & 236        & 236        \\
$\text{R}^2$            & 0.803      & 0.837      & 0.969      & 0.975      \\
Month FE                & No         & Yes        & No         & Yes        \\
\bottomrule
\end{tabular}
}
				\captionsetup{justification=justified, singlelinecheck=on, font=footnotesize}
				\caption{\textbf{FTX Event Study.}\footnotesizelarge 
		The table presents findings from difference-in-differences regression analyses of the trading activity on CEX and DEX exchanges around the collapse of the FTX exchange on November 10th, 2022.
		The sample is a panel at the day-exchange-type level, encompassing a period of two months before and after the event, ranging from September 9th, 2022 to January 9th, 2023.
		The dependent variable is the aggregate trading volume of CEXs (\CEXs) or DEXs (Uniswap v2 and v3), sampled at the daily frequency.
		This is regressed on 
		a treatment dummy (DEX) indicating DEXs, 
		a time dummy (FTX) indicating the period after the FTX collapse,
		and their interaction.
		In the first two specifications, the trading volume is expressed in billion US Dollars, while the next two specifications employ the logarithm of the trading volume as the dependent variable.
		In specifications (2) and (4) we saturate the regression model with month fixed effects.
		T-stats are adjusted for time-series auto-correlation and are reported in parentheses. 
		Asterisks denote significance levels (***$ =1\%$, **$ =5\%$, *$ =10\%$).
	}\label{tab:FTX_DD}
			\end{table}
		
	\clearpage\newpage

			\begin{table}[ht]
				\centering
				\resizebox{0.93\textwidth}{!}{\begin{tabular}{lcccc}
\toprule
{} &        (1) &        (2) &        (3) &        (4) \\
\midrule
Dependent Variable      &     Log(Liquidity) &     Log(Liquidity) &     Log(Liquidity) &     Log(Liquidity) \\
                        &            &            &            &            \\
Log(Predicted Liquidity)&    0.89*** & 0.61***     & 0.89***    & 0.53***         \\
                        &    (29.94)& (12.39)   & (30.14)   & (10.24) \\
\\
Constant                &    5.92*** &            &            &            \\
                        &    (21.71) &            &            &            \\
                        &            &            &            &            \\
Observations            & 42,293     & 42,293     & 42,293     & 42,293     \\
R-squared               & 0.92       & 0.97       & 0.92       & 0.98       \\
Pair Fixed Effects      &          - &        Yes &          - &        Yes \\
Date Fixed Effects      &          - &          - &        Yes &        Yes \\
\bottomrule
\end{tabular}}
				\captionsetup{justification=justified, singlelinecheck=on, font=footnotesize}
				\caption{\textbf{Model Fit.}\footnotesizelarge 
		The table reports results from a panel regression of
		log-observed liquidity levels
		onto log liquidity levels predicted by the equilibrium model outlined in Section \ref{sub:liquidity_model} and computed as in equation \eqref{eq:equilibrium}.
		Both the dependent and independent variables are computed at the pair-day level for $100$ exchange pairs from February 2021 to February 2022.
		We saturate the regression model with day- and pair fixed effects.
		T-stats are reported in parentheses, based on robust standard errors double-clustered at the pair- and day-level. 
		Asterisks denote significance levels (***$ =1\%$, **$ =5\%$, *$ =10\%$).
	}\label{tab:model_fit_new}
			\end{table}

	\vspace{5em}
	%
	\figurecap{FTX Bankruptcy}{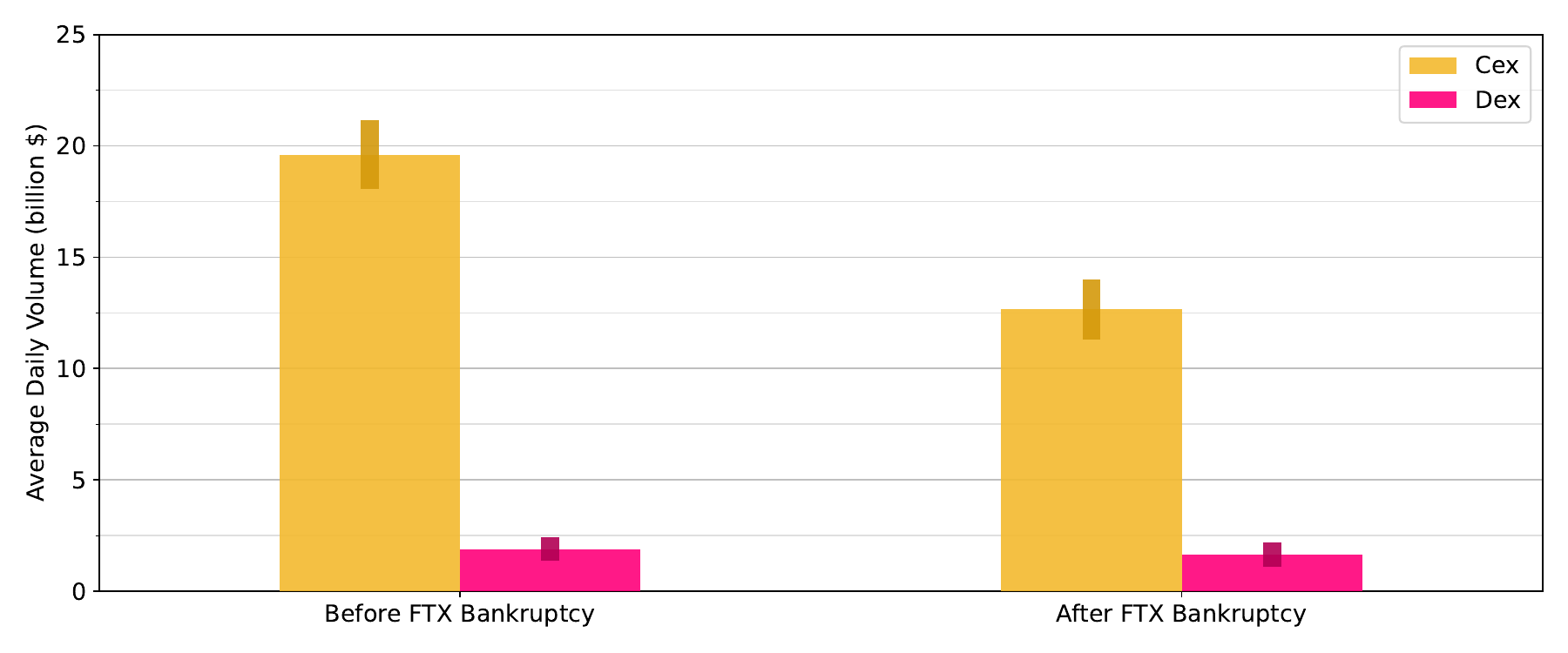}{0.95}{
		The figure shows the average daily volume of CEXs and DEXs during the two-month periods preceding and following the FTX bankruptcy, filed on November 10th, 2022. The vertical lines represent 95\% confidence intervals.
			The plot shows that CEX average daily volume decreased significantly by 35\%, from 16.69 to 10.86 million USD.
			On the contrary, the 13\% decrease in DEX daily volume from 1.88 to 1.63 million USD is not statistically significant.
	}

	%
	\figurecaph{Transaction Costs (Excluding Gas Fees)}{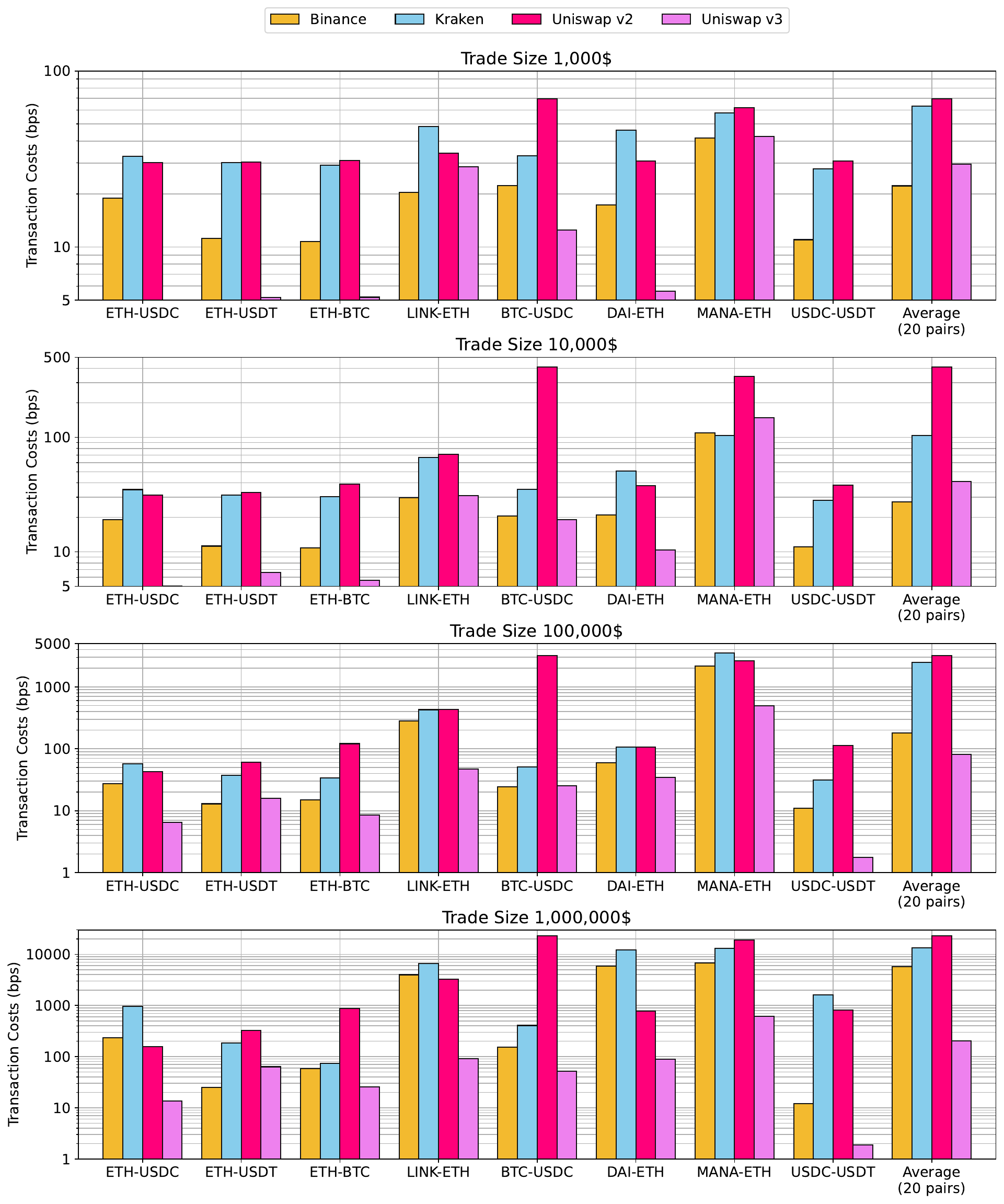}{1}{
		The figure presents average transaction costs, similar to Figure \ref{fig:TC_all_coins}. 
		For Uniswap v2 and Uniswap v3, these are based on equation \eqref{eq:TC_AMM}, but setting the gas fees component equal to zero.
		For Binance and Kraken, instead, the figures are computed as on equation \eqref{eq:TC_LOB}.
		The transaction costs are computed at the hourly frequency for the five pairs in our sample and for different trade sizes ($10^3, 10^4, 10^5$, and $10^6$ US dollars), then averaged from February 2021 to February 2023.
		For Uniswap v3, daily and for each trade size, we consider the liquidity pool offering the lowest transaction costs among those available for the specific exchange pair.
		The vertical axis employs a logarithmic scale and is reported in basis points.
	}

	%
	\figurecaph{Transaction Costs (All Exchanges)}{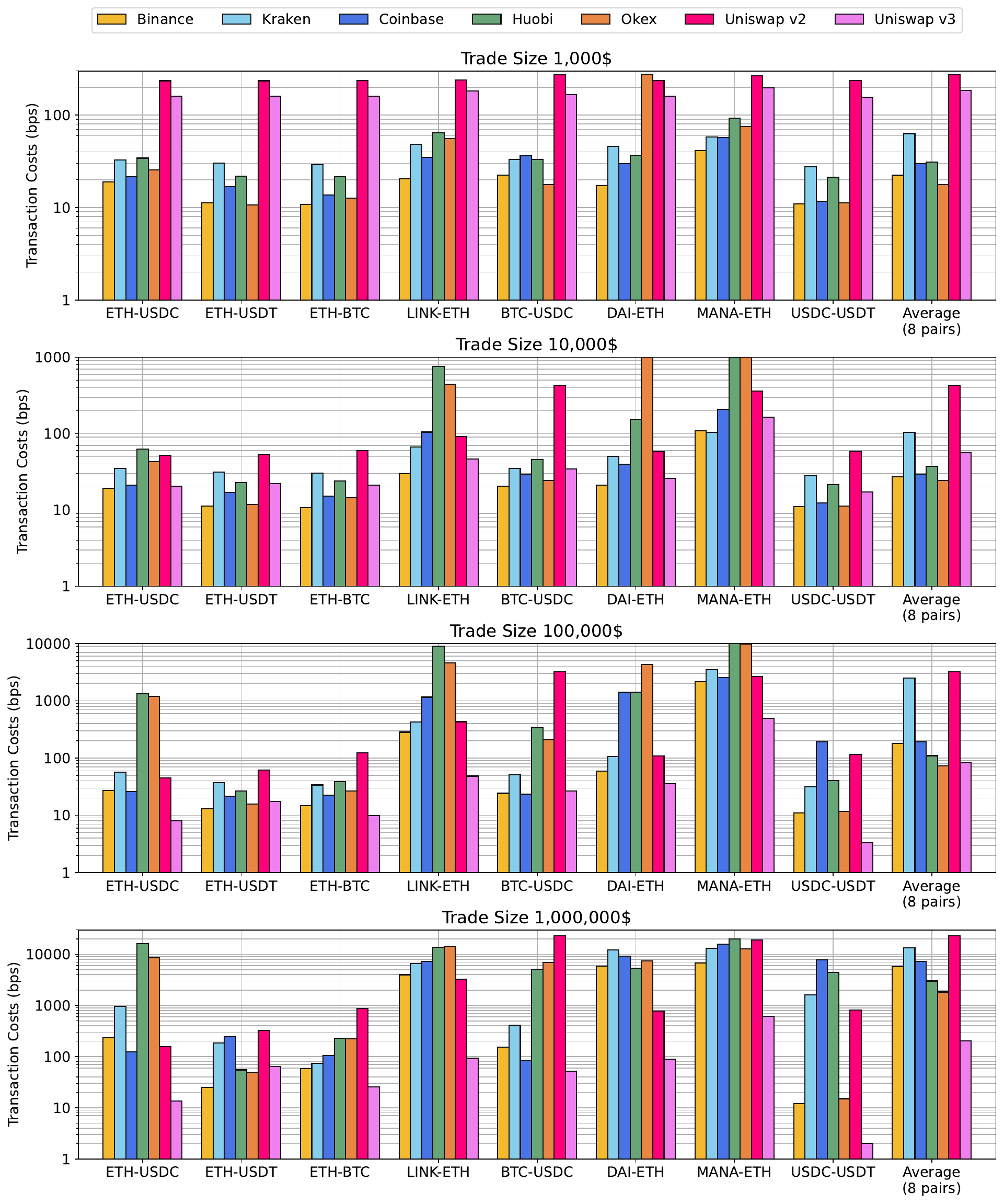}{1}{
		The figure presents average transaction costs, similar to Figure \ref{fig:TC_all_coins}. 
		The transaction costs are computed at the hourly frequency for the five pairs in our sample and for different trade sizes ($10^3, 10^4, 10^5$, and $10^6$ US dollars), then averaged from February 2021 to February 2023.
		For Uniswap v3, daily and for each trade size, we consider the liquidity pool offering the lowest transaction costs among those available for the specific exchange pair.
		The vertical axis employs a logarithmic scale and is reported in basis points.
	}

	%
	\figurecaph{Transaction Costs (Uniswap v3 Period)}{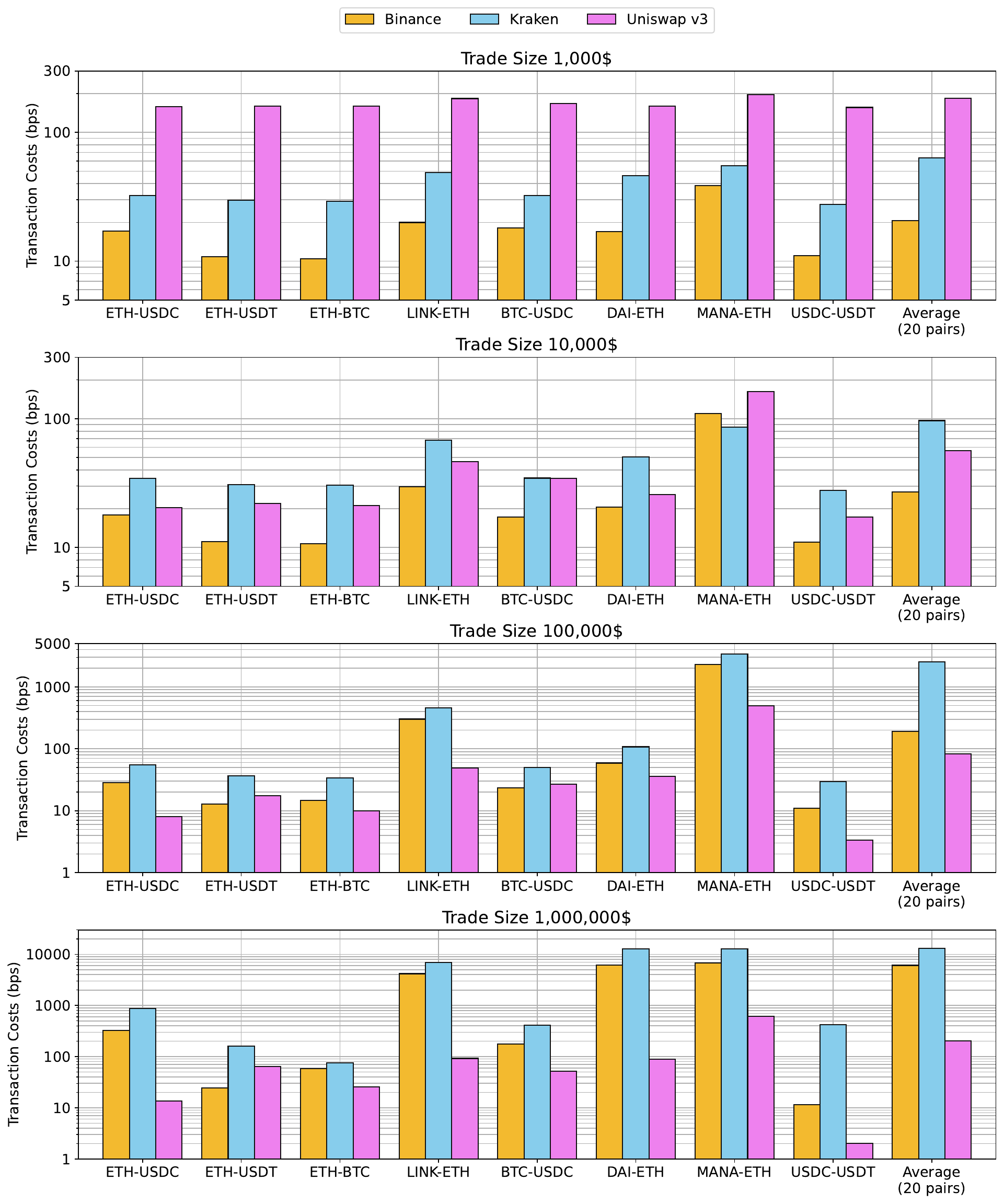}{1}{
		The figure is similar to Figure \ref{fig:TC_all_coins}, but based on the subsample in which Uniswap v3 is present.
		It presents transaction costs, computed as in equation \eqref{eq:TC_LOB} for the LOB-based Binance and Kraken and on equation \eqref{eq:TC_AMM} for the AMM-based Uniswap v3.
		The transaction costs are computed at the hourly frequency for the five pairs in our sample and for different trade sizes ($10^3, 10^4, 10^5$, and $10^6$ US dollars), then averaged from May, 2021 to February, 2023.
		As discussed in Section \ref{sec:TCs}, the displayed transaction costs include B/A spreads, exchange fees, and settlement fees (gas fees for Uniswap v3, and deposit and withdraw fees for the two CEXs).
		For Uniswap v3, daily and for each trade size, we consider the liquidity pool offering the lowest transaction costs among those available for the specific exchange pair.
		The vertical axis employs a logarithmic scale and is reported in basis points.
	}
	

	\figurecap{Model Fit}{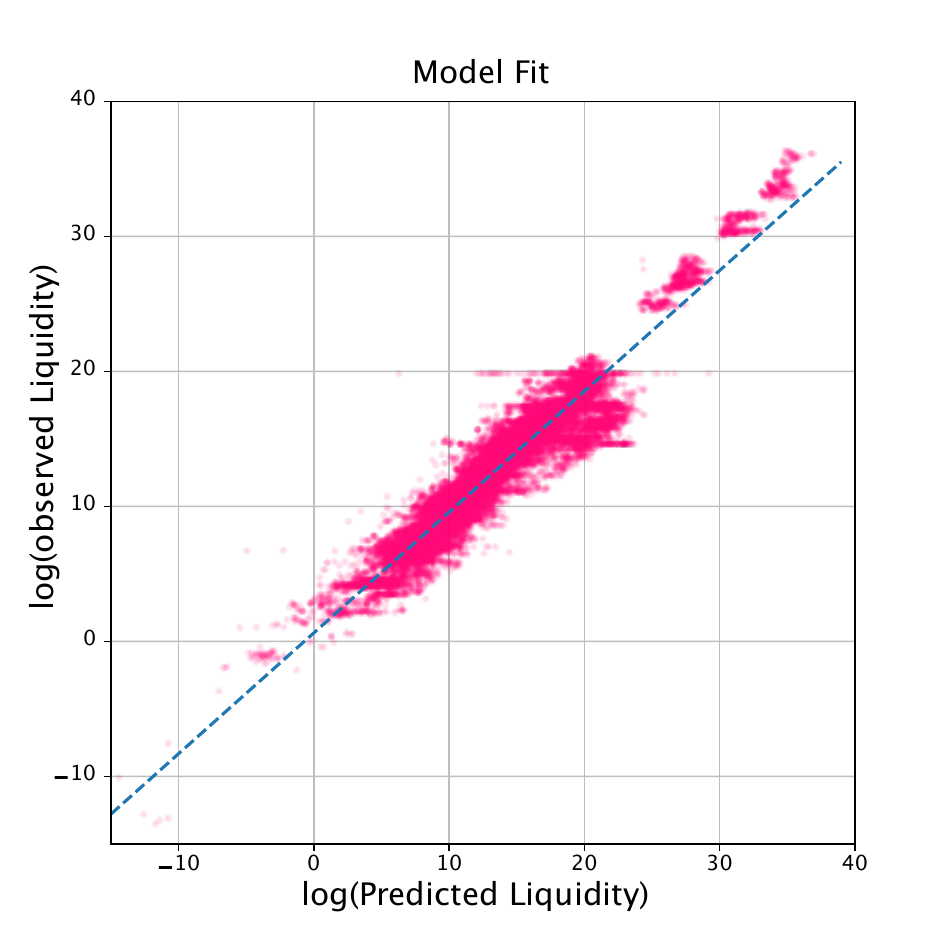}{0.7}{
		The figure presents a scatter plot of 
		The observed levels of liquidity (y-axis) 
		and those predicted by our model and computed as in equation \eqref{eq:equilibrium} (x-axis),
		based on $42,299$ daily observations of $100$ exchange pairs quoted in Uniswap v2. 
		from February 2021 to February 2022.
	}

	\figurecaph{The Impact of Gas Prices on DEX Price Deviations}{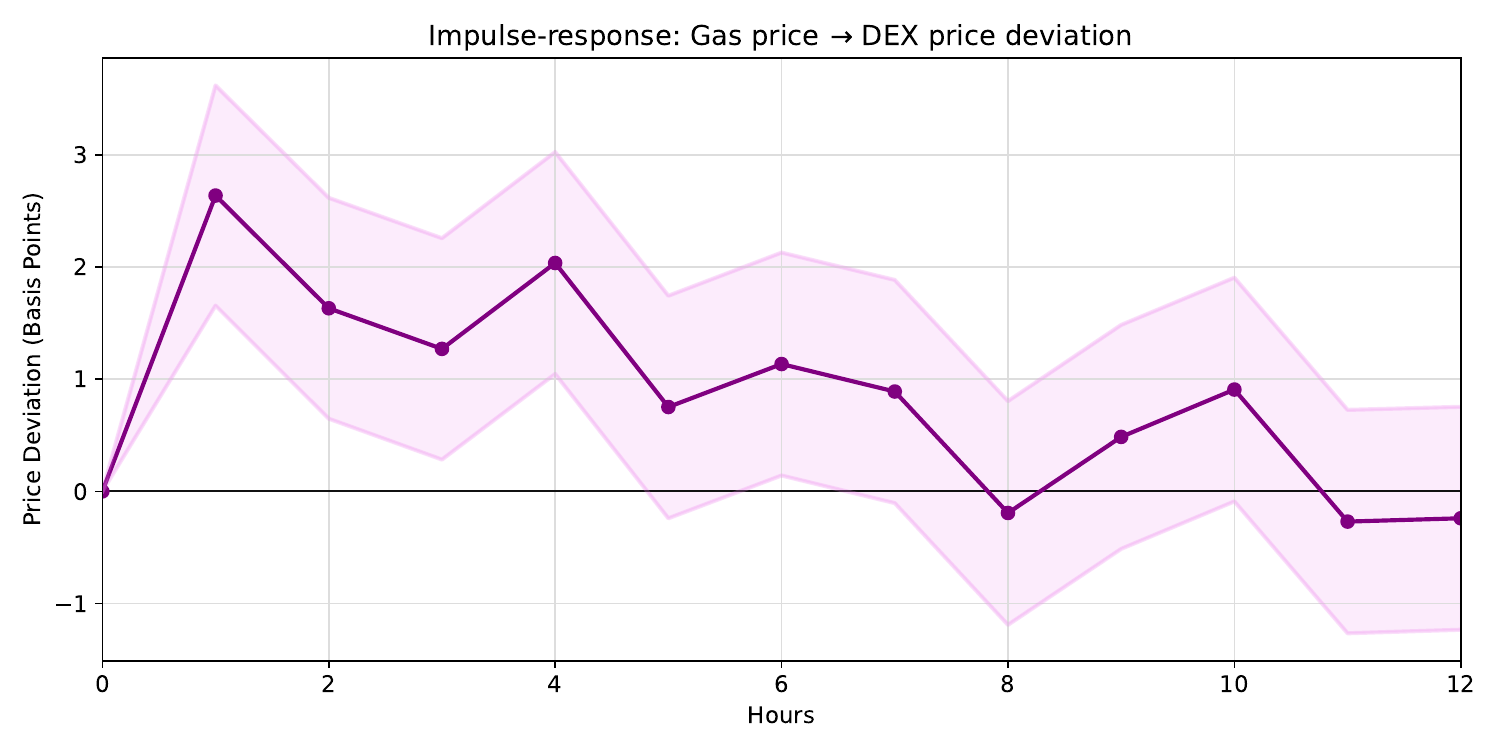}{1}{
		The figure presents the cumulative impulse-response function (IRF) of gas prices on DEX price deviations, with 95\% confidence intervals, indicating the response of DEX price deviations (in basis points) to a one standard deviation increase in gas prices on the Ethereum network.
		To obtain the IRF, for each triplet we fit a VAR model with 24 lags at the hourly frequency with the following endogenous variables: 
		the average gas price on the Ethereum network,
		the DEX average price deviation for the triplet,
		the CEX average price deviation for the triplet,
		the average DEX transaction costs of the three pairs,
		the average CEX-DEX spread of the three pairs,
		the average volatility of the three pairs, and
		the absolute value of ETH/USD return.
		Finally, we take the median of the resulting IRFs and the corresponding confidence intervals across the five triplets.
	}

\end{document}